\documentclass{bmcart}

\usepackage{soul}
\usepackage[version=3]{mhchem}
\usepackage{amsthm,amsmath}
\usepackage{physics}
\usepackage{graphicx}
\usepackage[utf8]{inputenc} 

\startlocaldefs
\endlocaldefs

\begin{document}

\begin{frontmatter}

\begin{fmbox}
\dochead{Research}


\title{VQE Method: A Short Survey and Recent Developments}


\author[
  addressref={aff1, aff3},             
  corref={aff3},                       
  email={fedorov@anl.gov}              
]{\inits{D.A.F.}\fnm{Dmitry A.} \snm{Fedorov}}
\author[
  addressref={aff2},
  corref={aff2},
  email={peng398@pnnl.gov}
]{\inits{B.P.}\fnm{Bo} \snm{Peng}}
\author[
  addressref={aff2},
  corref={aff2},
  email={niri.govind@pnnl.gov}
]{\inits{N.G.}\fnm{Niranjan} \snm{Govind}}
\author[
  addressref={aff3},
  corref={aff3},
  email={yuri@anl.gov}
]{\inits{Y.A.}\fnm{Yuri} \snm{Alexeev}}


\address[id=aff1]{%
  \orgdiv{},
  \orgname{Oak Ridge Associated Universities},
  \city{Oak Ridge},
  \cny{TN 37830}
}
\address[id=aff2]{
  \orgdiv{Physical and Computational Sciences Directorate},             
  \orgname{Pacific Northwest National Laboratory},          
  \city{Richland},                              
  \cny{WA 99354}                                    
}
\address[id=aff3]{
  \orgdiv{Computational Science Division},             
  \orgname{Argonne National Laboratory},          
  \city{Lemont},                              
  \cny{IL 60439}                                    
}



\end{fmbox}


\begin{abstractbox}

\begin{abstract} 
The variational quantum eigensolver (VQE) is a method that uses a hybrid quantum-classical computational approach to find eigenvalues  of a Hamiltonian. VQE has been proposed as an alternative to fully quantum algorithms such as quantum phase estimation because (QPE) fully quantum algorithms require quantum hardware that will not be accessible in the near future. VQE has been successfully applied to solve the electronic Schr\"{o}dinger equation for a variety of small molecules. However, the scalability of this method is limited by two factors: the complexity of the quantum circuits and the complexity of the classical optimization problem. Both of these factors are affected by the choice of the variational ansatz used to represent the trial wave function. Hence, the construction of an efficient ansatz is an active area of research. Put another way, modern quantum computers are not capable of executing deep quantum circuits produced by using currently available ansatzes for problems that map onto more than several qubits. 
In this review, we present recent developments in the field of designing efficient ansatzes that fall into two categories---chemistry--inspired and hardware--efficient---that produce quantum circuits that are easier to run on modern hardware. We discuss the shortfalls of ansatzes originally formulated for VQE simulations, how they are addressed in more sophisticated methods, and the potential ways for further improvements.


\end{abstract}




\end{abstractbox}
%

\end{frontmatter}




\section*{Introduction}

Quantum simulation of chemistry and materials is an important application for classical computing. For example, one can predict the rates of chemical reactions, determine molecular structure, and the properties of materials and molecules. It is achieved through the solution of the electronic structure Hamiltonian, which allows one to describe the properties of interacting electrons in the presence of stationary nuclei. The advent of quantum computing holds promise for significantly speeding up these calculations, which currently are done classically. This idea stems from the Feynman's postulate in the famous paper ``Simulating Physics with Computers,” published in 1982 \cite{feynman1982simulating}, that to simulate quantum systems, one would need to build quantum computers to perform quantum computations, and it is much more efficient than doing it classically for correlated systems. 
Indeed, one cannot simply scale the calculations on classical computers or apply massive parallelism because of the exponential growth of computational requirements with the size of a chemical system and complexity.

These ideas look attractive in theory, but in reality, there are severe limitations in the computational capabilities of the currently available small quantum devices, which are often referred to as NISQ (noisy intermediate-scale quantum) devices \cite{preskill2018quantum}. A second important aspect is the  inefficiency of existing quantum algorithms in terms of the resources that are needed to solve any useful problem  on a quantum computer faster than on a classical computer, which is often discussed in terms of the quantum advantage. For example, the resources estimated to perform chromium dimer calculations, which is a large enough molecular system to demonstrate the quantum advantage on a quantum computer, with existing algorithms require at least 1 million physical qubits \cite{elfving2020will,liu2021prospects} to make 60 virtual high-fidelity qubits. At the same time, these physical qubits need higher fidelity than do existing ones for error correction algorithms to work. This is far from the currently available 71 physical qubit quantum computers and will remain the case for the near future. 

Another attractive idea is to use quantum computing on quantum devices to improve quantum technology. Such a use will, in theory, allow systematic improvement of the technology. In fact, it is one of the research objectives of the national quantum center Q-NEXT \cite{qnext}, which is tasked with exploring the potential of this approach and developing new generations of methods.

The first algorithm that was proposed to solve the Schr\"{o}dinger equation on a quantum computer was the quantum phase estimation (QPE).\cite{kitaev_QPE_1995,QPE_Lloyd,QPE_Abrams_1997} It is a fully quantum algorithm that can extract the phase or eigenvalues of a unitary operator. By using the phase kickback trick and inverse quantum Fourier transform (IQFT),\cite{QFT} QPE obtains the binary representation of the phase or eigenvalue.  The QPE algorithm is versatile and is a part of other quantum algorithms, such as Shor's algorithm. QPE can achieve exponential speedup in finding the eigenspectrum of unitary operators, such as the electronic Hamiltonian, as long as an appropriate trial state with nonzero overlap with the real solution can be prepared. It is likely that quantum advantage can be demonstrated by using the QPE method when the first big enough fault-tolerant quantum computer is built.\cite{aharonov_1996} However, the problem with QPE, and IQFT in particular, is that it requires millions of qubits and gates even for relatively small systems, a requirement far beyond the capabilities of present NISQ hardware. 

To address this problem, a variational hybrid quantum method VQE was proposed~\cite{Peruzzo2014_VQE,McClean_2016,Romero_2018}. VQE is designed to utilize both quantum and classical resources to find variational solutions to eigenvalue problems. In a typical setup, the ground state trial wave function of the molecule is constructed from operators generating single- and double-excitation configurations from a Hartree-Fock wave function precomputed on a classical computer (UCCSD~ \cite{Pal1984use,Hoffmann1988,Kutzelnigg1991,Taube2006, Sur_2008,Cooper2010,Harsha2018,Evangelista2019} 
ansatz). Next, on a quantum computer, the trial wave function is prepared and the expectation value of the Hamiltonian is measured. Then, the parameters of the trial wave function are optimized iteratively on a classical computer using the variational principle. Although VQE simulations for small molecules have been performed on various quantum hardware architectures~\cite{Peruzzo2014_VQE, kandala_he_ansatz, OMalley_2016,nam2019groundstate,colless_2018,qcc,McCaskey2019,hf_on_qc_google,hempel2018, shen2017, Santagatieaap9646, gao2019, Gao2020, smart_mazziotti_exp2019}, major advances are needed to scale this approach to larger systems. VQE also suffers from the large size of the resulting quantum circuits (albeit not as large as QPE circuits) coupled with the need to perform a classical optimization on a large number of variational parameters, which can render calculations intractable. 

Trying to solve the electronic Schr\"{o}dinger equation on a quantum computer translates into massive quantum circuits, which are beyond the reach of NISQ computers. In the rest of the paper, we will look in detail at different approaches, which aim to reduce the size of circuits while maintaining high accuracy. Our paper is not a comprehensive review of all VQE methods. Several comprehensive reviews covering VQE have been published recently.\cite{quant_comp_chem_revmodphys, quant_chem_chemrev_2019, VQA_rev_2020, bharti2021noisy} The goal of this work is to present recent advances in methods that produce shorter VQE circuits or provide higher accuracy. In addition, we discuss these methods in more detail from a perspective of how they improve on the standard UCCSD and hardware--efficient ansatzes. 

The rest of the paper is organized as follows. In Section 2 we provide a brief description of the VQE method, required for understanding the rest of the paper, and introduce the concepts of chemistry-inspired and hardware-efficient ansatzes. In Section 3 we discuss the chemically inspired ansatzes for VQE simulations. Section 4 is devoted to hardware--efficient ansatzes. In Section 5 the conclusions are presented.

\section*{Introduction to VQE}

The VQE method was introduced to mitigate the significant hardware demands needed by the QPE approach on NISQ devices. VQE is a hybrid quantum-classical algorithm, where the computational workload is shared between the classical and quantum components of the hardware. It starts with a reasonable assumption about the form of the target wave function. The most common choice is to represent a wave function in a basis of atom-centered Gaussian basis functions. However, plane wave basis sets~\cite{plane_waves} can be used as well as the recently proposed ``basis-set free'' approach~\cite{basis_set_free}. A trial wave function or ansatz is constructed with adjustable parameters, followed by the design of a quantum circuit capable of realizing this ansatz. The ansatz parameters are then variationally adjusted until the expectation value of the electronic Hamiltonian 
\begin{equation}\label{ham_exp_value}
    E \leq \frac{\bra{\psi(\vec{\theta})} \hat{H}_{el} \ket{\psi(\vec{\theta})}}{\bra{\psi(\vec{\theta})}\ket{\psi(\vec{\theta})}} 
\end{equation}
is minimized.  In equation \ref{ham_exp_value} $\ket{\psi(\vec{\theta})}$ is the trial wave function that depends on the vector of variational parameters $\vec{\theta}$; $E$ is the ground state energy of $H_{el}$, an electronic Hamiltonian most commonly written in the second quantized form, although the first quantization representation has also been considered\cite{babbush_2019_first_quant}. In this work, we focus on the second quantized form of the Hamiltonian
\begin{equation}\label{el_hamiltonian}
    H_{el}=\sum_{p,q}h_{pq}a_p^\dagger a_q + \frac{1}{2}\sum_{p,q,r,s}h_{pqrs}a_p^\dagger a_q^\dagger a_r a_s ,
\end{equation}
where $a_p^{\dagger}$ and $a_p$ are fermionic creation and annihilation operators, which excite and deexcite electrons from orbital $p$. The first term in equation \ref{el_hamiltonian} corresponds to single-electron excitations, and the second term corresponds to two-electron excitations; $h_{pq}$ and $h_{pqrs}$ are one- and two-electron integrals that are easily computed on a classical computer. To evaluate the energy on a quantum computer, the Hamiltonian for indistinguishable fermions has to be mapped onto the Hamiltonian of distinguishable qubits by using one of three common mappings: Jordan--Wigner~\cite{Jordan1928} parity~\cite{bravyi2017_parity}, or Bravyi--Kitaev\cite{Bravyi2002_BK_mapping}. Regardless of the mapping choice, the resulting qubit Hamiltonian can be written as
\begin{equation}\label{qubit_hamiltonian}
    H=\sum_j\alpha_jP_j=\sum_j\alpha_j\prod_i\sigma_i^j ,
\end{equation}
where $\alpha_j$ are real scalar coefficients that depend on $h_{pq}$ and $h_{pqrs}$. $P_j$ are Pauli strings represented by a product of Pauli matrices $\sigma_i^j$, where $i$ denotes which qubit the Pauli operator acts on and $j$ denotes the term of the Hamiltonian. After the qubit Hamiltonian is prepared on a classical computer and the ansatz to represent the wave function is chosen, the trial wave function $\ket{\psi(\theta)}$ is prepared on a quantum computer. Then the quantum computer is used to measure the energy:
\begin{equation}\label{e_final}
    E(\vec{\theta})=\sum_j^N\alpha_j \bra{\psi(\vec{\theta})}\prod_i\sigma_i^j\ket{\psi(\vec{\theta})},
\end{equation}
where $N$ is the number of terms in the Hamiltonian and $\vec{\theta}$ is the vector of variational parameters. Depending on the chosen basis set and mapping type, the Hamiltonian can contain up to $M^4$ terms, where $M$ is the number of basis functions. Since these terms represented by the Pauli string operators $P_j$  are non-commutative in general, the state preparation step has to be performed repeatedly for each term that is measured separately. In addition, all the individual terms have to be measured enough times to build up sufficient expectation value statistics. This way of computing energy is known as Hamiltonian averaging~\cite{ham_averaging_McClean_2014}. Thus, the VQE approach trades long circuit depths typically found in QPE for shorter state preparation circuits, at the expense of a greater number of measurements, which scales as $O(\frac{1}{\epsilon^2})$, where $\epsilon$ is the desired precision. QPE converges quadratically faster with scaling of $O(\frac{1}{\epsilon})$~\cite{qpe_Wiebe}. 

Shorter circuits make VQE more amenable to NISQ devices. Despite these advantages, however, the performance of the algorithm depends largely on the quality and flexibility of the trial ansatz. The approaches for ansatz design can be divided into three categories. Chemistry-inspired ansatzes are designed by using the domain knowledge from traditional quantum chemistry in a way that every term in the ansatz describes a certain electron configuration. So-called hardware-efficient ansatzes are constructed from a limited set of gates that are easy to implement on quantum hardware, but a chemical interpretation of each term is not generally possible. The third kind lies between the chemistry-inspired and hardware-efficient and is called Hamiltonian variational ansatz~\cite{ham_var_ansatz_Wecker2015}. In the present work, we focus on the first two kinds.

The first VQE experiment by Peruzzo and co-workers~\cite{Peruzzo2014_VQE} utilized the commonly used unitary coupled cluster with singles and doubles (UCCSD) ansatz, which is a chemistry-inspired ansatz and represents a unitary version of the classical non-unitary CCSD method. The UCCSD trial state is prepared from the initial state $\ket{\phi}$, usually, a Hartree--Fock mean-field wave function, by applying the exponentiated excitation operator $U(\vec{\theta})=e^{\hat{T}-\hat{T}^{\dagger}}$:
\begin{equation}\label{uccsd_exc_op}
 \ket{\psi(\vec{\theta})}=e^{\hat{T}-\hat{T}^{\dagger}}\ket{\phi},
\end{equation}
where the excitation operator $\hat{T}=\sum_i\hat{T}_i$ is truncated at excitation level $i$. Truncation at $i=2$ yields the UCCSD ansatz, which includes single and double excitations:
    \begin{equation} \label{eq_uccsd}
    \hat{T}_{UCCSD}=\hat{T}_1 + \hat{T}_2= \sum_{i\in{virt},\alpha\in{occ}}t_i^\alpha\hat{a}^{\dagger}_i\hat{a}_\alpha+ \sum_{i,j\in{virt},\alpha,\beta\in{occ}}t_{ij}^{\alpha\beta}\hat{a}^{\dagger}_i\hat{a}^{\dagger}_j\hat{a}_\beta \hat{a}_\alpha .
    \end{equation}
In equation \ref{eq_uccsd}, $t_{i\alpha}$ and $t_{ij\alpha\beta}$ are cluster amplitudes,  $occ$ denotes orbitals that are occupied in the reference Hartree--Fock state, and $virt$ denotes the virtual (unoccupied) orbitals. The standard version of UCCSD has unfavorable scaling of the number of gates required for implementation with an increasing number of electrons and spin orbitals~\cite{kuhn_UCCSD_resources}. This scaling is a result of many terms with near-zero contributions to the correlation energy. In this work, we consider some of the methods designed to improve that scaling by choosing excitation operators that constitute the ansatz more efficiently.

Hardware-efficient ansatzes are composed of repeated, dense blocks of a limited selection of parametrized gates that are easy to implement with the available hardware. The main idea behind this approach is to build a trial state that is flexible with as few gates as possible. As a result, they are well adapted to the current quantum hardware. This approach has been used to compute the ground state energies of small molecular systems on quantum hardware~\cite{kandala_he_ansatz, kandala_2019_he_on_hardware}. Since this approach is agnostic to the chemical nature of the system being simulated, it has some significant drawbacks. The first problem shown by McClean and co-workers is the ``barren plateaus" of the variational parameter landscape~\cite{barren_plateaus}, where the derivative of the objective function is close to zero. In addition, hardware-efficient ansatzes require extra work to enforce the physical symmetries, such as electron parity. Not accounting for the symmetries increases the size of the solution space and complicates the variational parameter optimization. This suggests that an arbitrary, unstructured ansatz can lead to poor convergence of the algorithm. Several approaches have been proposed to mitigate this issue~\cite{Barkoutsos_2018, Ganzhorn2019, Grant2019initialization, par_corr_2021}. Further work is needed, however, to improve this approach beyond small systems. 

In the next sections, we discuss the recently proposed methods to construct more efficient ansatzes using chemistry-inspired and hardware-efficient approaches. These methods focus on improvements in different areas. They can be centered on constructing circuits with fewer CNOT gates that are easier to run on NISQ hardware, choosing the operators that are added to the ansatz based on their contributions to the correlation energy, or performing simulations with higher precision that goes beyond the standard UCCSD. However, each of them can be considered a way to increase the size of the molecules that can be accurately simulated on modern quantum computers.

\begin{figure}
    \includegraphics[width=0.95\textwidth]{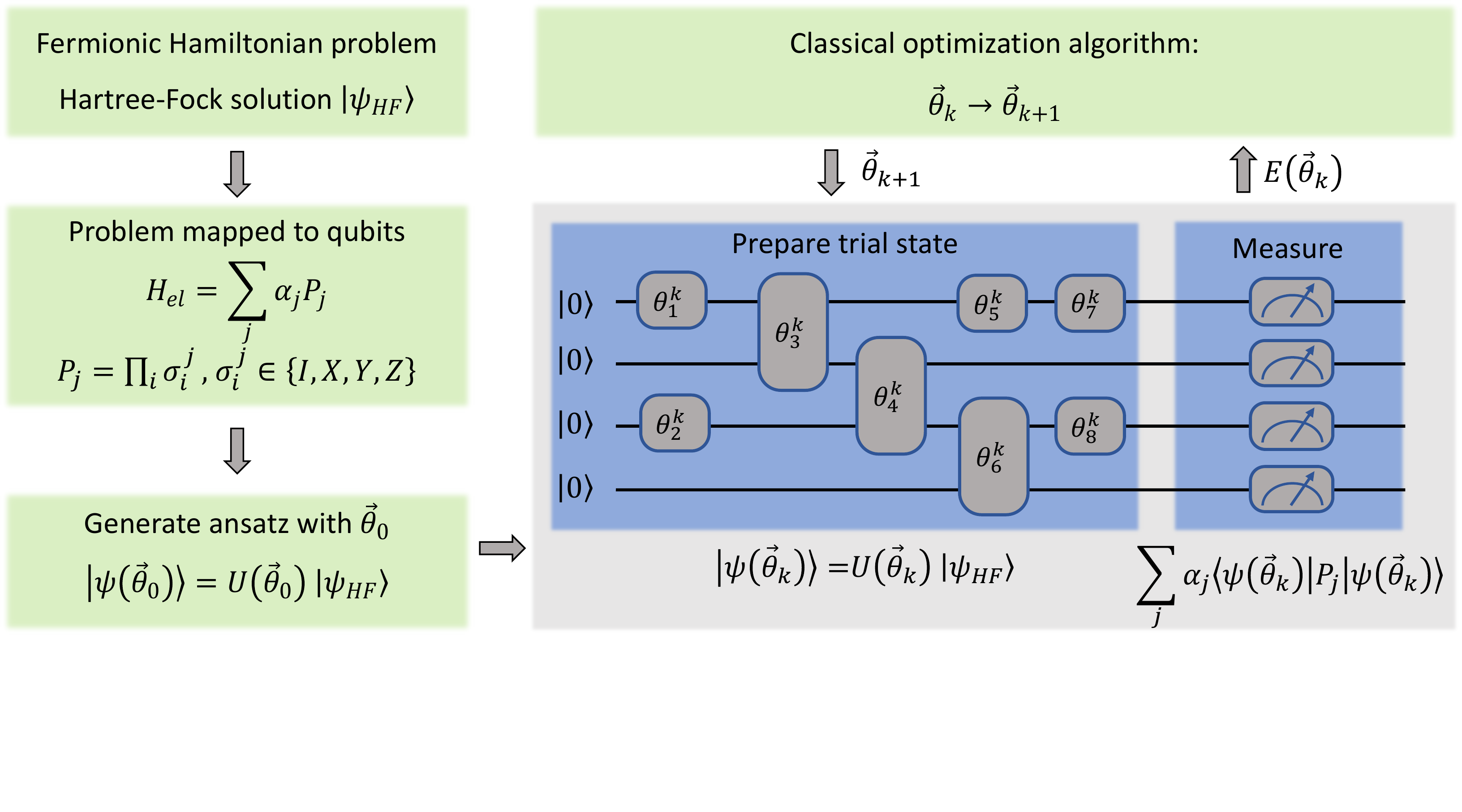}
    \caption{Schematic of the variational quantum eigensolver (VQE) method that minimizes the energy of the Hamiltonian $\bra{\psi(\vec{\theta})} \hat{H}_{el} \ket{\psi(\vec{\theta})}$ by adjusting variational parameters $\theta$. It uses classical computing resources denoted by green color and quantum computing resources denoted by blue. A simulation starts by constructing a fermionic Hamiltonian and finding mean-field solution $\ket{\psi_{HF}}$. Next, the fermionic Hamiltonian is mapped into qubit Hamiltonian, represented as a sum of Pauli strings $H=\sum_j\alpha_j\prod_i\sigma_i^j$. Then the ansatz to represent the wave function is chosen and initialized with the initial set of parameters $\vec{\theta}_0$.  The trial state is prepared on a quantum computer as a quantum circuit consisting of parametrized gates. The rest of the procedure is performed repeatedly until the convergence criterion is met. At iteration $k$ the energy of the Hamiltonian is computed by measuring every Hamiltonian term $\bra{\psi(\vec{\theta_k})}P_j\ket{\psi(\vec{\theta_k})}$ on a quantum computer and adding them on a classical computer. The energy $E(\vec{\theta_k})$ is fed into the classical algorithm that updates parameters for the next step of optimization $\vec{\theta}_{k+1}$ according to the chosen optimization algorithm.}
\end{figure}

\section*{Chemistry-Inspired Ansatze}

In this section, we discuss the ansatzes constructed in the fermionic space that use various techniques to improve the UCCSD method to obtain shorter circuits that are easier to run on NISQ hardware or obtain higher than UCCSD accuracy but with comparable circuit depth.

\subsection*{Unitary Pair Coupled Cluster with Generalized Singles and Doubles Product Wave Functions}
    One of the chemistry-inspired ansatzes designed to reduce circuit length is the Unitary Pair CC with Generalized Singles and Doubles (\textit{k}-UpCCGSD) method proposed by Lee et al.~\cite{k_up_uccsd}  The starting point for \textit{k}-UpCCGSD is the unitary pair coupled cluster double excitations method (UpCCD), which contains a significantly smaller number of operators compared with UCCSD because it includes only two-electron excitations from occupied spatial orbital $i$ to unoccupied spatial orbital $a$:
    \begin{equation} \label{eq_uccgsd}
    \hat{T}_2=\sum_{ia}t^{a_\alpha a_\beta}_{i_\alpha i_\beta} \hat{a}^{\dagger}_{a_\alpha}\hat{a}^{\dagger}_{a_\beta}\hat{a}_{i_\beta}\hat{a}_{i_\alpha},
    \end{equation}
    where excitations are performed from occupied orbitals $i$ to unoccupied orbitals $a$ and $\alpha$  and $\beta$ represent spin. It suffers less from non-variationality and is able to correctly break single bonds. However, it does not recover the dynamic correlation present in UCCD and loses invariance to rotations in occupied-occupied and virtual-virtual subspaces. To address these problems, in addition to pair double excitations,
    the full single excitations are included in the UpCCSD. For increased accuracy, generalized single and double excitations are included so that there is no distinction between occupied and virtual orbitals $p$ and $q$ when constructing excitations operators:
        \begin{equation}\label{UpCCGSD}
\hat{T}_{UpCCGSD}=\hat{T}_1 + \hat{T}_2= \sum_{pq}t^q_p\hat{a}^{\dagger}_q\hat{a}_p+ \sum_{pq}t^{q_\alpha q_\beta}_{p_\alpha p_\beta} \hat{a}^{\dagger}_{q_\alpha}\hat{a}^{\dagger}_{q_\beta}\hat{a}_{p_\beta}\hat{a}_{p_\alpha}.
    \end{equation}
  We note that UCCGSD always produces more accurate results than UCCSD but at a much higher computational cost. In UpCCGSD the cost is reduced by using only pair double excitations. UpCCGSD is inferior to UCCGSD because of the large number of missing double-excitation operators. For increased flexibility of ansatz, a product of \textit{k} unitary operators is included in the \textit{k}-UpCCGSD ansatz, resulting in the final expression:
    \begin{equation}
    \ket{\psi_{k-UpCCGSD}} = \prod_{\alpha=1}^k (e^{\hat{T}^{(\alpha)}-\hat{T}^{(\alpha)\dagger}}) \ket{\phi},
    \end{equation}
    where $\ket{\phi}$ is the reference wave function and each operator $\hat{T}^{(k)}$ contains an independent set of variational parameters for single and paired doubles CC amplitudes. Because of the sparsity of the UpCCGSD operator, the cost to prepare a \textit{k}-UpCCGSD state scales linearly with the system size with a prefactor \textit{k}. This structure also provides a way to systematically improve accuracy by increasing $k$.
    
    The \textit{k}-UpCCGSD method has been tested on \ce{H4}, \ce{H2O}, and \ce{N2} molecules. Compared with the UCCSD method, \textit{k}-UpCCGSD offers higher accuracy for the ground state energy with a smaller number of operators in the ansatz for all studied molecules. \textit{k}-UpCCGSD can achieve chemical accuracy with $k=2$ for \ce{H4} (STO-3G and 6-31G basis sets) and \ce{H2O} (STO-3G basis) molecules. UCCGSD is numerically exact for these systems, but it requires more operators in the ansatz. For \ce{N2} molecule with STO-3G basis, \textit{k}-UpCCGSD requires $k$=4 to achieve chemical accuracy, whereas UCCSD with more operators fails to come close to chemical accuracy and UCCGSD is within chemical accuracy but requires double the number of operators compared with \textit{k}-UpCCGSD.

\subsection*{Orbital Optimized UCC (OO-UCC) Ansatz}

OO-UCC is another variant of the UCC approach~\cite{Mizukami20_033421} where a single-particle orbital rotation operator $\hat{\kappa} = \sum_{pq} \kappa_{pq} (a_p^\dagger a_q - a_q^\dagger a_p) $ is introduced to the UCC energy functional,
\begin{align}
E(\hat{\tau},\hat{\kappa}) = \langle \Psi | e^{-\hat{\kappa}} \hat{H} e^{\hat{\kappa}} | \Psi \rangle = \langle \phi | e^{-\hat{\tau}}e^{-\hat{\kappa}} \hat{H} e^{\hat{\kappa}} e^{\hat{\tau}}| \phi \rangle,
\end{align}
(here $\hat{\tau} = \hat{T}-\hat{T}^\dagger$ is the UCC cluster operator) to variationally determine the coupled cluster amplitudes and also molecular orbital coefficients. Note that optimizing $\hat{\kappa}$ is essentially equivalent to minimizing a wave function with respect to orbital rotation parameters. When $\hat{\tau}$ is fixed, $\hat{\kappa}$ can be obtained through solving a linear equation, $\mathbf{H}\boldsymbol\kappa = -\mathbf{g}$, where $\mathbf{H}$ and $\mathbf{g}$ are the electronic Hessian and gradient, respectively. In practice, the OO-UCC ansatz can be easily derived from the UCCSD-alike ansatz (also similar to the DUCC ansatz discussed in the following section), except that only doubles are included,
\begin{align}
| \Phi \rangle_{\text{UCCSD}} = e^{\hat{\tau}_1+\hat{\tau}_2} | \phi \rangle \Rightarrow | \Phi \rangle_{\text{UCCSD}'} = e^{\hat{\tau}_2} e^{\hat{\tau}_1} | \phi \rangle \Rightarrow | \Phi \rangle_{\text{OO-UCCD}} = e^{\hat{\tilde{\tau}}_2} | \tilde{\phi} \rangle, 
\end{align}
where $\hat{\tau}_1$ and $\hat{\tau}_2$ don't commute, and the singles part $e^{\hat{\tau}_1}$ takes the role of carrying out orbital rotations that would be variationally optimized to obtain $|\tilde{\phi}\rangle = e^{\hat{\tau}_1} | \phi \rangle$ by using a classical computer, while the doubles part, $e^{\hat{\tilde{\tau}}_2}$, is optimized by the VQE. We note that the variational orbitals obtained from the  OO-UCCD approach coincide with the commonly known Brueckner orbitals. The reported numerical testing on small molecules  has  shown that in comparison with conventional UCCSD ansatzes, the qubitization of the OO-UCCD ansatz allows a slight reduction in the number of VQE parameters and the circuit depth. For example, for the \ce{NH3} molecule with the STO-3G basis set, the number of the VQE parameters and the circuit depth required in the OO-UCCD approach were 120 and 2,720, respectively, while those required by the conventional UCCSD were 135 and 2,780, respectively.
Regarding accuracy,  OO-UCCD maintains a similar level of accuracy to that of  UCCSD for most of the small molecules except for \ce{LiH}, where OO-UCCD results are closer to the full configuration interaction (FCI) results than to UCCSD.

\subsection*{Double Unitary Coupled Cluster (DUCC) Ansatz}

In modeling second-quantized problems, the major challenge comes from the sizable number of qubits that scales linearly with the size of the basis. To address this issue and enable more realistic simulations on NISQ computers, active space approximations are often employed. Along this line, Metcalf et al.~\cite{Metcalf20_6165} recently reported VQE applications based on DUCC ansatz that originate from the subsystem embedding sub-algebras coupled cluster studies~\cite{Kowalski18_094104} to constitute the effective form of the system Hamiltonian
\begin{align}
\bar{H}^{\text{eff}} &= e^{-\sigma_{\text{ext}}}He^{\sigma_{\text{ext}}}
\approx H + [H_\text{N},\sigma_{\text{ext}}] + \frac 1 2 [[F_{\text{N}},\sigma_{\text{ext}}], \sigma_{\text{ext}}], \label{ducc}
\end{align}
where $F_{\text{N}}$ and $H_{\text{N}}$ are normal product forms of Fock and Hamiltonian operators, respectively, and $\sigma_{\text{ext}} = \hat{T}_{\text{ext}} - \hat{T}_{\text{ext}}^\dagger$ is the anti-Hermitian cluster operator constructed from the cluster operators that produce excited configuration outside of the active space when acting on the reference $|\phi\rangle$. After the projection, the energy of the whole system, $E$, can be directly obtained by diagonalizing the effective Hamiltonian
\begin{align}
\bar{H}^{\text{eff}} e^{\sigma_{\text{int}}} |\phi\rangle &= E e^{\sigma_{\text{int}}} |\phi\rangle
\end{align}
with the eigenvector being the excited configurations within the active space.

Note that the DUCC formalism, just like the conventional CC formalism, is formally exact and independent of the choice of the active space, thus providing a systematically improvable hierarchy for implementation. Practically, the accuracy of the DUCC ansatzes depends on the approximate level of the effective Hamiltonian $\bar{H}^{\text{eff}}$ that can be controlled through either the length of the similarity expansion $e^{-\sigma_{\text{ext}}}He^{\sigma_{\text{ext}}}$ or the many-body terms included in $\sigma_{\text{ext}}$ (and $\hat{T}_{\text{ext}}$), or both. Also, when working with the VQE algorithm,  another source of approximation comes from the unitary ansatz for the active space.

For example, the reported DUCC-VQE algorithms include a DUCC effective Hamiltonian truncated at the second-order (see eq. (\ref{ducc})) and a UCCSD ansatz acting on the active space. In preliminary DUCC-VQE calculations on H$_2$, Li$_2$, and BeH$_2$ molecules~\cite{Metcalf20_6165}, the DUCC Hamiltonians have been shown to significantly outperform the bare Hamiltonian on the same active space in terms of approaching closer to the FCI results.

\subsection*{Quantum Subspace Expansion (QSE)}

The QSE scheme is similar to the generalized eigenvalue problem that is often encountered in quantum chemistry, where the Hamiltonian is diagonalized in a general non-orthogonal basis of many-body states,\cite{McClean2017Hybrid, colless_2018, Takeshita2020Quantum, Motta2019QITE, Huggins_2020, parrish2019quantum, Ollitrault2020, Stair2020Multireference} 
\begin{align}
\mathbf{H} \mathbf{c}  &= \mathbf{S}\mathbf{c}E.
\end{align}
Here the elements of the overlap matrix ($\mathbf{S}$) and Hamiltonian matrix ($\mathbf{H}$) in the non-orthogonal basis \{${\Phi_{i}, i=1,2,\cdots}$\} are given by
\begin{align}
S_{ij} &= \langle \Phi_i | \Phi_j \rangle, \notag \\
H_{ij} &= \langle \Phi_i | H |  \Phi_j \rangle.
\end{align}
Compared to the fully classical analog, the QSE quantum simulations utilize quantum devices to construct and store the arbitrarily complex states and measure the overlapping and Hamiltonian matrix elements, while leaving the corresponding eigenvalue problem to the classical machinery.

The many-body basis required for the QSE quantum simulations can be generated in many different ways. For example, analogous to the classical truncated configuration interaction expansion, McClean and co-workers proposed diagonalizing the Hamiltonian in the basis of states $a_i^\dagger a_j|\phi\rangle$ with $|\phi\rangle$ being a reference state obtained from a VQE run~\cite{McClean2017Hybrid, colless_2018, Takeshita2020Quantum}. The Hamiltonian elements in this basis are the three- and four-body density matrices. Following this line, the many-body basis can be employed to construct a Krylov subspace to improve efficiency and accuracy~\cite{Motta2019QITE, Stair2020Multireference}. One of the proposed approaches is the quantum Lanczos (QLanczos) algorithm~\cite{Motta2019QITE}, where the basis is the imaginary-time evolution of a single reference state sampled at regular intervals in imaginary time. In QLanczos, for a qubit-encoded Hamiltonian $H=\sum_{j}H_j$ (here $H_j=\alpha_jP_j$ with $\alpha_j$ complex scalars and $P_j$ Pauli string), the infinitesimal imaginary-time propagator $e^{\Delta_\tau H_j}$ is mirrored by a unitary evolution $e^{i\Delta_\tau A_j}$ acting on properly normalized states. One can show from the Taylor expansion of the time propagator that for an infinitesimal time step, the propagator would span a classical Krylov space. Similarly, via real-time evolution of a set of reference states, Stair et al.~\cite{Stair2020Multireference} also proposed a multireference selected quantum Krylov (MRSQK) algorithm based on QSE as an alternative to the quantum phase estimation algorithm and have shown that the proposed approach is able to capture the important multideterminantal features (if any) of the wave function and predict the energy of strongly correlated target states. 

An alternative approach for creating and utilizing the Krylov subspace was recently proposed by Kowalski and Peng~\cite{kowalski2020CMX,peng2021variational}. They found that the connected moments expansion, proposed and intensively developed in the 1980s and 1990s~\cite{horn1984t,cioslowski1987connected,peeters1984upper,soldatov1995generalized}, can be employed to re-engineer the energy functional to return a better energy estimate than the straightforward expectation value of the Hamiltonian operator for some trial wave function. Preliminary results on simple molecules and models exhibit high accuracy at finding the ground and excited states and their energies through the rotation of the trial wave function of modest quality,\cite{kowalski2020CMX} and potential capability to circumvent the `barren plateau' problem~\cite{peng2021variational}. 

We note that despite their higher accuracy and fewer numerical parameters to optimize, QSE methods also expose some practical issues. For example, almost all the above-mentioned QSE methods suffer from the linearly dependent basis that is generated from the procedure, which would cause numerical instabilities when solving the eigenvalue problem. When accounting for hardware noise, the instability would be potentially amplified, leading to worse gate fidelity and larger measurement errors. Furthermore, some QSE methods, in particular, the ones generating bases via real-time propagation,\cite{Huggins_2020, parrish2019quantum} may also require extra resources for the evaluation of off-diagonal matrix elements.

\subsection*{Quantum anti-Hermitian contracted Schr\"{o}dinger equation}

Contracted Schr\"{o}dinger equation (CSE) is an approach from the classical electronic structure theory that  is based on the reduced density matrix (RDM) theory and solves the contracted eigenvalue equation. The $N$-electron Schr\"{o}dinger equation is contracted onto the space of two electrons.\cite{cse:mazziotti1998, Mazziotti2002} The anti-Hermitian part of CSE, known as ACSE, has been used to find energies of the ground and excited states in strongly correlated systems on classical computers.\cite{acse_Mazziotti2004, exact_2body_exp_mazziotti2020, Mazziotti2006, mazziotti2007, acse_excited_states, Mukherjee2001}

The anti-Hermitian part of the CSE reads:
\begin{equation}\label{acse}
   \bra{\psi} [\hat{a}_i^\dagger \hat{a}_j^\dagger \hat{a}_k \hat{a}_l ,\hat{H}] \ket{\psi}=0.
\end{equation}
To solve the ACSE, the variational wave function ansatz is constructed iteratively by adding unitary two-body exponential operators:
\begin{equation}\label{acse_wf}
   \ket{\psi_{n+1}} = e^{\epsilon\hat{A}_n}\ket{\psi_n},
\end{equation}
where $\hat{A}_n$ is anti-hermitian two-body operator
\begin{equation}\label{An_operator}
    \hat{A}_n= \sum_{pqrs}A_n^{pqrs}\hat{a}_p^\dagger \hat{a}_q^\dagger \hat{a}_s \hat{a}_r.
\end{equation}
The energy at iteration $n$+1 is expressed as
\begin{equation}\label{acse_energy_n}
    E_{n+1} = E_n + \epsilon \bra{\psi_n} [\hat{H}, \hat{A}_n] \ket{\psi_n}+O(\epsilon^2).
\end{equation}
From Equation \ref{acse_energy_n} it is easy to derive the gradient of the energy with respect to variational parameter $A_n^{pqrs}$
\begin{equation}\label{acse_e_grad}
    \frac{\partial E}{\partial(A_n^{pqrs})}=-\epsilon\bra{\psi_n} [\hat{a}_p^\dagger \hat{a}_q^\dagger \hat{a}_s \hat{a}_r ,\hat{H}] \ket{\psi_n}+O(\epsilon^2).
\end{equation}
 The gradient in equation \ref{acse_e_grad} is the residual of the ACSE. It vanishes if and only if ACSE is satisfied, meaning that the wave function at iteration $n$ has converged and the minimum of energy is found. 
 On a classical computer, the solution of ACSE for the 2-particle RDM is indeterminant without the storage or reconstruction of 3-RDM, for which the cost scales exponentially. The quantum algorithm for solving the ACSE,\cite{qACSE_PRL2021} however, can be used to solve for 2-RDM without 3-RDM reconstruction. In QACSE the auxiliary states $\ket{\Lambda^{\pm}_n}=e^{\pm i\delta \hat{H}}\ket{\Psi_n}$ are prepared on a quantum computer. Then, the 2-RDMs of the auxiliary states are measured on a quantum computer to construct the $A$ matrix:
 \begin{equation}
     A^{pqrs}_n=\frac{1}{2i\delta}(\bra{\Lambda^+_n}\hat{a}_p^\dagger \hat{a}_q^\dagger \hat{a}_s \hat{a}_r\ket{\Lambda^+_n}-\bra{\Lambda^-_n}\hat{a}_p^\dagger \hat{a}_q^\dagger \hat{a}_s \hat{a}_r\ket{\Lambda^-_n})+O(\delta^2)
 \end{equation}
 This measurement of $A$ matrix on a quantum computer allows to avoid 3-RDM reconstruction. Thus, the QACSE algorithm has a potentially exponential speed-up compared to the classical one with full RDM reconstruction. The QACSE algorithm has been applied to \ce{H2}, \ce{H3}, and \ce{C6H4} molecules on quantum hardware.\cite{qACSE_PRL2021, Smart2021} 
 
It is important to note that the ACSE wave function also served as an inspiration for the development of the ADAPT-VQE ansatz and its different forms discussed in this article. Equation \ref{acse_e_grad} is central in constructing ADAPT-VQE ansatz as it allows to find the operator from the pool, which will result in the largest contribution to correlation energy.

\subsection*{Adaptive Derivative-Assembled Pseudo-Trotter Ansatz Variational Quantum Eigensolver (ADAPT-VQE)}

Grimsley et al. introduced the Adaptive Derivative-Assembled Pseudo-Trotter ansatz Variational Quantum Eigensolver, or ADAPT-VQE in short~\cite{Grimsley2019}. The key idea of the method is to construct an ansatz that would recover the most correlation energy with the least number of fermionic operators and variational parameters. It is inspired by the iterative algorithm for the ACSE solution (see previous section for details). The first step is to define a pool of operators that contains all fermionic excitation operators that can be added to the ansatz. In principle, this pool can be constructed from any set of operators, but the most straightforward choice is to use operators generated by the unitary coupled cluster ansatz. The use of the generalized version of UCCSD allows achieving higher than UCCSD accuracy with a smaller number of operators. In the generalized version (UCCGSD)~\cite{uccgsd} the operators are formed through single and double excitations over all occupied and virtual orbitals, which are not distinguished in this approach (eq. \ref{eq_uccgsd}).
Potentially, higher-order excitations can be included. After performing standard computations of one- and two-electron integrals along with reference Hartree--Fock wave function $\ket{\phi}$ on a classical computer the algorithm starts to gradually grow the variational ansatz with operators that would contribute the most correlation energy. Such operators are found through the computation of the following energy derivatives with respect to the variational parameters:
\begin{equation}\label{adapt_gradient}
    \frac{\partial E}{\partial \theta_i} = \bra{\psi} [\hat{H},\hat{\tau}_i] \ket{\psi},
\end{equation}
where $\hat{\tau}_i$ is a sum of excitation and de-excitation operators, for example, $\hat{\tau}_{pq}^{rs}=\hat{t}_{pq}^{rs}-\hat{t}^{pq}_{rs}$. Equation \ref{adapt_gradient} for computation of the gradient is equivalent to equation \ref{acse_e_grad} derive in the previous section of this article that describes ACSE. At every ADAPT-VQE step, one operator is added and full VQE optimization of all parameters is performed. The process is repeated until the convergence criterion is met when the norm of the gradient vector becomes smaller than the predefined threshold:
\begin{equation}
    \norm{\overrightarrow{g}} = \sqrt{\sum_i\left(\frac{\partial E}{\partial \theta_i}\right)^2} .
\end{equation}
When convergence is achieved, the algorithm produces the following ansatz:
\begin{equation}
    \ket{\psi^{ADAPT}} = (e^{\hat{\tau}_N}) (e^{\hat{\tau}_{N-1}})...(e^{\hat{\tau}_2}) (e^{\hat{\tau}_1}) \ket{\phi}.
\end{equation}

The ADAPT-VQE method has advantages compared with UCCSD for performing simulations on NISQ hardware. Compared with the UCCSD ansatz that produces many redundant terms contributing little to the correlation energy, ADAPT-VQE contains a much smaller number of operators. As a result, the quantum circuits are much shorter, which can enable simulations on quantum hardware for larger molecules than what is possible with UCCSD ansatz. In addition, it addresses perhaps the weakest quality of the VQE method, namely, that classical parameter optimization in VQE can become intractable because of barren plateaus\cite{barren_plateaus} or simply because the number of variational parameters is too large. ADAPT-VQE adds operators only with the highest contribution to the correlation energy and avoids the problem of having to optimize a large number of parameters with near-zero contributions. Another advantage is that parameters are optimized one at a time, so every iteration of ADAPT-VQE is started with all parameters preoptimized, except for the one newly added, which is likely to accelerate the rate of convergence. Another benefit of ADAPT-VQE is that the accuracy of the method can be controlled by adjusting the convergence criterion. When higher accuracy is needed, one can simply add operators with smaller contributions in the order of decreasing gradients (eq. \ref{adapt_gradient}) systematically converging to the solution provided by the full operator pool.

When considering the cost of simulations using a certain algorithm on NISQ hardware, among the most critical metrics are circuit depth, which is limited because of the short coherence times and large error rates, and the total shot count, which defines the time to solution. ADAPT-VQE is successful in identifying short and accurate ansatzes. However, the total shot count can drastically exceed the shot count for UCCSD ansatz. The reason is mainly that the number of energy gradients to be computed at each ADAPT iteration is equal to the number of operators in the pool, which grows quickly with system size. Another factor that increases the time to solution is that the number of VQE optimizations is equal to the final number of operators in the ansatz. Therefore, the time to solution can become unfeasible for large molecules, requiring a large number of operators to converge. We point out that for NISQ devices with limited coherence times, shorter circuits are more critical than total shot count and can be a defining factor if the simulation can be performed at all.

The examples of ADAPT-VQE implementation of ADAPT-VQE include potential energies of \ce{LiH}, \ce{BeH2}, and highly correlated \ce{H6} molecules using the STO-3G basis set.  For the \ce{LiH} molecule, the UCCSD ansatz produces results within chemical accuracy across the whole PES. ADAPT-VQE produces the same or better results with around 20 parameters instead of 60 required by UCCSD. With extra parameters, the errors can be reduced further. With an additional 5 parameters, the error is consistently under $10^{-3}$ kcal/mol. For the \ce{BeH2} molecule, UCCSD is within chemical accuracy around equilibrium; but for longer internuclear distances, the error gets larger than 1 kcal/mol. ADAPT-VQE achieves accuracies around 0.1 kcal/mol across the whole PES using around 30 parameters instead of 120 in UCCSD. For the \ce{H6} molecule, ADAPT-VQE also outperforms UCCSD. The number of required parameters grows significantly for longer distances to describe this highly correlated system. 

As with any adaptive method, estimating the resources required to perform a simulation is difficult. It strongly depends on the chemical system, initial operator pool, and specifics of the implementation. In the original implementation,\cite{Grimsley2019} operators are added one by one, and all parameters are reoptimized at every ADAPT-VQE iteration. Adding operators in batches can help reduce the cost, as can freezing some of the parameters for a certain number of iterations. Another way to reduce the shot count is to pre-screen the operators in the pool by using various techniques, for example, MP2 amplitude screening. Additionally, to reduce the time to solution, one can take advantage of the fact that the gradient calculations are independent and can be parallelized by using multiple quantum computers. Because of the limited availability of quantum hardware, however,  this can still be a problem. 

\subsection*{qubit-ADAPT-VQE}

Although the ADAPT-VQE algorithm allows one to significantly reduce the circuit depth compared with the UCCSD ansatz while achieving higher accuracy, the resulting quantum circuits are still beyond the reach of NISQ devices. Particularly problematic are the multi-qubit (e.g., CNOT) gates since they tend to have much higher error rates compared with 1-qubit gates. To reduce the number of CNOT gates in the circuits, Tang et al. proposed the qubit-ADAPT-VQE algorithm~\cite{qubit_adapt}. In this algorithm, the general concept is the same as in the original ADAPT-VQE, but instead of fermionic operators, the individual Pauli strings form the operator pool $\hat{\tau}=\hat{P}=i\prod_{i}p_i, p_i\in{\{X,Y,Z\}}$. These strings are obtained from the strings that are generated in the fermionic pool. This ansatz yields shallower circuits with fewer CNOT gates compared with the fermionic ADAPT-VQE, but at a cost of a larger number of variational parameters. Essentially, it is a way of offloading the computational effort from a quantum processor to a classical computer motivated by the limitations of NISQ hardware. 
	The qubit-ADAPT-VQE approach uses the same procedure to add operators from the operator pool to the ansatz by computing gradients of energy with respect to the variational parameters associated with the operators. Because of the larger number of parameters compared with the fermionic ADAPT, the ``qubit pool" requires more gradient calculations on a quantum computer. The number of operators in the pool scales exponentially as $2^n-1$. Therefore, it is crucial to eliminate redundant operators from the pool to reduce its size. First, the strings with an even number of Y operators are eliminated to ensure that the fermionic operator $\hat{P}=i\prod_{i}p_i, p_i\in{\{X,Y,Z\}}$ is real. In addition, chains of $Z$ gates are removed from the pool since they do not affect the performance of the method according to numerical simulations. However, the size of this ``qubit pool" remains large. Also, it is important to note that the chains of $Z$ gates are responsible for enforcing the anti-symmetry of the electronic wave function. The removal of these gates did not affect the numerical results for small test molecules but the effect of this symmetry removal on the accuracy of description of larger molecules is still unknown. 
	It has been shown that the size of the qubit-ADAPT-VQE pool can be reduced dramatically without sacrificing accuracy, and it has been analytically proven that complete pools of size $2n-2$ exist for any $n$~\cite{Grimsley2019}. The recipe for constructing such a minimal complete pool is a direction for future research.  If the extra measurement overhead in qubit-ADAPT-VQE can be reduced by an efficient algorithm for finding a complete pool, this method will be a good fit for applications on NISQ hardware, as long as the classical optimization part can be efficiently solved. The qubit-ADAPT-VQE method was tested on $H_4$, $H_6$, and $LiH$ molecules using STO-3G basis set. It can achieve the same accuracy as does the fermionic ADAPT with fewer CNOT gates but with a larger number of ADAPT iterations because of the larger number of variational parameters.

\section*{Hardware-Efficient Ansatzes}

In this section, we discuss hardware-efficient ansatzes that produce quantum circuits that are easier to run on modern quantum hardware.

\subsection*{Symmetry-Preserving State Preparation}

The idea of imposing symmetries associated with particle number, spin, and time-reversal symmetries goes along with the approach that bases the VQE ansatzes on the capabilities of the hardware and performs state preparation through combining parameterized gates available on the processor.\cite{Bian19_15, gard_sym_preserv}

The ansatzes in this category have the advantage of being compatible with the capabilities of the hardware, but the ad hoc ansatzes of this type can also cause the so-called ``barren plateaus" problem~\cite{barren_plateaus}, where gradients vanish exponentially in a sufficiently expressive parameterized quantum circuits, which in turn requires an exponentially large precision to navigate through the ``barren plateaus" landscape. Similar problems (i.e., gradient-vanishing) existed in the early studies of deep neural networks~\cite{Bradley09learningin, pmlr-v70-shalev-shwartz17a, Kremer2001Book} with mitigation techniques proposed later on.\cite{lecun2015deeplearning, pmlr-v37-ioffe15, Hinton2006, He2016} Nevertheless, for hardware-based ansatzes to be successful in solving the problems of interest, most of the approaches have focused on assuring that the hardware-based ansatzes span the part of the Hilbert space that includes the true solution.

So far,  two ways have been employed for facilitating the access of the parameterized quantum circuits to the ``right" part of the Hilbert space. On the one hand, penalty terms can be implemented in the VQE energy function for symmetry violations~\cite{McClean_2016, Ryabinkin2019}. On the other hand, the state preparation circuits need to be carefully designed to preserve appropriate symmetries regardless of variational parameter rotation. An early attempt in this direction~\cite{Wang2009}, in comparison with preparing a general state in the full Hilbert space of $n$ qubits that requires $\mathcal{O}(2^n)$ controlled-NOT gates, shows the actual number of the controlled-NOT gates required for exploring a smaller Hilbert space, where the true solution  scales only polynomially with the number of qubits, which can still be greatly reduced by several orders of magnitude by properly designing the quantum algorithm that accounts for additional symmetries.
Similar advantages have also been reported by Barkoutsos et al.~\cite{Barkoutsos_2018}, who found the reformulation of the molecular Hamiltonian in second quantization using the particle-hole picture in conjunction with a parameterized particle-conserving exchange-type gate~\cite{Ganzhorn2019, Roth2017, Egger2019, Sagastizabal2019} is able to improve the computational efficiency and accuracy for quantum chemistry simulations. Nevertheless, important open questions remain, including how other symmetries can be built into the circuits and whether more efficient circuits exist that contain the minimal number of parameters necessary to span the symmetry subspace.
To encode other symmetries into the circuits, while still balancing the efficiency and accuracy, Gard et al.~\cite{gard_sym_preserv} introduced generalized state preparation circuits that accommodate well-defined symmetries (including particle number, total spin, spin projection, and time  reversal) and require a minimal number of parameters to directly target the appropriate symmetry subspace. Although tested only for \ce{H2} and \ce{LiH} molecules, it is shown that the circuits are able to locate the true ground state within the subspace of states spanned by the circuit and reduce the complexity of the classical optimization step of the VQE algorithm, thus outperforming the standard state preparation ansatze.

Despite these efforts, as well as more recent ones~\cite{fontana2020optimizing, anand2020natural, arrasmith2020effect, uvarov2020barren, zhang2020trainability, pesah2020absence},  it is still too early to claim success since additional difficulties can come from hardware noise that can potentially modify the cost landscape associated with the parameter space. For example, a recent study~\cite{fontana2020optimizing} found that the hardware noise (specifically non-unital noise) can break the underlying symmetries in parameterized quantum circuits and lift the degeneracy of minima that falsifies local minima as global minima. 

\subsection*{Qubit Coupled Cluster Method}
The qubit coupled cluster (QCC) method introduced by Ryabinkin et al.~\cite{qcc} resembles the structure of coupled cluster; but instead of using fermionic excitations, the ansatz is built directly in the qubit space. The QCC wave function is of the form
\begin{equation}
    \ket{\Psi(\tau,\Omega)} = \hat{U}(\tau)\ket{\Omega}.
\end{equation}
The mean-field wave function $\ket{\Omega}$ is a product of single-qubit states
\begin{equation}
        \ket{\Omega}=\prod_{j=1}^{n}\ket{\Omega_j},
\end{equation}
where $n$ is the number of spin-orbitals. Each $\ket{\Omega_j}$ is parameterized with Bloch angles $\phi$ and $\theta$:
\begin{equation}
        \ket{\Omega_j}=\cos{\left(\frac{\theta_j}{2}\right)}\ket{\uparrow_j} + e^{i\phi_j}\sin{\frac{\theta_j}{2}}\ket{\downarrow_j} ,
\end{equation}
where $\ket{\uparrow_j}$ and $\ket{\downarrow_j}$ are eigenstates of $\hat{z}_j$.
Entanglement is introduced by the multi-qubit rotations with real amplitudes $\tau=\{\tau_k\}$:
\begin{equation}
        \hat{U}(\tau)=\prod_{k=1}^{N_{ent}} e^{-i\tau_k\hat{P}_k/2},
\end{equation}
where $\hat{P}_k$ are Pauli strings of length from 2 to a number of qubits $N_q$ and the number of entanglers $N_{ent}$ is less than or equal to the number of $\hat{P}_k$. The ground state energy of the system is obtained through minimization of the Hamiltonian expectation value
\begin{equation}\label{e_qcc}
        E_{QCC}=\underset{\Omega,\tau}{\min}\bra{\Psi(\tau,\Omega)}\hat{H}\ket{\Psi(\tau,\Omega)}=\underset{\Omega,\tau}{\min}\bra{\Omega}U(\tau)^{\dagger}\hat{H}U(\tau)\ket{\Omega}.
\end{equation}
    
Similar to other hardware-efficient methods working directly in the qubit space, QCC ansatz can result in unphysical results, for example breaking of symmetries, such as non-conservation of the total number of particles or obtaining a state with a wrong electronic spin. This problem  has been shown~\cite{Ryabinkin2019} to be more a rule than an exception. Therefore, additional constraints need to be used to ensure that results do not violate any physical laws. Yen et al.~\cite{yen2019_sym} proposed symmetry projectors that can enforce the symmetries and implemented these projectors with the QCC method.
    
The total number of operators in the transformed Hamiltonian $U(\tau)^{\dagger}\hat{H}U(\tau)$ scales exponentially and contains $~3^{N_{ent}}$ parameters. Therefore, the operators with the largest contribution to the correlation energy need to be chosen to avoid exponential scaling. In the QCC framework such screening of operators is performed by computing the energy derivative with respect to $\tau_k$ at $\tau_k=0$:
\begin{equation}
        \begin{aligned}
        \left|\left.\frac{dE[\hat{P}_k]}{d\tau_k}\right|_{\tau=0}\right| 
        & = \left|\left.\frac{d}{d\tau_k}\underset{\Omega}{\min}\bra{\Omega}e^{i\tau_k\hat{P}_k/2}\hat{H}e^{-i\tau_k\hat{P}_k/2}\ket{\Omega}\right|_{\tau_k=0}\right| \\
        & = \left|\bra{\Omega_{\min}}-\frac{i}{2}[\hat{H},\hat{P}_k]\ket{\Omega_{\min}}\right|>0 ,
        \end{aligned}
\end{equation}
where $\ket{\Omega_{\min}}$ is the qubit mean field (QMF) wave function at the point of minimum QMF energy. This expression for the gradient is analogous to one that is used in the ADAPT-VQE approach.  
    
The entanglers responsible for the largest energy reduction are ranked in order of decreasing first energy derivatives. If the first derivative vanishes, the second tier of entanglers can be formed by performing ranking using the second derivatives.  Thus, the QCC ansatz can be expanded systematically based on this ranking, with more terms recovering a larger amount of correlation energy. By including less important terms, one can systematically approach the exact solution. However, the computational complexity of this screening procedure is exponential with respect to the number of terms in the Hamiltonian. The screening procedure can be improved by taking advantage of the Hamiltonian symmetry and splitting operators into groups, which results in the polynomial complexity for screening the operators.\cite{it_QCC} 

Another shortcoming of QCC, an exponential number of operators, which restricts the application of QCC to small systems, was addressed in the iterative version of the QCC method (iQCC)~\cite{it_QCC}. In this method, the ``dressed" canonically transformed Hamiltonian is considered:
\begin{equation}\label{dressed_H}
    \hat{H}_d=\hat{U}^{\dagger}(\tau)\hat{H}\hat{U}(\tau).
\end{equation}
The dressed Hamiltonian is evaluated recursively according to
\begin{equation}\label{H_recursive}
    \begin{aligned}
    \hat{H}^{(k)}_d(\tau_k,...,\tau_1) & = e^{i\tau_k\hat{P}_k/2} \hat{H}^{(k-1)}_d(\tau_{k-1},...,\tau_1) e^{-i\tau_k\hat{P}_k/2} \\
    & =\hat{H}^{(k-1)}_d - \frac{i}{2}[\hat{H}^{(k-1)}_d,\hat{P}_k]\sin\tau_k \\
    & +\frac{1}{2}\hat{P}_k[\hat{H}^{(k-1)}_d,\hat{P}_k](1-\cos{\tau_k}),
    \end{aligned}
\end{equation}
where $k=1,..., N_{ent}$. Instead of optimizing all $N_{ent}$ amplitudes at once, the amplitudes can be optimized sequentially by introducing one or more operators at each iteration. During each iQCC iteration $N_{ent}\geq1$ amplitudes are optimized using the Hamiltonian from the previous step. Then the optimized amplitudes are used to construct the dressed Hamiltonian (eq. \ref{e_qcc}) for the next iteration according to equation \ref{H_recursive}. 

In general, this procedure introduces $3^{N_{ent}}$ distinct operators and shows the exponential complexity of the QCC ansatz. It has been demonstrated that if the amplitudes $\tau$ in eq. \ref{H_recursive} are fixed, the complexity of transforming the dressed Hamiltonian in eq. \ref{dressed_H} is $\sim M(3/2)^{N_{ent}}$, where $M$ is the number of terms in the Hamiltonian. The $(3/2)^{N_{ent}}$ factor comes from the canonical transformation of Hamiltonian (eq. \ref{dressed_H}) at each step of iQCC and for larger systems can make simulation prohibitively expensive. This problem can be partially solved by ``compressing" the dressed Hamiltonian. This procedure essentially removes the terms from the Hamiltonian with little contribution to the ground state energy in a controlled manner so that the energy changes by no more than desired accuracy $\epsilon$. Numerical examples have shown that the compressing procedure reduces the $(3/2)^{N_{ent}}$ factor. The benefits from this procedure increase for larger systems where the fraction of terms with a small contribution to energy in the Hamiltonian becomes large. In a more recent effort, aiming at reducing the number of operators in the dressed Hamiltonian, Eq. (\ref{H_recursive}), Izmaylov group introduced involutory linear combinations (ILC) of anti-commuting Pauli products (i.e. entanglers) in the QCC framework,\cite{izmaylov2021unitary} and the resulting QCC-ILC unitary dressing gives an exact quadratic truncation of the Baker-Campbell-Hausdorff expansion which, in comparison to the random QCC transformation, only results in quadratic growth of the number of Pauli strings in the dressed Hamiltonian while still yields accurate energy estimates in the strongly correlated regime. 

QCC has been benchmarked on \ce{LiH}, \ce{H2O}, and \ce{N2} molecules using the STO-3G basis. Chemical accuracy was achieved for these small molecules. However, performing QCC calculations on large molecules would require further improvements to the algorithm, in particular, the  $(3/2)^{N_{ent}}$ factor appearing in the dressed Hamiltonian transformation.

\section*{Summary}


The main shortcoming of the standard UCCSD ansatz for VQE simulations is rooted in the overall number of excitations. Although each excitation operator generally adds only one variational parameter, the total number of excitations grows rapidly with system size. A large number of amplitudes increases the length of circuits, thus increasing the load on the quantum computer. At the same time, a larger number of parameters have to be optimized within the classical VQE loop. Moreover, circuits to implement fermionic excitations contain a large number of multi-qubit gates, which are a problem for simulations on NISQ hardware because of the large error rates of multi-qubit gates. The methods described in this paper are designed to construct short-depth ansatzes that are easier to run on NISQ hardware. In general, chemistry-inspired ansatzes have a goal of a more careful choice of excitations to include in the ansatz, while hardware-efficient approaches focus on the reduction of the number of multi-qubit gates and using various techniques to choose compact blocks of entangling gates with single-qubit rotational gates more efficiently.

In adaptive methods that include ADAPT, qubit-ADAPT-VQE, and QCC methods, the addition of the operators to the ansatz is based on the calculation of the gradient of energy with respect to the parameter associated with this operator, where the largest gradient corresponds to the largest contribution to the correlation energy. It is straightforward to ensure that an ansatz does not contain redundant terms that complicate classical optimization and make quantum circuits longer. Doing so, however, comes at the cost of extra measurements associated with gradient calculations used to choose the operators contributing the most amount of correlation energy. QCC and qubit-ADAPT-VQE work directly in the qubit space and are considered more ``hardware-efficient" because they are generally more compact and easier to run on NISQ hardware. However, they require additional work to ensure the preservation of all physical symmetries. UCC-based methods such as ADAPT-VQE, \textit{k}-UpCCGSD, DUCC, OO-UCC, and ansatz are  constructed from fermionic operators and therefore avoid the problems with symmetry preservation.  However, the number of multi-qubit gates remains high in this method, which requires the error rates on quantum hardware to significantly improve before such methods can be used. On the other hand, QSE-based approaches exhibit higher accuracy from ansatzes prepared with relatively low circuit depth and require fewer numerical parameters to be optimized, but they need extra measurements for Hamiltonian powers, overlap matrix, and/or energy gradient, and may suffer from the linearly dependent many-body basis. Thus  many issues still are left to be resolved, and significant progress is expected in these directions in the post-VQE era.

\begin{backmatter}

\section*{Competing interests}
The authors declare that they have no competing interests

\section*{Funding}
This material is based upon work supported by the U.S. Department of Energy, Office of Science, National Quantum Information Science Research Centers. B.P.  also acknowledges support from the Laboratory Directed Research and Development (LDRD) Program at PNNL. This material is also based upon work supported by the U.S. Department of Energy, Office of Science, Office of Fusion Energy Sciences, under Award Number DE-SC0020249. Y.A.'s work at Argonne National Laboratory was supported by the U.S. Department of Energy, Office of Science, under contract DE-AC02-06CH11357.

\section*{Authors' contributions}
D.A.F. and B.P. contributed equally to the manuscript. All authors read and approved the final manuscript.


\bibliographystyle{bmc-mathphys} 

\begin{thebibliography}{107}
\ifx \bisbn   \undefined \def \bisbn  #1{ISBN #1}\fi
\ifx \binits  \undefined \def \binits#1{#1}\fi
\ifx \bauthor  \undefined \def \bauthor#1{#1}\fi
\ifx \batitle  \undefined \def \batitle#1{#1}\fi
\ifx \bjtitle  \undefined \def \bjtitle#1{#1}\fi
\ifx \bvolume  \undefined \def \bvolume#1{\textbf{#1}}\fi
\ifx \byear  \undefined \def \byear#1{#1}\fi
\ifx \bissue  \undefined \def \bissue#1{#1}\fi
\ifx \bfpage  \undefined \def \bfpage#1{#1}\fi
\ifx \blpage  \undefined \def \blpage #1{#1}\fi
\ifx \burl  \undefined \def \burl#1{\textsf{#1}}\fi
\ifx \doiurl  \undefined \def \doiurl#1{\textsf{#1}}\fi
\ifx \betal  \undefined \def \betal{\textit{et al.}}\fi
\ifx \binstitute  \undefined \def \binstitute#1{#1}\fi
\ifx \binstitutionaled  \undefined \def \binstitutionaled#1{#1}\fi
\ifx \bctitle  \undefined \def \bctitle#1{#1}\fi
\ifx \beditor  \undefined \def \beditor#1{#1}\fi
\ifx \bpublisher  \undefined \def \bpublisher#1{#1}\fi
\ifx \bbtitle  \undefined \def \bbtitle#1{#1}\fi
\ifx \bedition  \undefined \def \bedition#1{#1}\fi
\ifx \bseriesno  \undefined \def \bseriesno#1{#1}\fi
\ifx \blocation  \undefined \def \blocation#1{#1}\fi
\ifx \bsertitle  \undefined \def \bsertitle#1{#1}\fi
\ifx \bsnm \undefined \def \bsnm#1{#1}\fi
\ifx \bsuffix \undefined \def \bsuffix#1{#1}\fi
\ifx \bparticle \undefined \def \bparticle#1{#1}\fi
\ifx \barticle \undefined \def \barticle#1{#1}\fi
\ifx \bconfdate \undefined \def \bconfdate #1{#1}\fi
\ifx \botherref \undefined \def \botherref #1{#1}\fi
\ifx \url \undefined \def \url#1{\textsf{#1}}\fi
\ifx \bchapter \undefined \def \bchapter#1{#1}\fi
\ifx \bbook \undefined \def \bbook#1{#1}\fi
\ifx \bcomment \undefined \def \bcomment#1{#1}\fi
\ifx \oauthor \undefined \def \oauthor#1{#1}\fi
\ifx \citeauthoryear \undefined \def \citeauthoryear#1{#1}\fi
\ifx \endbibitem  \undefined \def \endbibitem {}\fi
\ifx \bconflocation  \undefined \def \bconflocation#1{#1}\fi
\ifx \arxivurl  \undefined \def \arxivurl#1{\textsf{#1}}\fi
\csname PreBibitemsHook\endcsname

\bibitem{feynman1982simulating}
\begin{botherref}
\oauthor{\bsnm{Feynman}, \binits{R.P.}}:
Simulating physics with computers.
Int. J. Theor. Phys
\textbf{21}(6/7)
(1982)
\end{botherref}
\endbibitem

\bibitem{preskill2018quantum}
\begin{barticle}
\bauthor{\bsnm{Preskill}, \binits{J.}}:
\batitle{Quantum computing in the {NISQ} era and beyond}.
\bjtitle{Quantum}
\bvolume{2},
\bfpage{79}
(\byear{2018})
\end{barticle}
\endbibitem

\bibitem{elfving2020will}
\begin{botherref}
\oauthor{\bsnm{Elfving}, \binits{V.E.}},
\oauthor{\bsnm{Broer}, \binits{B.W.}},
\oauthor{\bsnm{Webber}, \binits{M.}},
\oauthor{\bsnm{Gavartin}, \binits{J.}},
\oauthor{\bsnm{Halls}, \binits{M.D.}},
\oauthor{\bsnm{Lorton}, \binits{K.P.}},
\oauthor{\bsnm{Bochevarov}, \binits{A.}}:
How will quantum computers provide an industrially relevant computational
  advantage in quantum chemistry?
arXiv preprint arXiv:2009.12472
(2020)
\end{botherref}
\endbibitem

\bibitem{liu2021prospects}
\begin{botherref}
\oauthor{\bsnm{Liu}, \binits{H.}},
\oauthor{\bsnm{Low}, \binits{G.H.}},
\oauthor{\bsnm{Steiger}, \binits{D.S.}},
\oauthor{\bsnm{Häner}, \binits{T.}},
\oauthor{\bsnm{Reiher}, \binits{M.}},
\oauthor{\bsnm{Troyer}, \binits{M.}}:
Prospects of Quantum Computing for Molecular Sciences
(2021).
\arxivurl{2102.10081}
\end{botherref}
\endbibitem

\bibitem{qnext}
\begin{botherref}
{Q-NEXT National Quantum Center located in Argonne National Laboratory}.
Accessed: 2021-01-21
\end{botherref}
\endbibitem

\bibitem{kitaev_QPE_1995}
\begin{botherref}
\oauthor{\bsnm{Kitaev}, \binits{A.Y.}}:
{Quantum measurements and the Abelian stabilizer problem}.
arXiv
(1995).
\arxivurl{quant-ph/9511026}
\end{botherref}
\endbibitem

\bibitem{QPE_Lloyd}
\begin{barticle}
\bauthor{\bsnm{Abrams}, \binits{D.S.}},
\bauthor{\bsnm{Lloyd}, \binits{S.}}:
\batitle{Quantum algorithm providing exponential speed increase for finding
  eigenvalues and eigenvectors}.
\bjtitle{Phys. Rev. Lett.}
\bvolume{83},
\bfpage{5162}--\blpage{5165}
(\byear{1999}).
doi:\doiurl{10.1103/PhysRevLett.83.5162}
\end{barticle}
\endbibitem

\bibitem{QPE_Abrams_1997}
\begin{barticle}
\bauthor{\bsnm{Abrams}, \binits{D.S.}},
\bauthor{\bsnm{Lloyd}, \binits{S.}}:
\batitle{Simulation of many-body fermi systems on a universal quantum
  computer}.
\bjtitle{Phys. Rev. Lett.}
\bvolume{79},
\bfpage{2586}--\blpage{2589}
(\byear{1997}).
doi:\doiurl{10.1103/PhysRevLett.79.2586}
\end{barticle}
\endbibitem

\bibitem{QFT}
\begin{bchapter}
\bauthor{\bsnm{{Shor}}, \binits{P.W.}}:
\bctitle{Algorithms for quantum computation: discrete logarithms and
  factoring}.
In: \bbtitle{Proceedings 35th Annual Symposium on Foundations of Computer
  Science},
pp. \bfpage{124}--\blpage{134}
(\byear{1994}).
doi:\doiurl{10.1109/SFCS.1994.365700}
\end{bchapter}
\endbibitem

\bibitem{aharonov_1996}
\begin{botherref}
\oauthor{\bsnm{Aharonov}, \binits{D.}},
\oauthor{\bsnm{Ben-Or}, \binits{M.}}:
{Fault Tolerant Quantum Computation with Constant Error}.
arXiv
(1996).
\arxivurl{quant-ph/9611025}
\end{botherref}
\endbibitem

\bibitem{Peruzzo2014_VQE}
\begin{barticle}
\bauthor{\bsnm{Peruzzo}, \binits{A.}},
\bauthor{\bsnm{McClean}, \binits{J.}},
\bauthor{\bsnm{Shadbolt}, \binits{P.}},
\bauthor{\bsnm{Yung}, \binits{M.-H.}},
\bauthor{\bsnm{Zhou}, \binits{X.-Q.}},
\bauthor{\bsnm{Love}, \binits{P.J.}},
\bauthor{\bsnm{Aspuru-Guzik}, \binits{A.}},
\bauthor{\bsnm{O’Brien}, \binits{J.L.}}:
\batitle{A variational eigenvalue solver on a photonic quantum processor}.
\bjtitle{Nat. Commun.}
\bvolume{5}(\bissue{1}),
\bfpage{4213}
(\byear{2014}).
doi:\doiurl{10.1038/ncomms5213}
\end{barticle}
\endbibitem

\bibitem{McClean_2016}
\begin{barticle}
\bauthor{\bsnm{McClean}, \binits{J.R.}},
\bauthor{\bsnm{Romero}, \binits{J.}},
\bauthor{\bsnm{Babbush}, \binits{R.}},
\bauthor{\bsnm{Aspuru-Guzik}, \binits{A.}}:
\batitle{The theory of variational hybrid quantum-classical algorithms}.
\bjtitle{New J. Phys.}
\bvolume{18}(\bissue{2}),
\bfpage{023023}
(\byear{2016}).
doi:\doiurl{10.1088/1367-2630/18/2/023023}
\end{barticle}
\endbibitem

\bibitem{Romero_2018}
\begin{barticle}
\bauthor{\bsnm{Romero}, \binits{J.}},
\bauthor{\bsnm{Babbush}, \binits{R.}},
\bauthor{\bsnm{McClean}, \binits{J.R.}},
\bauthor{\bsnm{Hempel}, \binits{C.}},
\bauthor{\bsnm{Love}, \binits{P.J.}},
\bauthor{\bsnm{Aspuru-Guzik}, \binits{A.}}:
\batitle{Strategies for quantum computing molecular energies using the unitary
  coupled cluster ansatz}.
\bjtitle{Quantum Sci. Technol.}
\bvolume{4}(\bissue{1}),
\bfpage{014008}
(\byear{2018}).
doi:\doiurl{10.1088/2058-9565/aad3e4}
\end{barticle}
\endbibitem

\bibitem{Pal1984use}
\begin{barticle}
\bauthor{\bsnm{Pal}, \binits{S.}}:
\batitle{{Use of a unitary wavefunction in the calculation of static electronic
  properties}}.
\bjtitle{Theo. Chim. Acta}
\bvolume{66}(\bissue{3}),
\bfpage{207}--\blpage{215}
(\byear{1984}).
doi:\doiurl{10.1007/BF00549670}
\end{barticle}
\endbibitem

\bibitem{Hoffmann1988}
\begin{barticle}
\bauthor{\bsnm{Hoffmann}, \binits{M.R.}},
\bauthor{\bsnm{Simons}, \binits{J.}}:
\batitle{{A unitary multiconfigurational coupled‐cluster method: Theory and
  applications}}.
\bjtitle{J. Chem. Phys.}
\bvolume{88}(\bissue{2}),
\bfpage{993}--\blpage{1002}
(\byear{1988}).
doi:\doiurl{10.1063/1.454125}
\end{barticle}
\endbibitem

\bibitem{Kutzelnigg1991}
\begin{barticle}
\bauthor{\bsnm{Kutzelnigg}, \binits{W.}}:
\batitle{{Error analysis and improvements of coupled-cluster theory}}.
\bjtitle{Theo. Chim. Acta}
\bvolume{80}(\bissue{4}),
\bfpage{349}--\blpage{386}
(\byear{1991}).
doi:\doiurl{10.1007/BF01117418}
\end{barticle}
\endbibitem

\bibitem{Taube2006}
\begin{barticle}
\bauthor{\bsnm{Taube}, \binits{A.G.}},
\bauthor{\bsnm{Bartlett}, \binits{R.J.}}:
\batitle{{New perspectives on unitary coupled-cluster theory}}.
\bjtitle{Int. J. Quantum Chem.}
\bvolume{106}(\bissue{15}),
\bfpage{3393}--\blpage{3401}
(\byear{2006}).
doi:\doiurl{10.1002/qua.21198}
\end{barticle}
\endbibitem

\bibitem{Sur_2008}
\begin{barticle}
\bauthor{\bsnm{Sur}, \binits{C.}},
\bauthor{\bsnm{Chaudhuri}, \binits{R.K.}},
\bauthor{\bsnm{Sahoo}, \binits{B.K.}},
\bauthor{\bsnm{Das}, \binits{B.P.}},
\bauthor{\bsnm{Mukherjee}, \binits{D.}}:
\batitle{Relativistic unitary coupled cluster theory and applications}.
\bjtitle{J. Phys. B}
\bvolume{41}(\bissue{6}),
\bfpage{065001}
(\byear{2008}).
doi:\doiurl{10.1088/0953-4075/41/6/065001}
\end{barticle}
\endbibitem

\bibitem{Cooper2010}
\begin{barticle}
\bauthor{\bsnm{Cooper}, \binits{B.}},
\bauthor{\bsnm{Knowles}, \binits{P.J.}}:
\batitle{{Benchmark studies of variational, unitary and extended coupled
  cluster methods}}.
\bjtitle{J. Chem. Phys.}
\bvolume{133}(\bissue{23}),
\bfpage{234102}
(\byear{2010}).
doi:\doiurl{10.1063/1.3520564}
\end{barticle}
\endbibitem

\bibitem{Harsha2018}
\begin{barticle}
\bauthor{\bsnm{Harsha}, \binits{G.}},
\bauthor{\bsnm{Shiozaki}, \binits{T.}},
\bauthor{\bsnm{Scuseria}, \binits{G.E.}}:
\batitle{{On the difference between variational and unitary coupled cluster
  theories}}.
\bjtitle{J. Chem. Phys.}
\bvolume{148}(\bissue{4}),
\bfpage{044107}
(\byear{2018}).
doi:\doiurl{10.1063/1.5011033}
\end{barticle}
\endbibitem

\bibitem{Evangelista2019}
\begin{barticle}
\bauthor{\bsnm{Evangelista}, \binits{F.A.}},
\bauthor{\bsnm{Chan}, \binits{G.K.-L.}},
\bauthor{\bsnm{Scuseria}, \binits{G.E.}}:
\batitle{{Exact parameterization of fermionic wave functions via unitary
  coupled cluster theory}}.
\bjtitle{J. Chem. Phys.}
\bvolume{151}(\bissue{24}),
\bfpage{244112}
(\byear{2019}).
doi:\doiurl{10.1063/1.5133059}
\end{barticle}
\endbibitem

\bibitem{kandala_he_ansatz}
\begin{barticle}
\bauthor{\bsnm{Kandala}, \binits{A.}},
\bauthor{\bsnm{Mezzacapo}, \binits{A.}},
\bauthor{\bsnm{Temme}, \binits{K.}},
\bauthor{\bsnm{Takita}, \binits{M.}},
\bauthor{\bsnm{Brink}, \binits{M.}},
\bauthor{\bsnm{Chow}, \binits{J.M.}},
\bauthor{\bsnm{Gambetta}, \binits{J.M.}}:
\batitle{{Hardware-efficient variational quantum eigensolver for small
  molecules and quantum magnets}}.
\bjtitle{Nature}
\bvolume{549}(\bissue{7671}),
\bfpage{242}--\blpage{246}
(\byear{2017}).
doi:\doiurl{10.1038/nature23879}
\end{barticle}
\endbibitem

\bibitem{OMalley_2016}
\begin{barticle}
\bauthor{\bsnm{O'Malley}, \binits{P.J.J.}},
\bauthor{\bsnm{Babbush}, \binits{R.}},
\bauthor{\bsnm{Kivlichan}, \binits{I.D.}},
\bauthor{\bsnm{Romero}, \binits{J.}},
\bauthor{\bsnm{McClean}, \binits{J.R.}},
\bauthor{\bsnm{Barends}, \binits{R.}},
\bauthor{\bsnm{Kelly}, \binits{J.}},
\bauthor{\bsnm{Roushan}, \binits{P.}},
\bauthor{\bsnm{Tranter}, \binits{A.}},
\bauthor{\bsnm{Ding}, \binits{N.}},
\bauthor{\bsnm{Campbell}, \binits{B.}},
\bauthor{\bsnm{Chen}, \binits{Y.}},
\bauthor{\bsnm{Chen}, \binits{Z.}},
\bauthor{\bsnm{Chiaro}, \binits{B.}},
\bauthor{\bsnm{Dunsworth}, \binits{A.}},
\bauthor{\bsnm{Fowler}, \binits{A.G.}},
\bauthor{\bsnm{Jeffrey}, \binits{E.}},
\bauthor{\bsnm{Lucero}, \binits{E.}},
\bauthor{\bsnm{Megrant}, \binits{A.}},
\bauthor{\bsnm{Mutus}, \binits{J.Y.}},
\bauthor{\bsnm{Neeley}, \binits{M.}},
\bauthor{\bsnm{Neill}, \binits{C.}},
\bauthor{\bsnm{Quintana}, \binits{C.}},
\bauthor{\bsnm{Sank}, \binits{D.}},
\bauthor{\bsnm{Vainsencher}, \binits{A.}},
\bauthor{\bsnm{Wenner}, \binits{J.}},
\bauthor{\bsnm{White}, \binits{T.C.}},
\bauthor{\bsnm{Coveney}, \binits{P.V.}},
\bauthor{\bsnm{Love}, \binits{P.J.}},
\bauthor{\bsnm{Neven}, \binits{H.}},
\bauthor{\bsnm{Aspuru-Guzik}, \binits{A.}},
\bauthor{\bsnm{Martinis}, \binits{J.M.}}:
\batitle{Scalable quantum simulation of molecular energies}.
\bjtitle{Phys. Rev. X}
\bvolume{6},
\bfpage{031007}
(\byear{2016}).
doi:\doiurl{10.1103/PhysRevX.6.031007}
\end{barticle}
\endbibitem

\bibitem{nam2019groundstate}
\begin{botherref}
\oauthor{\bsnm{Nam}, \binits{Y.}},
\oauthor{\bsnm{Chen}, \binits{J.-S.}},
\oauthor{\bsnm{Pisenti}, \binits{N.C.}},
\oauthor{\bsnm{Wright}, \binits{K.}},
\oauthor{\bsnm{Delaney}, \binits{C.}},
\oauthor{\bsnm{Maslov}, \binits{D.}},
\oauthor{\bsnm{Brown}, \binits{K.R.}},
\oauthor{\bsnm{Allen}, \binits{S.}},
\oauthor{\bsnm{Amini}, \binits{J.M.}},
\oauthor{\bsnm{Apisdorf}, \binits{J.}},
\oauthor{\bsnm{Beck}, \binits{K.M.}},
\oauthor{\bsnm{Blinov}, \binits{A.}},
\oauthor{\bsnm{Chaplin}, \binits{V.}},
\oauthor{\bsnm{Chmielewski}, \binits{M.}},
\oauthor{\bsnm{Collins}, \binits{C.}},
\oauthor{\bsnm{Debnath}, \binits{S.}},
\oauthor{\bsnm{Ducore}, \binits{A.M.}},
\oauthor{\bsnm{Hudek}, \binits{K.M.}},
\oauthor{\bsnm{Keesan}, \binits{M.}},
\oauthor{\bsnm{Kreikemeier}, \binits{S.M.}},
\oauthor{\bsnm{Mizrahi}, \binits{J.}},
\oauthor{\bsnm{Solomon}, \binits{P.}},
\oauthor{\bsnm{Williams}, \binits{M.}},
\oauthor{\bsnm{Wong-Campos}, \binits{J.D.}},
\oauthor{\bsnm{Monroe}, \binits{C.}},
\oauthor{\bsnm{Kim}, \binits{J.}}:
Ground-state energy estimation of the water molecule on a trapped ion quantum
  computer.
arXiv
(2019).
\arxivurl{1902.10171}
\end{botherref}
\endbibitem

\bibitem{colless_2018}
\begin{barticle}
\bauthor{\bsnm{Colless}, \binits{J.I.}},
\bauthor{\bsnm{Ramasesh}, \binits{V.V.}},
\bauthor{\bsnm{Dahlen}, \binits{D.}},
\bauthor{\bsnm{Blok}, \binits{M.S.}},
\bauthor{\bsnm{Kimchi-Schwartz}, \binits{M.E.}},
\bauthor{\bsnm{McClean}, \binits{J.R.}},
\bauthor{\bsnm{Carter}, \binits{J.}},
\bauthor{\bparticle{de} \bsnm{Jong}, \binits{W.A.}},
\bauthor{\bsnm{Siddiqi}, \binits{I.}}:
\batitle{Computation of molecular spectra on a quantum processor with an
  error-resilient algorithm}.
\bjtitle{Phys. Rev. X}
\bvolume{8},
\bfpage{011021}
(\byear{2018}).
doi:\doiurl{10.1103/PhysRevX.8.011021}
\end{barticle}
\endbibitem

\bibitem{qcc}
\begin{barticle}
\bauthor{\bsnm{Ryabinkin}, \binits{I.G.}},
\bauthor{\bsnm{Yen}, \binits{T.-C.}},
\bauthor{\bsnm{Genin}, \binits{S.N.}},
\bauthor{\bsnm{Izmaylov}, \binits{A.F.}}:
\batitle{Qubit coupled cluster method: A systematic approach to quantum
  chemistry on a quantum computer}.
\bjtitle{J. Chem. Theory Comput.}
\bvolume{14}(\bissue{12}),
\bfpage{6317}--\blpage{6326}
(\byear{2018}).
doi:\doiurl{10.1021/acs.jctc.8b00932}
\end{barticle}
\endbibitem

\bibitem{McCaskey2019}
\begin{barticle}
\bauthor{\bsnm{McCaskey}, \binits{A.J.}},
\bauthor{\bsnm{Parks}, \binits{Z.P.}},
\bauthor{\bsnm{Jakowski}, \binits{J.}},
\bauthor{\bsnm{Moore}, \binits{S.V.}},
\bauthor{\bsnm{Morris}, \binits{T.D.}},
\bauthor{\bsnm{Humble}, \binits{T.S.}},
\bauthor{\bsnm{Pooser}, \binits{R.C.}}:
\batitle{{Quantum chemistry as a benchmark for near-term quantum computers}}.
\bjtitle{npj Quantum Inf.}
\bvolume{5}(\bissue{1}),
\bfpage{99}
(\byear{2019}).
doi:\doiurl{10.1038/s41534-019-0209-0}
\end{barticle}
\endbibitem

\bibitem{hf_on_qc_google}
\begin{barticle}
\bauthor{\bsnm{{Collaborators*†, Google AI Quantum and}}},
\bauthor{\bsnm{Arute}, \binits{F.}},
\bauthor{\bsnm{Arya}, \binits{K.}},
\bauthor{\bsnm{Babbush}, \binits{R.}},
\bauthor{\bsnm{Bacon}, \binits{D.}},
\bauthor{\bsnm{Bardin}, \binits{J.C.}},
\bauthor{\bsnm{Barends}, \binits{R.}},
\bauthor{\bsnm{Boixo}, \binits{S.}},
\bauthor{\bsnm{Broughton}, \binits{M.}},
\bauthor{\bsnm{Buckley}, \binits{B.B.}},
\bauthor{\bsnm{Buell}, \binits{D.A.}},
\bauthor{\bsnm{Burkett}, \binits{B.}},
\bauthor{\bsnm{Bushnell}, \binits{N.}},
\bauthor{\bsnm{Chen}, \binits{Y.}},
\bauthor{\bsnm{Chen}, \binits{Z.}},
\bauthor{\bsnm{Chiaro}, \binits{B.}},
\bauthor{\bsnm{Collins}, \binits{R.}},
\bauthor{\bsnm{Courtney}, \binits{W.}},
\bauthor{\bsnm{Demura}, \binits{S.}},
\bauthor{\bsnm{Dunsworth}, \binits{A.}},
\bauthor{\bsnm{Farhi}, \binits{E.}},
\bauthor{\bsnm{Fowler}, \binits{A.}},
\bauthor{\bsnm{Foxen}, \binits{B.}},
\bauthor{\bsnm{Gidney}, \binits{C.}},
\bauthor{\bsnm{Giustina}, \binits{M.}},
\bauthor{\bsnm{Graff}, \binits{R.}},
\bauthor{\bsnm{Habegger}, \binits{S.}},
\bauthor{\bsnm{Harrigan}, \binits{M.P.}},
\bauthor{\bsnm{Ho}, \binits{A.}},
\bauthor{\bsnm{Hong}, \binits{S.}},
\bauthor{\bsnm{Huang}, \binits{T.}},
\bauthor{\bsnm{Huggins}, \binits{W.J.}},
\bauthor{\bsnm{Ioffe}, \binits{L.}},
\bauthor{\bsnm{Isakov}, \binits{S.V.}},
\bauthor{\bsnm{Jeffrey}, \binits{E.}},
\bauthor{\bsnm{Jiang}, \binits{Z.}},
\bauthor{\bsnm{Jones}, \binits{C.}},
\bauthor{\bsnm{Kafri}, \binits{D.}},
\bauthor{\bsnm{Kechedzhi}, \binits{K.}},
\bauthor{\bsnm{Kelly}, \binits{J.}},
\bauthor{\bsnm{Kim}, \binits{S.}},
\bauthor{\bsnm{Klimov}, \binits{P.V.}},
\bauthor{\bsnm{Korotkov}, \binits{A.}},
\bauthor{\bsnm{Kostritsa}, \binits{F.}},
\bauthor{\bsnm{Landhuis}, \binits{D.}},
\bauthor{\bsnm{Laptev}, \binits{P.}},
\bauthor{\bsnm{Lindmark}, \binits{M.}},
\bauthor{\bsnm{Lucero}, \binits{E.}},
\bauthor{\bsnm{Martin}, \binits{O.}},
\bauthor{\bsnm{Martinis}, \binits{J.M.}},
\bauthor{\bsnm{McClean}, \binits{J.R.}},
\bauthor{\bsnm{McEwen}, \binits{M.}},
\bauthor{\bsnm{Megrant}, \binits{A.}},
\bauthor{\bsnm{Mi}, \binits{X.}},
\bauthor{\bsnm{Mohseni}, \binits{M.}},
\bauthor{\bsnm{Mruczkiewicz}, \binits{W.}},
\bauthor{\bsnm{Mutus}, \binits{J.}},
\bauthor{\bsnm{Naaman}, \binits{O.}},
\bauthor{\bsnm{Neeley}, \binits{M.}},
\bauthor{\bsnm{Neill}, \binits{C.}},
\bauthor{\bsnm{Neven}, \binits{H.}},
\bauthor{\bsnm{Niu}, \binits{M.Y.}},
\bauthor{\bsnm{O’Brien}, \binits{T.E.}},
\bauthor{\bsnm{Ostby}, \binits{E.}},
\bauthor{\bsnm{Petukhov}, \binits{A.}},
\bauthor{\bsnm{Putterman}, \binits{H.}},
\bauthor{\bsnm{Quintana}, \binits{C.}},
\bauthor{\bsnm{Roushan}, \binits{P.}},
\bauthor{\bsnm{Rubin}, \binits{N.C.}},
\bauthor{\bsnm{Sank}, \binits{D.}},
\bauthor{\bsnm{Satzinger}, \binits{K.J.}},
\bauthor{\bsnm{Smelyanskiy}, \binits{V.}},
\bauthor{\bsnm{Strain}, \binits{D.}},
\bauthor{\bsnm{Sung}, \binits{K.J.}},
\bauthor{\bsnm{Szalay}, \binits{M.}},
\bauthor{\bsnm{Takeshita}, \binits{T.Y.}},
\bauthor{\bsnm{Vainsencher}, \binits{A.}},
\bauthor{\bsnm{White}, \binits{T.}},
\bauthor{\bsnm{Wiebe}, \binits{N.}},
\bauthor{\bsnm{Yao}, \binits{Z.J.}},
\bauthor{\bsnm{Yeh}, \binits{P.}},
\bauthor{\bsnm{Zalcman}, \binits{A.}}:
\batitle{{Hartree-Fock on a superconducting qubit quantum computer}}.
\bjtitle{Science}
\bvolume{369}(\bissue{6507}),
\bfpage{1084}--\blpage{1089}
(\byear{2020}).
doi:\doiurl{10.1126/science.abb9811}.
\arxivurl{2004.04174}
\end{barticle}
\endbibitem

\bibitem{hempel2018}
\begin{barticle}
\bauthor{\bsnm{Hempel}, \binits{C.}},
\bauthor{\bsnm{Maier}, \binits{C.}},
\bauthor{\bsnm{Romero}, \binits{J.}},
\bauthor{\bsnm{McClean}, \binits{J.}},
\bauthor{\bsnm{Monz}, \binits{T.}},
\bauthor{\bsnm{Shen}, \binits{H.}},
\bauthor{\bsnm{Jurcevic}, \binits{P.}},
\bauthor{\bsnm{Lanyon}, \binits{B.P.}},
\bauthor{\bsnm{Love}, \binits{P.}},
\bauthor{\bsnm{Babbush}, \binits{R.}},
\bauthor{\bsnm{Aspuru-Guzik}, \binits{A.}},
\bauthor{\bsnm{Blatt}, \binits{R.}},
\bauthor{\bsnm{Roos}, \binits{C.F.}}:
\batitle{Quantum chemistry calculations on a trapped-ion quantum simulator}.
\bjtitle{Phys. Rev. X}
\bvolume{8},
\bfpage{031022}
(\byear{2018}).
doi:\doiurl{10.1103/PhysRevX.8.031022}
\end{barticle}
\endbibitem

\bibitem{shen2017}
\begin{barticle}
\bauthor{\bsnm{Shen}, \binits{Y.}},
\bauthor{\bsnm{Zhang}, \binits{X.}},
\bauthor{\bsnm{Zhang}, \binits{S.}},
\bauthor{\bsnm{Zhang}, \binits{J.-N.}},
\bauthor{\bsnm{Yung}, \binits{M.-H.}},
\bauthor{\bsnm{Kim}, \binits{K.}}:
\batitle{Quantum implementation of the unitary coupled cluster for simulating
  molecular electronic structure}.
\bjtitle{Phys. Rev. A}
\bvolume{95},
\bfpage{020501}
(\byear{2017}).
doi:\doiurl{10.1103/PhysRevA.95.020501}
\end{barticle}
\endbibitem

\bibitem{Santagatieaap9646}
\begin{botherref}
\oauthor{\bsnm{Santagati}, \binits{R.}},
\oauthor{\bsnm{Wang}, \binits{J.}},
\oauthor{\bsnm{Gentile}, \binits{A.A.}},
\oauthor{\bsnm{Paesani}, \binits{S.}},
\oauthor{\bsnm{Wiebe}, \binits{N.}},
\oauthor{\bsnm{McClean}, \binits{J.R.}},
\oauthor{\bsnm{Morley-Short}, \binits{S.}},
\oauthor{\bsnm{Shadbolt}, \binits{P.J.}},
\oauthor{\bsnm{Bonneau}, \binits{D.}},
\oauthor{\bsnm{Silverstone}, \binits{J.W.}},
\oauthor{\bsnm{Tew}, \binits{D.P.}},
\oauthor{\bsnm{Zhou}, \binits{X.}},
\oauthor{\bsnm{O{\textquoteright}Brien}, \binits{J.L.}},
\oauthor{\bsnm{Thompson}, \binits{M.G.}}:
Witnessing eigenstates for quantum simulation of hamiltonian spectra.
Sci. Adv.
\textbf{4}(1)
(2018).
doi:\doiurl{10.1126/sciadv.aap9646}.
\arxivurl{https://advances.sciencemag.org/content/4/1/eaap9646.full.pdf}
\end{botherref}
\endbibitem

\bibitem{gao2019}
\begin{barticle}
\bauthor{\bsnm{Gao}, \binits{Q.}},
\bauthor{\bsnm{Nakamura}, \binits{H.}},
\bauthor{\bsnm{Gujarati}, \binits{T.P.}},
\bauthor{\bsnm{Jones}, \binits{G.O.}},
\bauthor{\bsnm{Rice}, \binits{J.E.}},
\bauthor{\bsnm{Wood}, \binits{S.P.}},
\bauthor{\bsnm{Pistoia}, \binits{M.}},
\bauthor{\bsnm{Garcia}, \binits{J.M.}},
\bauthor{\bsnm{Yamamoto}, \binits{N.}}:
\batitle{Computational investigations of the lithium superoxide dimer
  rearrangement on noisy quantum devices}.
\bjtitle{J. Phys. Chem. A}
\bvolume{125}(\bissue{9}),
\bfpage{1827}--\blpage{1836}
(\byear{2021}).
doi:\doiurl{10.1021/acs.jpca.0c09530}.
\bcomment{PMID: 33635672}.
\arxivurl{https://doi.org/10.1021/acs.jpca.0c09530}
\end{barticle}
\endbibitem

\bibitem{Gao2020}
\begin{botherref}
\oauthor{\bsnm{Gao}, \binits{Q.}},
\oauthor{\bsnm{Jones}, \binits{G.O.}},
\oauthor{\bsnm{Motta}, \binits{M.}},
\oauthor{\bsnm{Sugawara}, \binits{M.}},
\oauthor{\bsnm{Watanabe}, \binits{H.C.}},
\oauthor{\bsnm{Kobayashi}, \binits{T.}},
\oauthor{\bsnm{Watanabe}, \binits{E.}},
\oauthor{\bsnm{Ohnishi}, \binits{Y.-y.}},
\oauthor{\bsnm{Nakamura}, \binits{H.}},
\oauthor{\bsnm{Yamamoto}, \binits{N.}}:
{Applications of Quantum Computing for Investigations of Electronic Transitions
  in Phenylsulfonyl-carbazole TADF Emitters}
(2020).
\arxivurl{2007.15795}
\end{botherref}
\endbibitem

\bibitem{smart_mazziotti_exp2019}
\begin{barticle}
\bauthor{\bsnm{Smart}, \binits{S.E.}},
\bauthor{\bsnm{Mazziotti}, \binits{D.A.}}:
\batitle{Quantum-classical hybrid algorithm using an error-mitigating
  $n$-representability condition to compute the mott metal-insulator
  transition}.
\bjtitle{Phys. Rev. A}
\bvolume{100},
\bfpage{022517}
(\byear{2019}).
doi:\doiurl{10.1103/PhysRevA.100.022517}
\end{barticle}
\endbibitem

\bibitem{quant_comp_chem_revmodphys}
\begin{barticle}
\bauthor{\bsnm{McArdle}, \binits{S.}},
\bauthor{\bsnm{Endo}, \binits{S.}}:
\batitle{Quantum computational chemistry}.
\bjtitle{Rev. Mod. Phys.}
\bvolume{92}(\bissue{1}),
\bfpage{015003}
(\byear{2020}).
doi:\doiurl{10.1103/revmodphys.92.015003}
\end{barticle}
\endbibitem

\bibitem{quant_chem_chemrev_2019}
\begin{barticle}
\bauthor{\bsnm{Cao}, \binits{Y.}},
\bauthor{\bsnm{Romero}, \binits{J.}},
\bauthor{\bsnm{Olson}, \binits{J.P.}},
\bauthor{\bsnm{Degroote}, \binits{M.}},
\bauthor{\bsnm{Johnson}, \binits{P.D.}},
\bauthor{\bsnm{Kieferova}, \binits{M.}},
\bauthor{\bsnm{Kivlichan}, \binits{I.D.}},
\bauthor{\bsnm{Menke}, \binits{T.}},
\bauthor{\bsnm{Peropadre}, \binits{B.}},
\bauthor{\bsnm{Sawaya}, \binits{N.P.D.}},
\bauthor{\bsnm{Sim}, \binits{S.}},
\bauthor{\bsnm{Veis}, \binits{L.}},
\bauthor{\bsnm{Aspuru-Guzik}, \binits{A.}}:
\batitle{Quantum chemistry in the age of quantum computing}.
\bjtitle{Chem. Rev.}
\bvolume{119}(\bissue{19}),
\bfpage{10856}--\blpage{10915}
(\byear{2019}).
doi:\doiurl{10.1021/acs.chemrev.8b00803}
\end{barticle}
\endbibitem

\bibitem{VQA_rev_2020}
\begin{botherref}
\oauthor{\bsnm{Cerezo}, \binits{M.}},
\oauthor{\bsnm{Arrasmith}, \binits{A.}},
\oauthor{\bsnm{Babbush}, \binits{R.}},
\oauthor{\bsnm{Benjamin}, \binits{S.C.}},
\oauthor{\bsnm{Endo}, \binits{S.}},
\oauthor{\bsnm{Fujii}, \binits{K.}},
\oauthor{\bsnm{McClean}, \binits{J.R.}},
\oauthor{\bsnm{Mitarai}, \binits{K.}},
\oauthor{\bsnm{Yuan}, \binits{X.}},
\oauthor{\bsnm{Cincio}, \binits{L.}},
\oauthor{\bsnm{Coles}, \binits{P.J.}}:
Variational quantum algorithms.
arXiv
(2020).
\arxivurl{2012.09265}
\end{botherref}
\endbibitem

\bibitem{bharti2021noisy}
\begin{botherref}
\oauthor{\bsnm{Bharti}, \binits{K.}},
\oauthor{\bsnm{Cervera-Lierta}, \binits{A.}},
\oauthor{\bsnm{Kyaw}, \binits{T.H.}},
\oauthor{\bsnm{Haug}, \binits{T.}},
\oauthor{\bsnm{Alperin-Lea}, \binits{S.}},
\oauthor{\bsnm{Anand}, \binits{A.}},
\oauthor{\bsnm{Degroote}, \binits{M.}},
\oauthor{\bsnm{Heimonen}, \binits{H.}},
\oauthor{\bsnm{Kottmann}, \binits{J.S.}},
\oauthor{\bsnm{Menke}, \binits{T.}},
\oauthor{\bsnm{Mok}, \binits{W.-K.}},
\oauthor{\bsnm{Sim}, \binits{S.}},
\oauthor{\bsnm{Kwek}, \binits{L.-C.}},
\oauthor{\bsnm{Aspuru-Guzik}, \binits{A.}}:
Noisy intermediate-scale quantum {(NISQ)} algorithms
(2021).
\arxivurl{2101.08448}
\end{botherref}
\endbibitem

\bibitem{plane_waves}
\begin{barticle}
\bauthor{\bsnm{Babbush}, \binits{R.}},
\bauthor{\bsnm{Wiebe}, \binits{N.}},
\bauthor{\bsnm{McClean}, \binits{J.}},
\bauthor{\bsnm{McClain}, \binits{J.}},
\bauthor{\bsnm{Neven}, \binits{H.}},
\bauthor{\bsnm{Chan}, \binits{G.K.-L.}}:
\batitle{Low-depth quantum simulation of materials}.
\bjtitle{Phys. Rev. X}
\bvolume{8},
\bfpage{011044}
(\byear{2018}).
doi:\doiurl{10.1103/PhysRevX.8.011044}
\end{barticle}
\endbibitem

\bibitem{basis_set_free}
\begin{barticle}
\bauthor{\bsnm{Kottmann}, \binits{J.S.}},
\bauthor{\bsnm{Schleich}, \binits{P.}},
\bauthor{\bsnm{Tamayo-Mendoza}, \binits{T.}},
\bauthor{\bsnm{Aspuru-Guzik}, \binits{A.}}:
\batitle{Reducing qubit requirements while maintaining numerical precision for
  the variational quantum eigensolver: A basis-set-free approach}.
\bjtitle{J. Phys. Chem. Lett.}
\bvolume{12}(\bissue{1}),
\bfpage{663}--\blpage{673}
(\byear{2021}).
doi:\doiurl{10.1021/acs.jpclett.0c03410}.
\bcomment{PMID: 33393305}.
\arxivurl{https://doi.org/10.1021/acs.jpclett.0c03410}
\end{barticle}
\endbibitem

\bibitem{babbush_2019_first_quant}
\begin{barticle}
\bauthor{\bsnm{Babbush}, \binits{R.}},
\bauthor{\bsnm{Berry}, \binits{D.W.}},
\bauthor{\bsnm{McClean}, \binits{J.R.}},
\bauthor{\bsnm{Neven}, \binits{H.}}:
\batitle{{Quantum simulation of chemistry with sublinear scaling in basis
  size}}.
\bjtitle{npj Quantum Inf.}
\bvolume{5}(\bissue{1}),
\bfpage{92}
(\byear{2019}).
doi:\doiurl{10.1038/s41534-019-0199-y}.
\arxivurl{1807.09802}
\end{barticle}
\endbibitem

\bibitem{Jordan1928}
\begin{barticle}
\bauthor{\bsnm{Jordan}, \binits{P.}},
\bauthor{\bsnm{Wigner}, \binits{E.}}:
\batitle{{{\"{U}}ber das Paulische {\"{A}}quivalenzverbot}}.
\bjtitle{Z. Phys.}
\bvolume{47}(\bissue{9-10}),
\bfpage{631}--\blpage{651}
(\byear{1928}).
doi:\doiurl{10.1007/BF01331938}
\end{barticle}
\endbibitem

\bibitem{bravyi2017_parity}
\begin{botherref}
\oauthor{\bsnm{Bravyi}, \binits{S.}},
\oauthor{\bsnm{Gambetta}, \binits{J.M.}},
\oauthor{\bsnm{Mezzacapo}, \binits{A.}},
\oauthor{\bsnm{Temme}, \binits{K.}}:
{Tapering off qubits to simulate fermionic Hamiltonians}.
arXiv
(2017).
\arxivurl{1701.08213}
\end{botherref}
\endbibitem

\bibitem{Bravyi2002_BK_mapping}
\begin{barticle}
\bauthor{\bsnm{Bravyi}, \binits{S.B.}},
\bauthor{\bsnm{Kitaev}, \binits{A.Y.}}:
\batitle{Fermionic quantum computation}.
\bjtitle{Ann. Phys.}
\bvolume{298}(\bissue{1}),
\bfpage{210}--\blpage{226}
(\byear{2002}).
doi:\doiurl{10.1006/aphy.2002.6254}
\end{barticle}
\endbibitem

\bibitem{ham_averaging_McClean_2014}
\begin{barticle}
\bauthor{\bsnm{McClean}, \binits{J.R.}},
\bauthor{\bsnm{Babbush}, \binits{R.}},
\bauthor{\bsnm{Love}, \binits{P.J.}},
\bauthor{\bsnm{Aspuru-Guzik}, \binits{A.}}:
\batitle{{Exploiting Locality in Quantum Computation for Quantum Chemistry}}.
\bjtitle{J. Phys. Chem. Lett.}
\bvolume{5}(\bissue{24}),
\bfpage{4368}--\blpage{4380}
(\byear{2014}).
doi:\doiurl{10.1021/jz501649m}
\end{barticle}
\endbibitem

\bibitem{qpe_Wiebe}
\begin{barticle}
\bauthor{\bsnm{Wiebe}, \binits{N.}},
\bauthor{\bsnm{Granade}, \binits{C.}}:
\batitle{Efficient {Bayesian} phase estimation}.
\bjtitle{Phys. Rev. Lett.}
\bvolume{117},
\bfpage{010503}
(\byear{2016}).
doi:\doiurl{10.1103/PhysRevLett.117.010503}
\end{barticle}
\endbibitem

\bibitem{ham_var_ansatz_Wecker2015}
\begin{barticle}
\bauthor{\bsnm{Wecker}, \binits{D.}},
\bauthor{\bsnm{Hastings}, \binits{M.B.}},
\bauthor{\bsnm{Troyer}, \binits{M.}}:
\batitle{Progress towards practical quantum variational algorithms}.
\bjtitle{Phys. Rev. A}
\bvolume{92},
\bfpage{042303}
(\byear{2015}).
doi:\doiurl{10.1103/PhysRevA.92.042303}
\end{barticle}
\endbibitem

\bibitem{kuhn_UCCSD_resources}
\begin{barticle}
\bauthor{\bsnm{Kühn}, \binits{M.}},
\bauthor{\bsnm{Zanker}, \binits{S.}},
\bauthor{\bsnm{Deglmann}, \binits{P.}},
\bauthor{\bsnm{Marthaler}, \binits{M.}},
\bauthor{\bsnm{Weiß}, \binits{H.}}:
\batitle{Accuracy and resource estimations for quantum chemistry on a near-term
  quantum computer}.
\bjtitle{J. Chem. Theory Comput.}
\bvolume{15}(\bissue{9}),
\bfpage{4764}--\blpage{4780}
(\byear{2019}).
doi:\doiurl{10.1021/acs.jctc.9b00236}.
\bcomment{PMID: 31403781}.
\arxivurl{https://doi.org/10.1021/acs.jctc.9b00236}
\end{barticle}
\endbibitem

\bibitem{kandala_2019_he_on_hardware}
\begin{barticle}
\bauthor{\bsnm{Kandala}, \binits{A.}},
\bauthor{\bsnm{Temme}, \binits{K.}},
\bauthor{\bsnm{Córcoles}, \binits{A.D.}},
\bauthor{\bsnm{Mezzacapo}, \binits{A.}},
\bauthor{\bsnm{Chow}, \binits{J.M.}},
\bauthor{\bsnm{Gambetta}, \binits{J.M.}}:
\batitle{{Error mitigation extends the computational reach of a noisy quantum
  processor}}.
\bjtitle{Nature}
\bvolume{567}(\bissue{7749}),
\bfpage{491}--\blpage{495}
(\byear{2019}).
doi:\doiurl{10.1038/s41586-019-1040-7}
\end{barticle}
\endbibitem

\bibitem{barren_plateaus}
\begin{barticle}
\bauthor{\bsnm{McClean}, \binits{J.R.}},
\bauthor{\bsnm{Boixo}, \binits{S.}},
\bauthor{\bsnm{Smelyanskiy}, \binits{V.N.}},
\bauthor{\bsnm{Babbush}, \binits{R.}},
\bauthor{\bsnm{Neven}, \binits{H.}}:
\batitle{{Barren plateaus in quantum neural network training landscapes}}.
\bjtitle{Nat. Commun.}
\bvolume{9}(\bissue{1}),
\bfpage{4812}
(\byear{2018}).
doi:\doiurl{10.1038/s41467-018-07090-4}.
\arxivurl{1803.11173}
\end{barticle}
\endbibitem

\bibitem{Barkoutsos_2018}
\begin{barticle}
\bauthor{\bsnm{Barkoutsos}, \binits{P.K.}},
\bauthor{\bsnm{Gonthier}, \binits{J.F.}},
\bauthor{\bsnm{Sokolov}, \binits{I.}},
\bauthor{\bsnm{Moll}, \binits{N.}},
\bauthor{\bsnm{Salis}, \binits{G.}},
\bauthor{\bsnm{Fuhrer}, \binits{A.}},
\bauthor{\bsnm{Ganzhorn}, \binits{M.}},
\bauthor{\bsnm{Egger}, \binits{D.J.}},
\bauthor{\bsnm{Troyer}, \binits{M.}},
\bauthor{\bsnm{Mezzacapo}, \binits{A.}},
\bauthor{\bsnm{Filipp}, \binits{S.}},
\bauthor{\bsnm{Tavernelli}, \binits{I.}}:
\batitle{{Quantum algorithms for electronic structure calculations:
  Particle-hole Hamiltonian and optimized wave-function expansions}}.
\bjtitle{Phys. Rev. A}
\bvolume{98}(\bissue{2}),
\bfpage{022322}
(\byear{2018}).
doi:\doiurl{10.1103/physreva.98.022322}.
\arxivurl{1805.04340}
\end{barticle}
\endbibitem

\bibitem{Ganzhorn2019}
\begin{barticle}
\bauthor{\bsnm{Ganzhorn}, \binits{M.}},
\bauthor{\bsnm{Egger}, \binits{D.J.}},
\bauthor{\bsnm{Barkoutsos}, \binits{P.}},
\bauthor{\bsnm{Ollitrault}, \binits{P.}},
\bauthor{\bsnm{Salis}, \binits{G.}},
\bauthor{\bsnm{Moll}, \binits{N.}},
\bauthor{\bsnm{Roth}, \binits{M.}},
\bauthor{\bsnm{Fuhrer}, \binits{A.}},
\bauthor{\bsnm{Mueller}, \binits{P.}},
\bauthor{\bsnm{Woerner}, \binits{S.}},
\bauthor{\bsnm{Tavernelli}, \binits{I.}},
\bauthor{\bsnm{Filipp}, \binits{S.}}:
\batitle{Gate-efficient simulation of molecular eigenstates on a quantum
  computer}.
\bjtitle{Phys. Rev. Applied}
\bvolume{11},
\bfpage{044092}
(\byear{2019}).
doi:\doiurl{10.1103/PhysRevApplied.11.044092}
\end{barticle}
\endbibitem

\bibitem{Grant2019initialization}
\begin{barticle}
\bauthor{\bsnm{Grant}, \binits{E.}},
\bauthor{\bsnm{Wossnig}, \binits{L.}},
\bauthor{\bsnm{Ostaszewski}, \binits{M.}},
\bauthor{\bsnm{Benedetti}, \binits{M.}}:
\batitle{An initialization strategy for addressing barren plateaus in
  parametrized quantum circuits}.
\bjtitle{{Quantum}}
\bvolume{3},
\bfpage{214}
(\byear{2019}).
doi:\doiurl{10.22331/q-2019-12-09-214}
\end{barticle}
\endbibitem

\bibitem{par_corr_2021}
\begin{barticle}
\bauthor{\bsnm{Volkoff}, \binits{T.}},
\bauthor{\bsnm{Coles}, \binits{P.J.}}:
\batitle{Large gradients via correlation in random parameterized quantum
  circuits}.
\bjtitle{Quantum Sci. Technol.}
\bvolume{6}(\bissue{2}),
\bfpage{025008}
(\byear{2021}).
doi:\doiurl{10.1088/2058-9565/abd891}.
\arxivurl{2005.12200}
\end{barticle}
\endbibitem

\bibitem{k_up_uccsd}
\begin{barticle}
\bauthor{\bsnm{Lee}, \binits{J.}},
\bauthor{\bsnm{Huggins}, \binits{W.J.}},
\bauthor{\bsnm{Head-Gordon}, \binits{M.}},
\bauthor{\bsnm{Whaley}, \binits{K.B.}}:
\batitle{Generalized unitary coupled cluster wave functions for quantum
  computation}.
\bjtitle{J. Chem. Theory Comput.}
\bvolume{15}(\bissue{1}),
\bfpage{311}--\blpage{324}
(\byear{2018}).
doi:\doiurl{10.1021/acs.jctc.8b01004}
\end{barticle}
\endbibitem

\bibitem{Mizukami20_033421}
\begin{barticle}
\bauthor{\bsnm{Mizukami}, \binits{W.}},
\bauthor{\bsnm{Mitarai}, \binits{K.}},
\bauthor{\bsnm{Nakagawa}, \binits{Y.O.}},
\bauthor{\bsnm{Yamamoto}, \binits{T.}},
\bauthor{\bsnm{Yan}, \binits{T.}},
\bauthor{\bsnm{Ohnishi}, \binits{Y.-y.}}:
\batitle{Orbital optimized unitary coupled cluster theory for quantum
  computer}.
\bjtitle{Phys. Rev. Research}
\bvolume{2},
\bfpage{033421}
(\byear{2020}).
doi:\doiurl{10.1103/PhysRevResearch.2.033421}
\end{barticle}
\endbibitem

\bibitem{Metcalf20_6165}
\begin{barticle}
\bauthor{\bsnm{Metcalf}, \binits{M.}},
\bauthor{\bsnm{Bauman}, \binits{N.P.}},
\bauthor{\bsnm{Kowalski}, \binits{K.}},
\bauthor{\bparticle{de} \bsnm{Jong}, \binits{W.A.}}:
\batitle{Resource-efficient chemistry on quantum computers with the variational
  quantum eigensolver and the double unitary coupled-cluster approach}.
\bjtitle{J. Chem. Theory Comput.}
\bvolume{16}(\bissue{10}),
\bfpage{6165}--\blpage{6175}
(\byear{2020}).
doi:\doiurl{10.1021/acs.jctc.0c00421}.
\bcomment{PMID: 32915568}.
\arxivurl{https://doi.org/10.1021/acs.jctc.0c00421}
\end{barticle}
\endbibitem

\bibitem{Kowalski18_094104}
\begin{barticle}
\bauthor{\bsnm{Kowalski}, \binits{K.}}:
\batitle{Properties of coupled-cluster equations originating in excitation
  sub-algebras}.
\bjtitle{J. Chem. Phys.}
\bvolume{148}(\bissue{9}),
\bfpage{094104}
(\byear{2018}).
doi:\doiurl{10.1063/1.5010693}.
\arxivurl{https://doi.org/10.1063/1.5010693}
\end{barticle}
\endbibitem

\bibitem{McClean2017Hybrid}
\begin{barticle}
\bauthor{\bsnm{McClean}, \binits{J.R.}},
\bauthor{\bsnm{Kimchi-Schwartz}, \binits{M.E.}},
\bauthor{\bsnm{Carter}, \binits{J.}},
\bauthor{\bparticle{de} \bsnm{Jong}, \binits{W.A.}}:
\batitle{Hybrid quantum-classical hierarchy for mitigation of decoherence and
  determination of excited states}.
\bjtitle{Phys. Rev. A}
\bvolume{95},
\bfpage{042308}
(\byear{2017}).
doi:\doiurl{10.1103/PhysRevA.95.042308}
\end{barticle}
\endbibitem

\bibitem{Takeshita2020Quantum}
\begin{barticle}
\bauthor{\bsnm{Takeshita}, \binits{T.}},
\bauthor{\bsnm{Rubin}, \binits{N.C.}},
\bauthor{\bsnm{Jiang}, \binits{Z.}},
\bauthor{\bsnm{Lee}, \binits{E.}},
\bauthor{\bsnm{Babbush}, \binits{R.}},
\bauthor{\bsnm{McClean}, \binits{J.R.}}:
\batitle{Increasing the representation accuracy of quantum simulations of
  chemistry without extra quantum resources}.
\bjtitle{Phys. Rev. X}
\bvolume{10},
\bfpage{011004}
(\byear{2020}).
doi:\doiurl{10.1103/PhysRevX.10.011004}
\end{barticle}
\endbibitem

\bibitem{Motta2019QITE}
\begin{botherref}
\oauthor{\bsnm{Motta}, \binits{M.}},
\oauthor{\bsnm{Sun}, \binits{C.}},
\oauthor{\bsnm{Tan}, \binits{A.T.}},
\oauthor{\bsnm{O’Rourke}, \binits{M.J.}},
\oauthor{\bsnm{Ye}, \binits{E.}},
\oauthor{\bsnm{Minnich}, \binits{A.J.}},
\oauthor{\bsnm{Brandão}, \binits{F.G.}},
\oauthor{\bsnm{Chan}, \binits{G.K.-L.}}:
Determining eigenstates and thermal states on a quantum computer using quantum
  imaginary time evolution.
Nat. Phys.
\textbf{16}(2)
(2019).
doi:\doiurl{10.1038/s41567-019-0704-4}
\end{botherref}
\endbibitem

\bibitem{Huggins_2020}
\begin{barticle}
\bauthor{\bsnm{Huggins}, \binits{W.J.}},
\bauthor{\bsnm{Lee}, \binits{J.}},
\bauthor{\bsnm{Baek}, \binits{U.}},
\bauthor{\bsnm{O'Gorman}, \binits{B.}},
\bauthor{\bsnm{Whaley}, \binits{K.B.}}:
\batitle{A non-orthogonal variational quantum eigensolver}.
\bjtitle{New J. Phys.}
\bvolume{22}(\bissue{7}),
\bfpage{073009}
(\byear{2020}).
doi:\doiurl{10.1088/1367-2630/ab867b}
\end{barticle}
\endbibitem

\bibitem{parrish2019quantum}
\begin{botherref}
\oauthor{\bsnm{Parrish}, \binits{R.M.}},
\oauthor{\bsnm{McMahon}, \binits{P.L.}}:
Quantum filter diagonalization: Quantum eigendecomposition without full quantum
  phase estimation.
arXiv
(2019).
\arxivurl{1909.08925}
\end{botherref}
\endbibitem

\bibitem{Ollitrault2020}
\begin{barticle}
\bauthor{\bsnm{Ollitrault}, \binits{P.J.}},
\bauthor{\bsnm{Kandala}, \binits{A.}},
\bauthor{\bsnm{Chen}, \binits{C.-F.}},
\bauthor{\bsnm{Barkoutsos}, \binits{P.K.}},
\bauthor{\bsnm{Mezzacapo}, \binits{A.}},
\bauthor{\bsnm{Pistoia}, \binits{M.}},
\bauthor{\bsnm{Sheldon}, \binits{S.}},
\bauthor{\bsnm{Woerner}, \binits{S.}},
\bauthor{\bsnm{Gambetta}, \binits{J.M.}},
\bauthor{\bsnm{Tavernelli}, \binits{I.}}:
\batitle{Quantum equation of motion for computing molecular excitation energies
  on a noisy quantum processor}.
\bjtitle{Phys. Rev. Research}
\bvolume{2},
\bfpage{043140}
(\byear{2020}).
doi:\doiurl{10.1103/PhysRevResearch.2.043140}
\end{barticle}
\endbibitem

\bibitem{Stair2020Multireference}
\begin{barticle}
\bauthor{\bsnm{Stair}, \binits{N.H.}},
\bauthor{\bsnm{Huang}, \binits{R.}},
\bauthor{\bsnm{Evangelista}, \binits{F.A.}}:
\batitle{A multireference quantum {Krylov} algorithm for strongly correlated
  electrons}.
\bjtitle{J. Chem. Theory Comput.}
\bvolume{16}(\bissue{4}),
\bfpage{2236}--\blpage{2245}
(\byear{2020}).
doi:\doiurl{10.1021/acs.jctc.9b01125}.
\bcomment{PMID: 32091895}.
\arxivurl{https://doi.org/10.1021/acs.jctc.9b01125}
\end{barticle}
\endbibitem

\bibitem{kowalski2020CMX}
\begin{barticle}
\bauthor{\bsnm{Kowalski}, \binits{K.}},
\bauthor{\bsnm{Peng}, \binits{B.}}:
\batitle{Quantum simulations employing connected moments expansions}.
\bjtitle{J. Chem. Phys.}
\bvolume{153}(\bissue{20}),
\bfpage{201102}
(\byear{2020}).
doi:\doiurl{10.1063/5.0030688}.
\arxivurl{https://doi.org/10.1063/5.0030688}
\end{barticle}
\endbibitem

\bibitem{peng2021variational}
\begin{botherref}
\oauthor{\bsnm{Peng}, \binits{B.}},
\oauthor{\bsnm{Kowalski}, \binits{K.}}:
Variational quantum solver employing the {PDS} energy functional.
arXiv
(2021).
\arxivurl{2101.08526}
\end{botherref}
\endbibitem

\bibitem{horn1984t}
\begin{barticle}
\bauthor{\bsnm{Horn}, \binits{D.}},
\bauthor{\bsnm{Weinstein}, \binits{M.}}:
\batitle{The t expansion: a nonperturbative analytic tool for hamiltonian
  systems}.
\bjtitle{Phys. Rev. D}
\bvolume{30}(\bissue{6}),
\bfpage{1256}
(\byear{1984})
\end{barticle}
\endbibitem

\bibitem{cioslowski1987connected}
\begin{barticle}
\bauthor{\bsnm{Cioslowski}, \binits{J.}}:
\batitle{Connected moments expansion: a new tool for quantum many-body theory}.
\bjtitle{Phys. Rev. Lett.}
\bvolume{58}(\bissue{2}),
\bfpage{83}
(\byear{1987})
\end{barticle}
\endbibitem

\bibitem{peeters1984upper}
\begin{barticle}
\bauthor{\bsnm{Peeters}, \binits{F.}},
\bauthor{\bsnm{Devreese}, \binits{J.}}:
\batitle{Upper bounds for the free energy. a generalisation of the {Bogolubov}
  inequality and the feynman inequality}.
\bjtitle{J. Phys. A}
\bvolume{17}(\bissue{3}),
\bfpage{625}
(\byear{1984})
\end{barticle}
\endbibitem

\bibitem{soldatov1995generalized}
\begin{barticle}
\bauthor{\bsnm{Soldatov}, \binits{A.}}:
\batitle{Generalized variational principle in quantum mechanics}.
\bjtitle{Int. J. Mod. Phys. B}
\bvolume{9}(\bissue{22}),
\bfpage{2899}--\blpage{2936}
(\byear{1995})
\end{barticle}
\endbibitem

\bibitem{cse:mazziotti1998}
\begin{barticle}
\bauthor{\bsnm{Mazziotti}, \binits{D.A.}}:
\batitle{Contracted schr\"odinger equation: Determining quantum energies and
  two-particle density matrices without wave functions}.
\bjtitle{Phys. Rev. A}
\bvolume{57},
\bfpage{4219}--\blpage{4234}
(\byear{1998}).
doi:\doiurl{10.1103/PhysRevA.57.4219}
\end{barticle}
\endbibitem

\bibitem{Mazziotti2002}
\begin{barticle}
\bauthor{\bsnm{Mazziotti}, \binits{D.A.}}:
\batitle{{Variational method for solving the contracted Schr{\"{o}}dinger
  equation through a projection of the N -particle power method onto the
  two-particle space}}.
\bjtitle{J. Chem. Phys.}
\bvolume{116}(\bissue{4}),
\bfpage{1239}--\blpage{1249}
(\byear{2002}).
doi:\doiurl{10.1063/1.1430257}
\end{barticle}
\endbibitem

\bibitem{acse_Mazziotti2004}
\begin{barticle}
\bauthor{\bsnm{Mazziotti}, \binits{D.A.}}:
\batitle{Exactness of wave functions from two-body exponential transformations
  in many-body quantum theory}.
\bjtitle{Phys. Rev. A}
\bvolume{69},
\bfpage{012507}
(\byear{2004}).
doi:\doiurl{10.1103/PhysRevA.69.012507}
\end{barticle}
\endbibitem

\bibitem{exact_2body_exp_mazziotti2020}
\begin{barticle}
\bauthor{\bsnm{Mazziotti}, \binits{D.A.}}:
\batitle{Exact two-body expansion of the many-particle wave function}.
\bjtitle{Phys. Rev. A}
\bvolume{102},
\bfpage{030802}
(\byear{2020}).
doi:\doiurl{10.1103/PhysRevA.102.030802}
\end{barticle}
\endbibitem

\bibitem{Mazziotti2006}
\begin{barticle}
\bauthor{\bsnm{Mazziotti}, \binits{D.A.}}:
\batitle{Anti-hermitian contracted schr\"odinger equation: Direct determination
  of the two-electron reduced density matrices of many-electron molecules}.
\bjtitle{Phys. Rev. Lett.}
\bvolume{97},
\bfpage{143002}
(\byear{2006}).
doi:\doiurl{10.1103/PhysRevLett.97.143002}
\end{barticle}
\endbibitem

\bibitem{mazziotti2007}
\begin{barticle}
\bauthor{\bsnm{Mazziotti}, \binits{D.A.}}:
\batitle{Anti-hermitian part of the contracted schr\"odinger equation for the
  direct calculation of two-electron reduced density matrices}.
\bjtitle{Phys. Rev. A}
\bvolume{75},
\bfpage{022505}
(\byear{2007}).
doi:\doiurl{10.1103/PhysRevA.75.022505}
\end{barticle}
\endbibitem

\bibitem{acse_excited_states}
\begin{barticle}
\bauthor{\bsnm{Gidofalvi}, \binits{G.}},
\bauthor{\bsnm{Mazziotti}, \binits{D.A.}}:
\batitle{Direct calculation of excited-state electronic energies and
  two-electron reduced density matrices from the anti-hermitian contracted
  schr\"odinger equation}.
\bjtitle{Phys. Rev. A}
\bvolume{80},
\bfpage{022507}
(\byear{2009}).
doi:\doiurl{10.1103/PhysRevA.80.022507}
\end{barticle}
\endbibitem

\bibitem{Mukherjee2001}
\begin{barticle}
\bauthor{\bsnm{Mukherjee}, \binits{D.}},
\bauthor{\bsnm{Kutzelnigg}, \binits{W.}}:
\batitle{Irreducible brillouin conditions and contracted schr{\"{o}}dinger
  equations for n -electron systems. i. the equations satisfied by the density
  cumulants}.
\bjtitle{J. Chem. Phys.}
\bvolume{114}(\bissue{5}),
\bfpage{2047}--\blpage{2061}
(\byear{2001}).
doi:\doiurl{10.1063/1.1337058}
\end{barticle}
\endbibitem

\bibitem{qACSE_PRL2021}
\begin{barticle}
\bauthor{\bsnm{Smart}, \binits{S.E.}},
\bauthor{\bsnm{Mazziotti}, \binits{D.A.}}:
\batitle{{Quantum Solver of Contracted Eigenvalue Equations for Scalable
  Molecular Simulations on Quantum Computing Devices}}.
\bjtitle{Phys. Rev. Lett.}
\bvolume{126}(\bissue{7}),
\bfpage{070504}
(\byear{2021}).
doi:\doiurl{10.1103/physrevlett.126.070504}.
\arxivurl{2004.11416}
\end{barticle}
\endbibitem

\bibitem{Smart2021}
\begin{botherref}
\oauthor{\bsnm{Smart}, \binits{S.E.}},
\oauthor{\bsnm{Boyn}, \binits{J.-N.}},
\oauthor{\bsnm{Mazziotti}, \binits{D.A.}}:
{Resolving Correlated States of Benzyne on a Quantum Computer with an
  Error-Mitigated Quantum Contracted Eigenvalue Solver}
(2021).
\arxivurl{2103.06876}
\end{botherref}
\endbibitem

\bibitem{Grimsley2019}
\begin{barticle}
\bauthor{\bsnm{Grimsley}, \binits{H.R.}},
\bauthor{\bsnm{Economou}, \binits{S.E.}},
\bauthor{\bsnm{Barnes}, \binits{E.}},
\bauthor{\bsnm{Mayhall}, \binits{N.J.}}:
\batitle{{An adaptive variational algorithm for exact molecular simulations on
  a quantum computer}}.
\bjtitle{Nat. Commun.}
\bvolume{10}(\bissue{1}),
\bfpage{3007}
(\byear{2019}).
doi:\doiurl{10.1038/s41467-019-10988-2}
\end{barticle}
\endbibitem

\bibitem{uccgsd}
\begin{barticle}
\bauthor{\bsnm{Lee}, \binits{J.}},
\bauthor{\bsnm{Huggins}, \binits{W.J.}},
\bauthor{\bsnm{Head-Gordon}, \binits{M.}},
\bauthor{\bsnm{Whaley}, \binits{K.B.}}:
\batitle{Generalized unitary coupled cluster wave functions for quantum
  computation}.
\bjtitle{J. Chem. Theory Comput.}
\bvolume{15}(\bissue{1}),
\bfpage{311}--\blpage{324}
(\byear{2018}).
doi:\doiurl{10.1021/acs.jctc.8b01004}
\end{barticle}
\endbibitem

\bibitem{qubit_adapt}
\begin{botherref}
\oauthor{\bsnm{Tang}, \binits{H.L.}},
\oauthor{\bsnm{Shkolnikov}, \binits{V.O.}},
\oauthor{\bsnm{Barron}, \binits{G.S.}},
\oauthor{\bsnm{Grimsley}, \binits{H.R.}},
\oauthor{\bsnm{Mayhall}, \binits{N.J.}},
\oauthor{\bsnm{Barnes}, \binits{E.}},
\oauthor{\bsnm{Economou}, \binits{S.E.}}:
{qubit-ADAPT-VQE: An adaptive algorithm for constructing hardware-efficient
  ansatze on a quantum processor}.
arXiv
(2019).
\arxivurl{1911.10205}
\end{botherref}
\endbibitem

\bibitem{Bian19_15}
\begin{barticle}
\bauthor{\bsnm{Bian}, \binits{T.}},
\bauthor{\bsnm{Murphy}, \binits{D.}},
\bauthor{\bsnm{Xia}, \binits{R.}},
\bauthor{\bsnm{Daskin}, \binits{A.}},
\bauthor{\bsnm{Kais}, \binits{S.}}:
\batitle{Quantum computing methods for electronic states of the water
  molecule}.
\bjtitle{Mol. Phys.}
\bvolume{117}(\bissue{15--16}),
\bfpage{2069}--\blpage{2082}
(\byear{2019}).
doi:\doiurl{10.1080/00268976.2019.1580392}.
\arxivurl{https://doi.org/10.1080/00268976.2019.1580392}
\end{barticle}
\endbibitem

\bibitem{gard_sym_preserv}
\begin{barticle}
\bauthor{\bsnm{Gard}, \binits{B.T.}},
\bauthor{\bsnm{Zhu}, \binits{L.}},
\bauthor{\bsnm{Barron}, \binits{G.S.}},
\bauthor{\bsnm{Mayhall}, \binits{N.J.}},
\bauthor{\bsnm{Economou}, \binits{S.E.}},
\bauthor{\bsnm{Barnes}, \binits{E.}}:
\batitle{Efficient symmetry-preserving state preparation circuits for the
  variational quantum eigensolver algorithm}.
\bjtitle{npj Quantum Inf.}
\bvolume{6}(\bissue{1}),
\bfpage{10}
(\byear{2020}).
doi:\doiurl{10.1038/s41534-019-0240-1}
\end{barticle}
\endbibitem

\bibitem{Bradley09learningin}
\begin{botherref}
\oauthor{\bsnm{Bradley}, \binits{D.M.}},
\oauthor{\bsnm{Bagnell}, \binits{J.A.}},
\oauthor{\bsnm{Bengio}, \binits{Y.}},
\oauthor{\bsnm{Hebert}, \binits{M.}},
\oauthor{\bsnm{De}, \binits{F.}},
\oauthor{\bsnm{Torre}, \binits{L.}}:
Learning in modular systems.
Technical report
(2009)
\end{botherref}
\endbibitem

\bibitem{pmlr-v70-shalev-shwartz17a}
\begin{bchapter}
\bauthor{\bsnm{Shalev-Shwartz}, \binits{S.}},
\bauthor{\bsnm{Shamir}, \binits{O.}},
\bauthor{\bsnm{Shammah}, \binits{S.}}:
\bctitle{Failures of gradient-based deep learning}.
In: \beditor{\bsnm{Precup}, \binits{D.}},
\beditor{\bsnm{Teh}, \binits{Y.W.}} (eds.)
\bbtitle{Proceedings of the 34th International Conference on Machine Learning}.
\bsertitle{Proceedings of Machine Learning Research},
vol. \bseriesno{70},
pp. \bfpage{3067}--\blpage{3075}.
\bpublisher{PMLR},
\blocation{International Convention Centre, Sydney, Australia}
(\byear{2017}).
\burl{http://proceedings.mlr.press/v70/shalev-shwartz17a.html}
\end{bchapter}
\endbibitem

\bibitem{Kremer2001Book}
\begin{bbook}
\bauthor{\bsnm{Kremer}, \binits{S.C.}},
\bauthor{\bsnm{Kolen}, \binits{J.F.}}:
\bbtitle{Field Guide to Dynamical Recurrent Networks},
\bedition{1st} edn.
\bpublisher{Wiley-IEEE Press}, \blocation{???}
(\byear{2001})
\end{bbook}
\endbibitem

\bibitem{lecun2015deeplearning}
\begin{barticle}
\bauthor{\bsnm{LeCun}, \binits{Y.}},
\bauthor{\bsnm{Bengio}, \binits{Y.}},
\bauthor{\bsnm{Hinton}, \binits{G.}}:
\batitle{Deep learning}.
\bjtitle{Nature}
\bvolume{521}(\bissue{7553}),
\bfpage{436}--\blpage{444}
(\byear{2015}).
doi:\doiurl{10.1038/nature14539}
\end{barticle}
\endbibitem

\bibitem{pmlr-v37-ioffe15}
\begin{bchapter}
\bauthor{\bsnm{Ioffe}, \binits{S.}},
\bauthor{\bsnm{Szegedy}, \binits{C.}}:
\bctitle{Batch normalization: Accelerating deep network training by reducing
  internal covariate shift}.
In: \beditor{\bsnm{Bach}, \binits{F.}},
\beditor{\bsnm{Blei}, \binits{D.}} (eds.)
\bbtitle{Proceedings of the 32nd International Conference on Machine Learning}.
\bsertitle{Proceedings of Machine Learning Research},
vol. \bseriesno{37},
pp. \bfpage{448}--\blpage{456}.
\bpublisher{PMLR},
\blocation{Lille, France}
(\byear{2015}).
\burl{http://proceedings.mlr.press/v37/ioffe15.html}
\end{bchapter}
\endbibitem

\bibitem{Hinton2006}
\begin{barticle}
\bauthor{\bsnm{Hinton}, \binits{G.E.}},
\bauthor{\bsnm{Osindero}, \binits{S.}},
\bauthor{\bsnm{Teh}, \binits{Y.-W.}}:
\batitle{A fast learning algorithm for deep belief nets}.
\bjtitle{Neural Comput.}
\bvolume{18}(\bissue{7}),
\bfpage{1527}--\blpage{1554}
(\byear{2006}).
doi:\doiurl{10.1162/neco.2006.18.7.1527}.
\bcomment{PMID: 16764513}.
\arxivurl{https://doi.org/10.1162/neco.2006.18.7.1527}
\end{barticle}
\endbibitem

\bibitem{He2016}
\begin{bchapter}
\bauthor{\bsnm{{He}}, \binits{K.}},
\bauthor{\bsnm{{Zhang}}, \binits{X.}},
\bauthor{\bsnm{{Ren}}, \binits{S.}},
\bauthor{\bsnm{{Sun}}, \binits{J.}}:
\bctitle{Deep residual learning for image recognition}.
In: \bbtitle{2016 IEEE Conference on Computer Vision and Pattern Recognition
  (CVPR)},
pp. \bfpage{770}--\blpage{778}
(\byear{2016}).
doi:\doiurl{10.1109/CVPR.2016.90}
\end{bchapter}
\endbibitem

\bibitem{Ryabinkin2019}
\begin{barticle}
\bauthor{\bsnm{Ryabinkin}, \binits{I.G.}},
\bauthor{\bsnm{Genin}, \binits{S.N.}},
\bauthor{\bsnm{Izmaylov}, \binits{A.F.}}:
\batitle{Constrained variational quantum eigensolver: Quantum computer search
  engine in the {Fock space}}.
\bjtitle{J. Chem. Theory Comput.}
\bvolume{15}(\bissue{1}),
\bfpage{249}--\blpage{255}
(\byear{2019}).
doi:\doiurl{10.1021/acs.jctc.8b00943}
\end{barticle}
\endbibitem

\bibitem{Wang2009}
\begin{barticle}
\bauthor{\bsnm{Wang}, \binits{H.}},
\bauthor{\bsnm{Ashhab}, \binits{S.}},
\bauthor{\bsnm{Nori}, \binits{F.}}:
\batitle{Efficient quantum algorithm for preparing molecular-system-like states
  on a quantum computer}.
\bjtitle{Phys. Rev. A}
\bvolume{79},
\bfpage{042335}
(\byear{2009}).
doi:\doiurl{10.1103/PhysRevA.79.042335}
\end{barticle}
\endbibitem

\bibitem{Roth2017}
\begin{barticle}
\bauthor{\bsnm{Roth}, \binits{M.}},
\bauthor{\bsnm{Ganzhorn}, \binits{M.}},
\bauthor{\bsnm{Moll}, \binits{N.}},
\bauthor{\bsnm{Filipp}, \binits{S.}},
\bauthor{\bsnm{Salis}, \binits{G.}},
\bauthor{\bsnm{Schmidt}, \binits{S.}}:
\batitle{Analysis of a parametrically driven exchange-type gate and a
  two-photon excitation gate between superconducting qubits}.
\bjtitle{Phys. Rev. A}
\bvolume{96},
\bfpage{062323}
(\byear{2017}).
doi:\doiurl{10.1103/PhysRevA.96.062323}
\end{barticle}
\endbibitem

\bibitem{Egger2019}
\begin{barticle}
\bauthor{\bsnm{Egger}, \binits{D.J.}},
\bauthor{\bsnm{Ganzhorn}, \binits{M.}},
\bauthor{\bsnm{Salis}, \binits{G.}},
\bauthor{\bsnm{Fuhrer}, \binits{A.}},
\bauthor{\bsnm{M\"uller}, \binits{P.}},
\bauthor{\bsnm{Barkoutsos}, \binits{P.K.}},
\bauthor{\bsnm{Moll}, \binits{N.}},
\bauthor{\bsnm{Tavernelli}, \binits{I.}},
\bauthor{\bsnm{Filipp}, \binits{S.}}:
\batitle{Entanglement generation in superconducting qubits using holonomic
  operations}.
\bjtitle{Phys. Rev. Applied}
\bvolume{11},
\bfpage{014017}
(\byear{2019}).
doi:\doiurl{10.1103/PhysRevApplied.11.014017}
\end{barticle}
\endbibitem

\bibitem{Sagastizabal2019}
\begin{barticle}
\bauthor{\bsnm{Sagastizabal}, \binits{R.}},
\bauthor{\bsnm{Bonet-Monroig}, \binits{X.}},
\bauthor{\bsnm{Singh}, \binits{M.}},
\bauthor{\bsnm{Rol}, \binits{M.A.}},
\bauthor{\bsnm{Bultink}, \binits{C.C.}},
\bauthor{\bsnm{Fu}, \binits{X.}},
\bauthor{\bsnm{Price}, \binits{C.H.}},
\bauthor{\bsnm{Ostroukh}, \binits{V.P.}},
\bauthor{\bsnm{Muthusubramanian}, \binits{N.}},
\bauthor{\bsnm{Bruno}, \binits{A.}},
\bauthor{\bsnm{Beekman}, \binits{M.}},
\bauthor{\bsnm{Haider}, \binits{N.}},
\bauthor{\bsnm{O'Brien}, \binits{T.E.}},
\bauthor{\bsnm{DiCarlo}, \binits{L.}}:
\batitle{Experimental error mitigation via symmetry verification in a
  variational quantum eigensolver}.
\bjtitle{Phys. Rev. A}
\bvolume{100},
\bfpage{010302}
(\byear{2019}).
doi:\doiurl{10.1103/PhysRevA.100.010302}
\end{barticle}
\endbibitem

\bibitem{fontana2020optimizing}
\begin{botherref}
\oauthor{\bsnm{Fontana}, \binits{E.}},
\oauthor{\bsnm{Cerezo}, \binits{M.}},
\oauthor{\bsnm{Arrasmith}, \binits{A.}},
\oauthor{\bsnm{Rungger}, \binits{I.}},
\oauthor{\bsnm{Coles}, \binits{P.J.}}:
Optimizing parametrized quantum circuits via noise-induced breaking of
  symmetries.
arXiv
(2020).
\arxivurl{2011.08763}
\end{botherref}
\endbibitem

\bibitem{anand2020natural}
\begin{botherref}
\oauthor{\bsnm{Anand}, \binits{A.}},
\oauthor{\bsnm{Degroote}, \binits{M.}},
\oauthor{\bsnm{Aspuru-Guzik}, \binits{A.}}:
Natural evolutionary strategies for variational quantum computation.
arXiv
(2020).
\arxivurl{2012.00101}
\end{botherref}
\endbibitem

\bibitem{arrasmith2020effect}
\begin{botherref}
\oauthor{\bsnm{Arrasmith}, \binits{A.}},
\oauthor{\bsnm{Cerezo}, \binits{M.}},
\oauthor{\bsnm{Czarnik}, \binits{P.}},
\oauthor{\bsnm{Cincio}, \binits{L.}},
\oauthor{\bsnm{Coles}, \binits{P.J.}}:
Effect of barren plateaus on gradient-free optimization.
arXiv
(2020).
\arxivurl{2011.12245}
\end{botherref}
\endbibitem

\bibitem{uvarov2020barren}
\begin{botherref}
\oauthor{\bsnm{Uvarov}, \binits{A.}},
\oauthor{\bsnm{Biamonte}, \binits{J.}}:
On barren plateaus and cost function locality in variational quantum
  algorithms.
arXiv
(2020).
\arxivurl{2011.10530}
\end{botherref}
\endbibitem

\bibitem{zhang2020trainability}
\begin{botherref}
\oauthor{\bsnm{Zhang}, \binits{K.}},
\oauthor{\bsnm{Hsieh}, \binits{M.-H.}},
\oauthor{\bsnm{Liu}, \binits{L.}},
\oauthor{\bsnm{Tao}, \binits{D.}}:
Toward trainability of quantum neural networks.
arXiv
(2020).
\arxivurl{2011.06258}
\end{botherref}
\endbibitem

\bibitem{pesah2020absence}
\begin{botherref}
\oauthor{\bsnm{Pesah}, \binits{A.}},
\oauthor{\bsnm{Cerezo}, \binits{M.}},
\oauthor{\bsnm{Wang}, \binits{S.}},
\oauthor{\bsnm{Volkoff}, \binits{T.}},
\oauthor{\bsnm{Sornborger}, \binits{A.T.}},
\oauthor{\bsnm{Coles}, \binits{P.J.}}:
Absence of barren plateaus in quantum convolutional neural networks.
arXiv
(2020).
\arxivurl{2011.02966}
\end{botherref}
\endbibitem

\bibitem{yen2019_sym}
\begin{barticle}
\bauthor{\bsnm{Yen}, \binits{T.-C.}},
\bauthor{\bsnm{Lang}, \binits{R.A.}},
\bauthor{\bsnm{Izmaylov}, \binits{A.F.}}:
\batitle{{Exact and approximate symmetry projectors for the electronic
  structure problem on a quantum computer}}.
\bjtitle{J. Chem. Phys.}
\bvolume{151}(\bissue{16}),
\bfpage{164111}
(\byear{2019}).
doi:\doiurl{10.1063/1.5110682}.
\arxivurl{1905.08109}
\end{barticle}
\endbibitem

\bibitem{it_QCC}
\begin{barticle}
\bauthor{\bsnm{Ryabinkin}, \binits{I.G.}},
\bauthor{\bsnm{Lang}, \binits{R.A.}},
\bauthor{\bsnm{Genin}, \binits{S.N.}},
\bauthor{\bsnm{Izmaylov}, \binits{A.F.}}:
\batitle{Iterative qubit coupled cluster approach with efficient screening of
  generators}.
\bjtitle{J. Chem. Theory Comput.}
\bvolume{16}(\bissue{2}),
\bfpage{1055}--\blpage{1063}
(\byear{2020}).
doi:\doiurl{10.1021/acs.jctc.9b01084}
\end{barticle}
\endbibitem

\bibitem{izmaylov2021unitary}
\begin{barticle}
\bauthor{\bsnm{Lang}, \binits{R.A.}},
\bauthor{\bsnm{Ryabinkin}, \binits{I.G.}},
\bauthor{\bsnm{Izmaylov}, \binits{A.F.}}:
\batitle{Unitary transformation of the electronic hamiltonian with an exact
  quadratic truncation of the baker-campbell-hausdorff expansion}.
\bjtitle{J. Chem. Theory Comput.}
\bvolume{17}(\bissue{1}),
\bfpage{66}--\blpage{78}
(\byear{2021}).
doi:\doiurl{10.1021/acs.jctc.0c00170}
\end{barticle}
\endbibitem

\end{thebibliography}

\newcommand{\BMCxmlcomment}[1]{}

\BMCxmlcomment{

<refgrp>

<bibl id="B1">
  <title><p>Simulating physics with computers</p></title>
  <aug>
    <au><snm>Feynman</snm><fnm>RP</fnm></au>
  </aug>
  <source>Int. J. Theor. Phys</source>
  <pubdate>1982</pubdate>
  <volume>21</volume>
  <issue>6/7</issue>
</bibl>

<bibl id="B2">
  <title><p>Quantum Computing in the {NISQ} era and beyond</p></title>
  <aug>
    <au><snm>Preskill</snm><fnm>J</fnm></au>
  </aug>
  <source>Quantum</source>
  <publisher>Verein zur F{\"o}rderung des Open Access Publizierens in den
  Quantenwissenschaften</publisher>
  <pubdate>2018</pubdate>
  <volume>2</volume>
  <fpage>79</fpage>
</bibl>

<bibl id="B3">
  <title><p>How will quantum computers provide an industrially relevant
  computational advantage in quantum chemistry?</p></title>
  <aug>
    <au><snm>Elfving</snm><fnm>VE</fnm></au>
    <au><snm>Broer</snm><fnm>BW</fnm></au>
    <au><snm>Webber</snm><fnm>M</fnm></au>
    <au><snm>Gavartin</snm><fnm>J</fnm></au>
    <au><snm>Halls</snm><fnm>MD</fnm></au>
    <au><snm>Lorton</snm><fnm>KP</fnm></au>
    <au><snm>Bochevarov</snm><fnm>A</fnm></au>
  </aug>
  <source>arXiv preprint arXiv:2009.12472</source>
  <pubdate>2020</pubdate>
</bibl>

<bibl id="B4">
  <title><p>Prospects of Quantum Computing for Molecular Sciences</p></title>
  <aug>
    <au><snm>Liu</snm><fnm>H</fnm></au>
    <au><snm>Low</snm><fnm>GH</fnm></au>
    <au><snm>Steiger</snm><fnm>DS</fnm></au>
    <au><snm>Häner</snm><fnm>T</fnm></au>
    <au><snm>Reiher</snm><fnm>M</fnm></au>
    <au><snm>Troyer</snm><fnm>M</fnm></au>
  </aug>
  <pubdate>2021</pubdate>
</bibl>

<bibl id="B5">
  <title><p>{Q-NEXT National Quantum Center located in Argonne National
  Laboratory}</p></title>
  <source>\url{http://https://www.q-next.org/}</source>
  <note>Accessed: 2021-01-21</note>
</bibl>

<bibl id="B6">
  <title><p>{Quantum measurements and the Abelian stabilizer
  problem}</p></title>
  <aug>
    <au><snm>Kitaev</snm><fnm>AY</fnm></au>
  </aug>
  <source>arXiv</source>
  <pubdate>1995</pubdate>
</bibl>

<bibl id="B7">
  <title><p>Quantum Algorithm Providing Exponential Speed Increase for Finding
  Eigenvalues and Eigenvectors</p></title>
  <aug>
    <au><snm>Abrams</snm><fnm>DS</fnm></au>
    <au><snm>Lloyd</snm><fnm>S</fnm></au>
  </aug>
  <source>Phys. Rev. Lett.</source>
  <publisher>American Physical Society</publisher>
  <pubdate>1999</pubdate>
  <volume>83</volume>
  <fpage>5162</fpage>
  <lpage>-5165</lpage>
  <url>https://link.aps.org/doi/10.1103/PhysRevLett.83.5162</url>
</bibl>

<bibl id="B8">
  <title><p>Simulation of Many-Body Fermi Systems on a Universal Quantum
  Computer</p></title>
  <aug>
    <au><snm>Abrams</snm><fnm>DS</fnm></au>
    <au><snm>Lloyd</snm><fnm>S</fnm></au>
  </aug>
  <source>Phys. Rev. Lett.</source>
  <publisher>American Physical Society</publisher>
  <pubdate>1997</pubdate>
  <volume>79</volume>
  <fpage>2586</fpage>
  <lpage>-2589</lpage>
  <url>https://link.aps.org/doi/10.1103/PhysRevLett.79.2586</url>
</bibl>

<bibl id="B9">
  <title><p>Algorithms for quantum computation: discrete logarithms and
  factoring</p></title>
  <aug>
    <au><snm>{Shor}</snm><fnm>P. W.</fnm></au>
  </aug>
  <source>Proceedings 35th Annual Symposium on Foundations of Computer
  Science</source>
  <pubdate>1994</pubdate>
  <fpage>124</fpage>
  <lpage>-134</lpage>
</bibl>

<bibl id="B10">
  <title><p>{Fault Tolerant Quantum Computation with Constant
  Error}</p></title>
  <aug>
    <au><snm>Aharonov</snm><fnm>D</fnm></au>
    <au><snm>Ben Or</snm><fnm>M</fnm></au>
  </aug>
  <source>arXiv</source>
  <pubdate>1996</pubdate>
</bibl>

<bibl id="B11">
  <title><p>A variational eigenvalue solver on a photonic quantum
  processor</p></title>
  <aug>
    <au><snm>Peruzzo</snm><fnm>A</fnm></au>
    <au><snm>McClean</snm><fnm>J</fnm></au>
    <au><snm>Shadbolt</snm><fnm>P</fnm></au>
    <au><snm>Yung</snm><fnm>MH</fnm></au>
    <au><snm>Zhou</snm><fnm>XQ</fnm></au>
    <au><snm>Love</snm><fnm>PJ</fnm></au>
    <au><snm>Aspuru Guzik</snm><fnm>A</fnm></au>
    <au><snm>O’Brien</snm><fnm>JL</fnm></au>
  </aug>
  <source>Nat. Commun.</source>
  <pubdate>2014</pubdate>
  <volume>5</volume>
  <issue>1</issue>
  <fpage>4213</fpage>
</bibl>

<bibl id="B12">
  <title><p>The theory of variational hybrid quantum-classical
  algorithms</p></title>
  <aug>
    <au><snm>McClean</snm><fnm>JR</fnm></au>
    <au><snm>Romero</snm><fnm>J</fnm></au>
    <au><snm>Babbush</snm><fnm>R</fnm></au>
    <au><snm>Aspuru Guzik</snm><fnm>A</fnm></au>
  </aug>
  <source>New J. Phys.</source>
  <publisher>{IOP} Publishing</publisher>
  <pubdate>2016</pubdate>
  <volume>18</volume>
  <issue>2</issue>
  <fpage>023023</fpage>
  <url>https://doi.org/10.1088/1367-2630/18/2/023023</url>
</bibl>

<bibl id="B13">
  <title><p>Strategies for quantum computing molecular energies using the
  unitary coupled cluster ansatz</p></title>
  <aug>
    <au><snm>Romero</snm><fnm>J</fnm></au>
    <au><snm>Babbush</snm><fnm>R</fnm></au>
    <au><snm>McClean</snm><fnm>JR</fnm></au>
    <au><snm>Hempel</snm><fnm>C</fnm></au>
    <au><snm>Love</snm><fnm>PJ</fnm></au>
    <au><snm>Aspuru Guzik</snm><fnm>A</fnm></au>
  </aug>
  <source>Quantum Sci. Technol.</source>
  <publisher>{IOP} Publishing</publisher>
  <pubdate>2018</pubdate>
  <volume>4</volume>
  <issue>1</issue>
  <fpage>014008</fpage>
  <url>https://doi.org/10.1088/2058-9565/aad3e4</url>
</bibl>

<bibl id="B14">
  <title><p>{Use of a unitary wavefunction in the calculation of static
  electronic properties}</p></title>
  <aug>
    <au><snm>Pal</snm><fnm>S</fnm></au>
  </aug>
  <source>Theo. Chim. Acta</source>
  <pubdate>1984</pubdate>
  <volume>66</volume>
  <issue>3</issue>
  <fpage>207</fpage>
  <lpage>-215</lpage>
  <url>https://doi.org/10.1007/BF00549670</url>
</bibl>

<bibl id="B15">
  <title><p>{A unitary multiconfigurational coupled‐cluster method: Theory
  and applications}</p></title>
  <aug>
    <au><snm>Hoffmann</snm><fnm>MR</fnm></au>
    <au><snm>Simons</snm><fnm>J</fnm></au>
  </aug>
  <source>J. Chem. Phys.</source>
  <pubdate>1988</pubdate>
  <volume>88</volume>
  <issue>2</issue>
  <fpage>993</fpage>
  <lpage>-1002</lpage>
  <url>http://aip.scitation.org/doi/10.1063/1.454125</url>
</bibl>

<bibl id="B16">
  <title><p>{Error analysis and improvements of coupled-cluster
  theory}</p></title>
  <aug>
    <au><snm>Kutzelnigg</snm><fnm>W</fnm></au>
  </aug>
  <source>Theo. Chim. Acta</source>
  <pubdate>1991</pubdate>
  <volume>80</volume>
  <issue>4</issue>
  <fpage>349</fpage>
  <lpage>-386</lpage>
  <url>https://doi.org/10.1007/BF01117418</url>
</bibl>

<bibl id="B17">
  <title><p>{New perspectives on unitary coupled-cluster theory}</p></title>
  <aug>
    <au><snm>Taube</snm><fnm>AG</fnm></au>
    <au><snm>Bartlett</snm><fnm>RJ</fnm></au>
  </aug>
  <source>Int. J. Quantum Chem.</source>
  <pubdate>2006</pubdate>
  <volume>106</volume>
  <issue>15</issue>
  <fpage>3393</fpage>
  <lpage>-3401</lpage>
  <url>https://onlinelibrary.wiley.com/doi/10.1002/qua.21198</url>
</bibl>

<bibl id="B18">
  <title><p>Relativistic unitary coupled cluster theory and
  applications</p></title>
  <aug>
    <au><snm>Sur</snm><fnm>C</fnm></au>
    <au><snm>Chaudhuri</snm><fnm>RK</fnm></au>
    <au><snm>Sahoo</snm><fnm>BK</fnm></au>
    <au><snm>Das</snm><fnm>B P</fnm></au>
    <au><snm>Mukherjee</snm><fnm>D</fnm></au>
  </aug>
  <source>J. Phys. B</source>
  <publisher>{IOP} Publishing</publisher>
  <pubdate>2008</pubdate>
  <volume>41</volume>
  <issue>6</issue>
  <fpage>065001</fpage>
  <url>https://doi.org/10.1088/0953-4075/41/6/065001</url>
</bibl>

<bibl id="B19">
  <title><p>{Benchmark studies of variational, unitary and extended coupled
  cluster methods}</p></title>
  <aug>
    <au><snm>Cooper</snm><fnm>B</fnm></au>
    <au><snm>Knowles</snm><fnm>PJ</fnm></au>
  </aug>
  <source>J. Chem. Phys.</source>
  <pubdate>2010</pubdate>
  <volume>133</volume>
  <issue>23</issue>
  <fpage>234102</fpage>
  <url>http://aip.scitation.org/doi/10.1063/1.3520564</url>
</bibl>

<bibl id="B20">
  <title><p>{On the difference between variational and unitary coupled cluster
  theories}</p></title>
  <aug>
    <au><snm>Harsha</snm><fnm>G</fnm></au>
    <au><snm>Shiozaki</snm><fnm>T</fnm></au>
    <au><snm>Scuseria</snm><fnm>GE</fnm></au>
  </aug>
  <source>J. Chem. Phys.</source>
  <pubdate>2018</pubdate>
  <volume>148</volume>
  <issue>4</issue>
  <fpage>044107</fpage>
  <url>http://aip.scitation.org/doi/10.1063/1.5011033</url>
</bibl>

<bibl id="B21">
  <title><p>{Exact parameterization of fermionic wave functions via unitary
  coupled cluster theory}</p></title>
  <aug>
    <au><snm>Evangelista</snm><fnm>FA</fnm></au>
    <au><snm>Chan</snm><fnm>GKL</fnm></au>
    <au><snm>Scuseria</snm><fnm>GE</fnm></au>
  </aug>
  <source>J. Chem. Phys.</source>
  <pubdate>2019</pubdate>
  <volume>151</volume>
  <issue>24</issue>
  <fpage>244112</fpage>
  <url>http://aip.scitation.org/doi/10.1063/1.5133059</url>
</bibl>

<bibl id="B22">
  <title><p>{Hardware-efficient variational quantum eigensolver for small
  molecules and quantum magnets}</p></title>
  <aug>
    <au><snm>Kandala</snm><fnm>A</fnm></au>
    <au><snm>Mezzacapo</snm><fnm>A</fnm></au>
    <au><snm>Temme</snm><fnm>K</fnm></au>
    <au><snm>Takita</snm><fnm>M</fnm></au>
    <au><snm>Brink</snm><fnm>M</fnm></au>
    <au><snm>Chow</snm><fnm>JM</fnm></au>
    <au><snm>Gambetta</snm><fnm>JM</fnm></au>
  </aug>
  <source>Nature</source>
  <pubdate>2017</pubdate>
  <volume>549</volume>
  <issue>7671</issue>
  <fpage>242</fpage>
  <lpage>-246</lpage>
</bibl>

<bibl id="B23">
  <title><p>Scalable Quantum Simulation of Molecular Energies</p></title>
  <aug>
    <au><snm>O'Malley</snm><fnm>P. J. J.</fnm></au>
    <au><snm>Babbush</snm><fnm>R.</fnm></au>
    <au><snm>Kivlichan</snm><fnm>I. D.</fnm></au>
    <au><snm>Romero</snm><fnm>J.</fnm></au>
    <au><snm>McClean</snm><fnm>J. R.</fnm></au>
    <au><snm>Barends</snm><fnm>R.</fnm></au>
    <au><snm>Kelly</snm><fnm>J.</fnm></au>
    <au><snm>Roushan</snm><fnm>P.</fnm></au>
    <au><snm>Tranter</snm><fnm>A.</fnm></au>
    <au><snm>Ding</snm><fnm>N.</fnm></au>
    <au><snm>Campbell</snm><fnm>B.</fnm></au>
    <au><snm>Chen</snm><fnm>Y.</fnm></au>
    <au><snm>Chen</snm><fnm>Z.</fnm></au>
    <au><snm>Chiaro</snm><fnm>B.</fnm></au>
    <au><snm>Dunsworth</snm><fnm>A.</fnm></au>
    <au><snm>Fowler</snm><fnm>A. G.</fnm></au>
    <au><snm>Jeffrey</snm><fnm>E.</fnm></au>
    <au><snm>Lucero</snm><fnm>E.</fnm></au>
    <au><snm>Megrant</snm><fnm>A.</fnm></au>
    <au><snm>Mutus</snm><fnm>J. Y.</fnm></au>
    <au><snm>Neeley</snm><fnm>M.</fnm></au>
    <au><snm>Neill</snm><fnm>C.</fnm></au>
    <au><snm>Quintana</snm><fnm>C.</fnm></au>
    <au><snm>Sank</snm><fnm>D.</fnm></au>
    <au><snm>Vainsencher</snm><fnm>A.</fnm></au>
    <au><snm>Wenner</snm><fnm>J.</fnm></au>
    <au><snm>White</snm><fnm>T. C.</fnm></au>
    <au><snm>Coveney</snm><fnm>P. V.</fnm></au>
    <au><snm>Love</snm><fnm>P. J.</fnm></au>
    <au><snm>Neven</snm><fnm>H.</fnm></au>
    <au><snm>Aspuru Guzik</snm><fnm>A.</fnm></au>
    <au><snm>Martinis</snm><fnm>J. M.</fnm></au>
  </aug>
  <source>Phys. Rev. X</source>
  <publisher>American Physical Society</publisher>
  <pubdate>2016</pubdate>
  <volume>6</volume>
  <fpage>031007</fpage>
  <url>https://link.aps.org/doi/10.1103/PhysRevX.6.031007</url>
</bibl>

<bibl id="B24">
  <title><p>Ground-state energy estimation of the water molecule on a trapped
  ion quantum computer</p></title>
  <aug>
    <au><snm>Nam</snm><fnm>Y</fnm></au>
    <au><snm>Chen</snm><fnm>JS</fnm></au>
    <au><snm>Pisenti</snm><fnm>NC</fnm></au>
    <au><snm>Wright</snm><fnm>K</fnm></au>
    <au><snm>Delaney</snm><fnm>C</fnm></au>
    <au><snm>Maslov</snm><fnm>D</fnm></au>
    <au><snm>Brown</snm><fnm>KR</fnm></au>
    <au><snm>Allen</snm><fnm>S</fnm></au>
    <au><snm>Amini</snm><fnm>JM</fnm></au>
    <au><snm>Apisdorf</snm><fnm>J</fnm></au>
    <au><snm>Beck</snm><fnm>KM</fnm></au>
    <au><snm>Blinov</snm><fnm>A</fnm></au>
    <au><snm>Chaplin</snm><fnm>V</fnm></au>
    <au><snm>Chmielewski</snm><fnm>M</fnm></au>
    <au><snm>Collins</snm><fnm>C</fnm></au>
    <au><snm>Debnath</snm><fnm>S</fnm></au>
    <au><snm>Ducore</snm><fnm>AM</fnm></au>
    <au><snm>Hudek</snm><fnm>KM</fnm></au>
    <au><snm>Keesan</snm><fnm>M</fnm></au>
    <au><snm>Kreikemeier</snm><fnm>SM</fnm></au>
    <au><snm>Mizrahi</snm><fnm>J</fnm></au>
    <au><snm>Solomon</snm><fnm>P</fnm></au>
    <au><snm>Williams</snm><fnm>M</fnm></au>
    <au><snm>Wong Campos</snm><fnm>JD</fnm></au>
    <au><snm>Monroe</snm><fnm>C</fnm></au>
    <au><snm>Kim</snm><fnm>J</fnm></au>
  </aug>
  <source>arXiv</source>
  <pubdate>2019</pubdate>
</bibl>

<bibl id="B25">
  <title><p>Computation of Molecular Spectra on a Quantum Processor with an
  Error-Resilient Algorithm</p></title>
  <aug>
    <au><snm>Colless</snm><fnm>J. I.</fnm></au>
    <au><snm>Ramasesh</snm><fnm>V. V.</fnm></au>
    <au><snm>Dahlen</snm><fnm>D.</fnm></au>
    <au><snm>Blok</snm><fnm>M. S.</fnm></au>
    <au><snm>Kimchi Schwartz</snm><fnm>M. E.</fnm></au>
    <au><snm>McClean</snm><fnm>J. R.</fnm></au>
    <au><snm>Carter</snm><fnm>J.</fnm></au>
    <au><snm>Jong</snm><fnm>W. A.</fnm></au>
    <au><snm>Siddiqi</snm><fnm>I.</fnm></au>
  </aug>
  <source>Phys. Rev. X</source>
  <publisher>American Physical Society</publisher>
  <pubdate>2018</pubdate>
  <volume>8</volume>
  <fpage>011021</fpage>
  <url>https://link.aps.org/doi/10.1103/PhysRevX.8.011021</url>
</bibl>

<bibl id="B26">
  <title><p>Qubit Coupled Cluster Method: A Systematic Approach to Quantum
  Chemistry on a Quantum Computer</p></title>
  <aug>
    <au><snm>Ryabinkin</snm><fnm>IG</fnm></au>
    <au><snm>Yen</snm><fnm>TC</fnm></au>
    <au><snm>Genin</snm><fnm>SN</fnm></au>
    <au><snm>Izmaylov</snm><fnm>AF</fnm></au>
  </aug>
  <source>J. Chem. Theory Comput.</source>
  <pubdate>2018</pubdate>
  <volume>14</volume>
  <issue>12</issue>
  <fpage>6317</fpage>
  <lpage>-6326</lpage>
</bibl>

<bibl id="B27">
  <title><p>{Quantum chemistry as a benchmark for near-term quantum
  computers}</p></title>
  <aug>
    <au><snm>McCaskey</snm><fnm>AJ</fnm></au>
    <au><snm>Parks</snm><fnm>ZP</fnm></au>
    <au><snm>Jakowski</snm><fnm>J</fnm></au>
    <au><snm>Moore</snm><fnm>SV</fnm></au>
    <au><snm>Morris</snm><fnm>TD</fnm></au>
    <au><snm>Humble</snm><fnm>TS</fnm></au>
    <au><snm>Pooser</snm><fnm>RC</fnm></au>
  </aug>
  <source>npj Quantum Inf.</source>
  <pubdate>2019</pubdate>
  <volume>5</volume>
  <issue>1</issue>
  <fpage>99</fpage>
</bibl>

<bibl id="B28">
  <title><p>{Hartree-Fock on a superconducting qubit quantum
  computer}</p></title>
  <aug>
    <au><cnm>{Collaborators*†, Google AI Quantum and}</cnm></au>
    <au><snm>Arute</snm><fnm>F</fnm></au>
    <au><snm>Arya</snm><fnm>K</fnm></au>
    <au><snm>Babbush</snm><fnm>R</fnm></au>
    <au><snm>Bacon</snm><fnm>D</fnm></au>
    <au><snm>Bardin</snm><fnm>JC</fnm></au>
    <au><snm>Barends</snm><fnm>R</fnm></au>
    <au><snm>Boixo</snm><fnm>S</fnm></au>
    <au><snm>Broughton</snm><fnm>M</fnm></au>
    <au><snm>Buckley</snm><fnm>BB</fnm></au>
    <au><snm>Buell</snm><fnm>DA</fnm></au>
    <au><snm>Burkett</snm><fnm>B</fnm></au>
    <au><snm>Bushnell</snm><fnm>N</fnm></au>
    <au><snm>Chen</snm><fnm>Y</fnm></au>
    <au><snm>Chen</snm><fnm>Z</fnm></au>
    <au><snm>Chiaro</snm><fnm>B</fnm></au>
    <au><snm>Collins</snm><fnm>R</fnm></au>
    <au><snm>Courtney</snm><fnm>W</fnm></au>
    <au><snm>Demura</snm><fnm>S</fnm></au>
    <au><snm>Dunsworth</snm><fnm>A</fnm></au>
    <au><snm>Farhi</snm><fnm>E</fnm></au>
    <au><snm>Fowler</snm><fnm>A</fnm></au>
    <au><snm>Foxen</snm><fnm>B</fnm></au>
    <au><snm>Gidney</snm><fnm>C</fnm></au>
    <au><snm>Giustina</snm><fnm>M</fnm></au>
    <au><snm>Graff</snm><fnm>R</fnm></au>
    <au><snm>Habegger</snm><fnm>S</fnm></au>
    <au><snm>Harrigan</snm><fnm>MP</fnm></au>
    <au><snm>Ho</snm><fnm>A</fnm></au>
    <au><snm>Hong</snm><fnm>S</fnm></au>
    <au><snm>Huang</snm><fnm>T</fnm></au>
    <au><snm>Huggins</snm><fnm>WJ</fnm></au>
    <au><snm>Ioffe</snm><fnm>L</fnm></au>
    <au><snm>Isakov</snm><fnm>SV</fnm></au>
    <au><snm>Jeffrey</snm><fnm>E</fnm></au>
    <au><snm>Jiang</snm><fnm>Z</fnm></au>
    <au><snm>Jones</snm><fnm>C</fnm></au>
    <au><snm>Kafri</snm><fnm>D</fnm></au>
    <au><snm>Kechedzhi</snm><fnm>K</fnm></au>
    <au><snm>Kelly</snm><fnm>J</fnm></au>
    <au><snm>Kim</snm><fnm>S</fnm></au>
    <au><snm>Klimov</snm><fnm>PV</fnm></au>
    <au><snm>Korotkov</snm><fnm>A</fnm></au>
    <au><snm>Kostritsa</snm><fnm>F</fnm></au>
    <au><snm>Landhuis</snm><fnm>D</fnm></au>
    <au><snm>Laptev</snm><fnm>P</fnm></au>
    <au><snm>Lindmark</snm><fnm>M</fnm></au>
    <au><snm>Lucero</snm><fnm>E</fnm></au>
    <au><snm>Martin</snm><fnm>O</fnm></au>
    <au><snm>Martinis</snm><fnm>JM</fnm></au>
    <au><snm>McClean</snm><fnm>JR</fnm></au>
    <au><snm>McEwen</snm><fnm>M</fnm></au>
    <au><snm>Megrant</snm><fnm>A</fnm></au>
    <au><snm>Mi</snm><fnm>X</fnm></au>
    <au><snm>Mohseni</snm><fnm>M</fnm></au>
    <au><snm>Mruczkiewicz</snm><fnm>W</fnm></au>
    <au><snm>Mutus</snm><fnm>J</fnm></au>
    <au><snm>Naaman</snm><fnm>O</fnm></au>
    <au><snm>Neeley</snm><fnm>M</fnm></au>
    <au><snm>Neill</snm><fnm>C</fnm></au>
    <au><snm>Neven</snm><fnm>H</fnm></au>
    <au><snm>Niu</snm><fnm>MY</fnm></au>
    <au><snm>O’Brien</snm><fnm>TE</fnm></au>
    <au><snm>Ostby</snm><fnm>E</fnm></au>
    <au><snm>Petukhov</snm><fnm>A</fnm></au>
    <au><snm>Putterman</snm><fnm>H</fnm></au>
    <au><snm>Quintana</snm><fnm>C</fnm></au>
    <au><snm>Roushan</snm><fnm>P</fnm></au>
    <au><snm>Rubin</snm><fnm>NC</fnm></au>
    <au><snm>Sank</snm><fnm>D</fnm></au>
    <au><snm>Satzinger</snm><fnm>KJ</fnm></au>
    <au><snm>Smelyanskiy</snm><fnm>V</fnm></au>
    <au><snm>Strain</snm><fnm>D</fnm></au>
    <au><snm>Sung</snm><fnm>KJ</fnm></au>
    <au><snm>Szalay</snm><fnm>M</fnm></au>
    <au><snm>Takeshita</snm><fnm>TY</fnm></au>
    <au><snm>Vainsencher</snm><fnm>A</fnm></au>
    <au><snm>White</snm><fnm>T</fnm></au>
    <au><snm>Wiebe</snm><fnm>N</fnm></au>
    <au><snm>Yao</snm><fnm>ZJ</fnm></au>
    <au><snm>Yeh</snm><fnm>P</fnm></au>
    <au><snm>Zalcman</snm><fnm>A</fnm></au>
  </aug>
  <source>Science</source>
  <pubdate>2020</pubdate>
  <volume>369</volume>
  <issue>6507</issue>
  <fpage>1084</fpage>
  <lpage>-1089</lpage>
</bibl>

<bibl id="B29">
  <title><p>Quantum Chemistry Calculations on a Trapped-Ion Quantum
  Simulator</p></title>
  <aug>
    <au><snm>Hempel</snm><fnm>C</fnm></au>
    <au><snm>Maier</snm><fnm>C</fnm></au>
    <au><snm>Romero</snm><fnm>J</fnm></au>
    <au><snm>McClean</snm><fnm>J</fnm></au>
    <au><snm>Monz</snm><fnm>T</fnm></au>
    <au><snm>Shen</snm><fnm>H</fnm></au>
    <au><snm>Jurcevic</snm><fnm>P</fnm></au>
    <au><snm>Lanyon</snm><fnm>BP</fnm></au>
    <au><snm>Love</snm><fnm>P</fnm></au>
    <au><snm>Babbush</snm><fnm>R</fnm></au>
    <au><snm>Aspuru Guzik</snm><fnm>A</fnm></au>
    <au><snm>Blatt</snm><fnm>R</fnm></au>
    <au><snm>Roos</snm><fnm>CF</fnm></au>
  </aug>
  <source>Phys. Rev. X</source>
  <publisher>American Physical Society</publisher>
  <pubdate>2018</pubdate>
  <volume>8</volume>
  <fpage>031022</fpage>
  <url>https://link.aps.org/doi/10.1103/PhysRevX.8.031022</url>
</bibl>

<bibl id="B30">
  <title><p>Quantum implementation of the unitary coupled cluster for
  simulating molecular electronic structure</p></title>
  <aug>
    <au><snm>Shen</snm><fnm>Y</fnm></au>
    <au><snm>Zhang</snm><fnm>X</fnm></au>
    <au><snm>Zhang</snm><fnm>S</fnm></au>
    <au><snm>Zhang</snm><fnm>JN</fnm></au>
    <au><snm>Yung</snm><fnm>MH</fnm></au>
    <au><snm>Kim</snm><fnm>K</fnm></au>
  </aug>
  <source>Phys. Rev. A</source>
  <publisher>American Physical Society</publisher>
  <pubdate>2017</pubdate>
  <volume>95</volume>
  <fpage>020501</fpage>
  <url>https://link.aps.org/doi/10.1103/PhysRevA.95.020501</url>
</bibl>

<bibl id="B31">
  <title><p>Witnessing eigenstates for quantum simulation of Hamiltonian
  spectra</p></title>
  <aug>
    <au><snm>Santagati</snm><fnm>R</fnm></au>
    <au><snm>Wang</snm><fnm>J</fnm></au>
    <au><snm>Gentile</snm><fnm>AA</fnm></au>
    <au><snm>Paesani</snm><fnm>S</fnm></au>
    <au><snm>Wiebe</snm><fnm>N</fnm></au>
    <au><snm>McClean</snm><fnm>JR</fnm></au>
    <au><snm>Morley Short</snm><fnm>S</fnm></au>
    <au><snm>Shadbolt</snm><fnm>PJ</fnm></au>
    <au><snm>Bonneau</snm><fnm>D</fnm></au>
    <au><snm>Silverstone</snm><fnm>JW</fnm></au>
    <au><snm>Tew</snm><fnm>DP</fnm></au>
    <au><snm>Zhou</snm><fnm>X</fnm></au>
    <au><snm>O{\textquoteright}Brien</snm><fnm>JL</fnm></au>
    <au><snm>Thompson</snm><fnm>MG</fnm></au>
  </aug>
  <source>Sci. Adv.</source>
  <publisher>American Association for the Advancement of Science</publisher>
  <pubdate>2018</pubdate>
  <volume>4</volume>
  <issue>1</issue>
  <url>https://advances.sciencemag.org/content/4/1/eaap9646</url>
</bibl>

<bibl id="B32">
  <title><p>Computational Investigations of the Lithium Superoxide Dimer
  Rearrangement on Noisy Quantum Devices</p></title>
  <aug>
    <au><snm>Gao</snm><fnm>Q</fnm></au>
    <au><snm>Nakamura</snm><fnm>H</fnm></au>
    <au><snm>Gujarati</snm><fnm>TP</fnm></au>
    <au><snm>Jones</snm><fnm>GO</fnm></au>
    <au><snm>Rice</snm><fnm>JE</fnm></au>
    <au><snm>Wood</snm><fnm>SP</fnm></au>
    <au><snm>Pistoia</snm><fnm>M</fnm></au>
    <au><snm>Garcia</snm><fnm>JM</fnm></au>
    <au><snm>Yamamoto</snm><fnm>N</fnm></au>
  </aug>
  <source>J. Phys. Chem. A</source>
  <pubdate>2021</pubdate>
  <volume>125</volume>
  <issue>9</issue>
  <fpage>1827</fpage>
  <lpage>1836</lpage>
  <url>https://doi.org/10.1021/acs.jpca.0c09530</url>
  <note>PMID: 33635672</note>
</bibl>

<bibl id="B33">
  <title><p>{Applications of Quantum Computing for Investigations of Electronic
  Transitions in Phenylsulfonyl-carbazole TADF Emitters}</p></title>
  <aug>
    <au><snm>Gao</snm><fnm>Q</fnm></au>
    <au><snm>Jones</snm><fnm>GO</fnm></au>
    <au><snm>Motta</snm><fnm>M</fnm></au>
    <au><snm>Sugawara</snm><fnm>M</fnm></au>
    <au><snm>Watanabe</snm><fnm>HC</fnm></au>
    <au><snm>Kobayashi</snm><fnm>T</fnm></au>
    <au><snm>Watanabe</snm><fnm>E</fnm></au>
    <au><snm>Ohnishi</snm><fnm>Yy</fnm></au>
    <au><snm>Nakamura</snm><fnm>H</fnm></au>
    <au><snm>Yamamoto</snm><fnm>N</fnm></au>
  </aug>
  <pubdate>2020</pubdate>
  <url>http://arxiv.org/abs/2007.15795</url>
</bibl>

<bibl id="B34">
  <title><p>Quantum-classical hybrid algorithm using an error-mitigating
  $N$-representability condition to compute the Mott metal-insulator
  transition</p></title>
  <aug>
    <au><snm>Smart</snm><fnm>SE</fnm></au>
    <au><snm>Mazziotti</snm><fnm>DA</fnm></au>
  </aug>
  <source>Phys. Rev. A</source>
  <publisher>American Physical Society</publisher>
  <pubdate>2019</pubdate>
  <volume>100</volume>
  <fpage>022517</fpage>
  <url>https://link.aps.org/doi/10.1103/PhysRevA.100.022517</url>
</bibl>

<bibl id="B35">
  <title><p>Quantum computational chemistry</p></title>
  <aug>
    <au><snm>McArdle</snm><fnm>S</fnm></au>
    <au><snm>Endo</snm><fnm>S</fnm></au>
  </aug>
  <source>Rev. Mod. Phys.</source>
  <pubdate>2020</pubdate>
  <volume>92</volume>
  <issue>1</issue>
  <fpage>015003</fpage>
</bibl>

<bibl id="B36">
  <title><p>Quantum Chemistry in the Age of Quantum Computing</p></title>
  <aug>
    <au><snm>Cao</snm><fnm>Y</fnm></au>
    <au><snm>Romero</snm><fnm>J</fnm></au>
    <au><snm>Olson</snm><fnm>JP</fnm></au>
    <au><snm>Degroote</snm><fnm>M</fnm></au>
    <au><snm>Johnson</snm><fnm>PD</fnm></au>
    <au><snm>Kieferova</snm><fnm>M</fnm></au>
    <au><snm>Kivlichan</snm><fnm>ID</fnm></au>
    <au><snm>Menke</snm><fnm>T</fnm></au>
    <au><snm>Peropadre</snm><fnm>B</fnm></au>
    <au><snm>Sawaya</snm><fnm>NPD</fnm></au>
    <au><snm>Sim</snm><fnm>S</fnm></au>
    <au><snm>Veis</snm><fnm>L</fnm></au>
    <au><snm>Aspuru Guzik</snm><fnm>A</fnm></au>
  </aug>
  <source>Chem. Rev.</source>
  <pubdate>2019</pubdate>
  <volume>119</volume>
  <issue>19</issue>
  <fpage>10856</fpage>
  <lpage>-10915</lpage>
</bibl>

<bibl id="B37">
  <title><p>Variational quantum algorithms</p></title>
  <aug>
    <au><snm>Cerezo</snm><fnm>M</fnm></au>
    <au><snm>Arrasmith</snm><fnm>A</fnm></au>
    <au><snm>Babbush</snm><fnm>R</fnm></au>
    <au><snm>Benjamin</snm><fnm>SC</fnm></au>
    <au><snm>Endo</snm><fnm>S</fnm></au>
    <au><snm>Fujii</snm><fnm>K</fnm></au>
    <au><snm>McClean</snm><fnm>JR</fnm></au>
    <au><snm>Mitarai</snm><fnm>K</fnm></au>
    <au><snm>Yuan</snm><fnm>X</fnm></au>
    <au><snm>Cincio</snm><fnm>L</fnm></au>
    <au><snm>Coles</snm><fnm>PJ</fnm></au>
  </aug>
  <source>arXiv</source>
  <pubdate>2020</pubdate>
</bibl>

<bibl id="B38">
  <title><p>Noisy intermediate-scale quantum {(NISQ)} algorithms</p></title>
  <aug>
    <au><snm>Bharti</snm><fnm>K</fnm></au>
    <au><snm>Cervera Lierta</snm><fnm>A</fnm></au>
    <au><snm>Kyaw</snm><fnm>TH</fnm></au>
    <au><snm>Haug</snm><fnm>T</fnm></au>
    <au><snm>Alperin Lea</snm><fnm>S</fnm></au>
    <au><snm>Anand</snm><fnm>A</fnm></au>
    <au><snm>Degroote</snm><fnm>M</fnm></au>
    <au><snm>Heimonen</snm><fnm>H</fnm></au>
    <au><snm>Kottmann</snm><fnm>JS</fnm></au>
    <au><snm>Menke</snm><fnm>T</fnm></au>
    <au><snm>Mok</snm><fnm>WK</fnm></au>
    <au><snm>Sim</snm><fnm>S</fnm></au>
    <au><snm>Kwek</snm><fnm>LC</fnm></au>
    <au><snm>Aspuru Guzik</snm><fnm>A</fnm></au>
  </aug>
  <pubdate>2021</pubdate>
</bibl>

<bibl id="B39">
  <title><p>Low-Depth Quantum Simulation of Materials</p></title>
  <aug>
    <au><snm>Babbush</snm><fnm>R</fnm></au>
    <au><snm>Wiebe</snm><fnm>N</fnm></au>
    <au><snm>McClean</snm><fnm>J</fnm></au>
    <au><snm>McClain</snm><fnm>J</fnm></au>
    <au><snm>Neven</snm><fnm>H</fnm></au>
    <au><snm>Chan</snm><fnm>GKL</fnm></au>
  </aug>
  <source>Phys. Rev. X</source>
  <publisher>American Physical Society</publisher>
  <pubdate>2018</pubdate>
  <volume>8</volume>
  <fpage>011044</fpage>
  <url>https://link.aps.org/doi/10.1103/PhysRevX.8.011044</url>
</bibl>

<bibl id="B40">
  <title><p>Reducing Qubit Requirements while Maintaining Numerical Precision
  for the Variational Quantum Eigensolver: A Basis-Set-Free
  Approach</p></title>
  <aug>
    <au><snm>Kottmann</snm><fnm>JS</fnm></au>
    <au><snm>Schleich</snm><fnm>P</fnm></au>
    <au><snm>Tamayo Mendoza</snm><fnm>T</fnm></au>
    <au><snm>Aspuru Guzik</snm><fnm>A</fnm></au>
  </aug>
  <source>J. Phys. Chem. Lett.</source>
  <pubdate>2021</pubdate>
  <volume>12</volume>
  <issue>1</issue>
  <fpage>663</fpage>
  <lpage>-673</lpage>
  <url>https://doi.org/10.1021/acs.jpclett.0c03410</url>
  <note>PMID: 33393305</note>
</bibl>

<bibl id="B41">
  <title><p>{Quantum simulation of chemistry with sublinear scaling in basis
  size}</p></title>
  <aug>
    <au><snm>Babbush</snm><fnm>R</fnm></au>
    <au><snm>Berry</snm><fnm>DW</fnm></au>
    <au><snm>McClean</snm><fnm>JR</fnm></au>
    <au><snm>Neven</snm><fnm>H</fnm></au>
  </aug>
  <source>npj Quantum Inf.</source>
  <pubdate>2019</pubdate>
  <volume>5</volume>
  <issue>1</issue>
  <fpage>92</fpage>
</bibl>

<bibl id="B42">
  <title><p>{{\"{U}}ber das Paulische {\"{A}}quivalenzverbot}</p></title>
  <aug>
    <au><snm>Jordan</snm><fnm>P.</fnm></au>
    <au><snm>Wigner</snm><fnm>E.</fnm></au>
  </aug>
  <source>Z. Phys.</source>
  <pubdate>1928</pubdate>
  <volume>47</volume>
  <issue>9-10</issue>
  <fpage>631</fpage>
  <lpage>-651</lpage>
  <url>http://link.springer.com/10.1007/BF01331938</url>
</bibl>

<bibl id="B43">
  <title><p>{Tapering off qubits to simulate fermionic
  Hamiltonians}</p></title>
  <aug>
    <au><snm>Bravyi</snm><fnm>S</fnm></au>
    <au><snm>Gambetta</snm><fnm>JM</fnm></au>
    <au><snm>Mezzacapo</snm><fnm>A</fnm></au>
    <au><snm>Temme</snm><fnm>K</fnm></au>
  </aug>
  <source>arXiv</source>
  <pubdate>2017</pubdate>
</bibl>

<bibl id="B44">
  <title><p>Fermionic Quantum Computation</p></title>
  <aug>
    <au><snm>Bravyi</snm><fnm>SB</fnm></au>
    <au><snm>Kitaev</snm><fnm>AY</fnm></au>
  </aug>
  <source>Ann. Phys.</source>
  <pubdate>2002</pubdate>
  <volume>298</volume>
  <issue>1</issue>
  <fpage>210</fpage>
  <lpage>-226</lpage>
  <url>https://linkinghub.elsevier.com/retrieve/pii/S0003491602962548</url>
</bibl>

<bibl id="B45">
  <title><p>{Exploiting Locality in Quantum Computation for Quantum
  Chemistry}</p></title>
  <aug>
    <au><snm>McClean</snm><fnm>JR</fnm></au>
    <au><snm>Babbush</snm><fnm>R</fnm></au>
    <au><snm>Love</snm><fnm>PJ</fnm></au>
    <au><snm>Aspuru Guzik</snm><fnm>A</fnm></au>
  </aug>
  <source>J. Phys. Chem. Lett.</source>
  <pubdate>2014</pubdate>
  <volume>5</volume>
  <issue>24</issue>
  <fpage>4368</fpage>
  <lpage>-4380</lpage>
</bibl>

<bibl id="B46">
  <title><p>Efficient {Bayesian} Phase Estimation</p></title>
  <aug>
    <au><snm>Wiebe</snm><fnm>N</fnm></au>
    <au><snm>Granade</snm><fnm>C</fnm></au>
  </aug>
  <source>Phys. Rev. Lett.</source>
  <publisher>American Physical Society</publisher>
  <pubdate>2016</pubdate>
  <volume>117</volume>
  <fpage>010503</fpage>
  <url>https://link.aps.org/doi/10.1103/PhysRevLett.117.010503</url>
</bibl>

<bibl id="B47">
  <title><p>Progress towards practical quantum variational
  algorithms</p></title>
  <aug>
    <au><snm>Wecker</snm><fnm>D</fnm></au>
    <au><snm>Hastings</snm><fnm>MB</fnm></au>
    <au><snm>Troyer</snm><fnm>M</fnm></au>
  </aug>
  <source>Phys. Rev. A</source>
  <publisher>American Physical Society</publisher>
  <pubdate>2015</pubdate>
  <volume>92</volume>
  <fpage>042303</fpage>
  <url>https://link.aps.org/doi/10.1103/PhysRevA.92.042303</url>
</bibl>

<bibl id="B48">
  <title><p>Accuracy and Resource Estimations for Quantum Chemistry on a
  Near-Term Quantum Computer</p></title>
  <aug>
    <au><snm>Kühn</snm><fnm>M</fnm></au>
    <au><snm>Zanker</snm><fnm>S</fnm></au>
    <au><snm>Deglmann</snm><fnm>P</fnm></au>
    <au><snm>Marthaler</snm><fnm>M</fnm></au>
    <au><snm>Weiß</snm><fnm>H</fnm></au>
  </aug>
  <source>J. Chem. Theory Comput.</source>
  <pubdate>2019</pubdate>
  <volume>15</volume>
  <issue>9</issue>
  <fpage>4764</fpage>
  <lpage>-4780</lpage>
  <url>https://doi.org/10.1021/acs.jctc.9b00236</url>
  <note>PMID: 31403781</note>
</bibl>

<bibl id="B49">
  <title><p>{Error mitigation extends the computational reach of a noisy
  quantum processor}</p></title>
  <aug>
    <au><snm>Kandala</snm><fnm>A</fnm></au>
    <au><snm>Temme</snm><fnm>K</fnm></au>
    <au><snm>Córcoles</snm><fnm>AD</fnm></au>
    <au><snm>Mezzacapo</snm><fnm>A</fnm></au>
    <au><snm>Chow</snm><fnm>JM</fnm></au>
    <au><snm>Gambetta</snm><fnm>JM</fnm></au>
  </aug>
  <source>Nature</source>
  <pubdate>2019</pubdate>
  <volume>567</volume>
  <issue>7749</issue>
  <fpage>491</fpage>
  <lpage>-495</lpage>
</bibl>

<bibl id="B50">
  <title><p>{Barren plateaus in quantum neural network training
  landscapes}</p></title>
  <aug>
    <au><snm>McClean</snm><fnm>JR</fnm></au>
    <au><snm>Boixo</snm><fnm>S</fnm></au>
    <au><snm>Smelyanskiy</snm><fnm>VN</fnm></au>
    <au><snm>Babbush</snm><fnm>R</fnm></au>
    <au><snm>Neven</snm><fnm>H</fnm></au>
  </aug>
  <source>Nat. Commun.</source>
  <pubdate>2018</pubdate>
  <volume>9</volume>
  <issue>1</issue>
  <fpage>4812</fpage>
</bibl>

<bibl id="B51">
  <title><p>{Quantum algorithms for electronic structure calculations:
  Particle-hole Hamiltonian and optimized wave-function expansions}</p></title>
  <aug>
    <au><snm>Barkoutsos</snm><fnm>PK</fnm></au>
    <au><snm>Gonthier</snm><fnm>JF</fnm></au>
    <au><snm>Sokolov</snm><fnm>I</fnm></au>
    <au><snm>Moll</snm><fnm>N</fnm></au>
    <au><snm>Salis</snm><fnm>G</fnm></au>
    <au><snm>Fuhrer</snm><fnm>A</fnm></au>
    <au><snm>Ganzhorn</snm><fnm>M</fnm></au>
    <au><snm>Egger</snm><fnm>DJ</fnm></au>
    <au><snm>Troyer</snm><fnm>M</fnm></au>
    <au><snm>Mezzacapo</snm><fnm>A</fnm></au>
    <au><snm>Filipp</snm><fnm>S</fnm></au>
    <au><snm>Tavernelli</snm><fnm>I</fnm></au>
  </aug>
  <source>Phys. Rev. A</source>
  <pubdate>2018</pubdate>
  <volume>98</volume>
  <issue>2</issue>
  <fpage>022322</fpage>
</bibl>

<bibl id="B52">
  <title><p>Gate-Efficient Simulation of Molecular Eigenstates on a Quantum
  Computer</p></title>
  <aug>
    <au><snm>Ganzhorn</snm><fnm>M.</fnm></au>
    <au><snm>Egger</snm><fnm>D.J.</fnm></au>
    <au><snm>Barkoutsos</snm><fnm>P.</fnm></au>
    <au><snm>Ollitrault</snm><fnm>P.</fnm></au>
    <au><snm>Salis</snm><fnm>G.</fnm></au>
    <au><snm>Moll</snm><fnm>N.</fnm></au>
    <au><snm>Roth</snm><fnm>M.</fnm></au>
    <au><snm>Fuhrer</snm><fnm>A.</fnm></au>
    <au><snm>Mueller</snm><fnm>P.</fnm></au>
    <au><snm>Woerner</snm><fnm>S.</fnm></au>
    <au><snm>Tavernelli</snm><fnm>I.</fnm></au>
    <au><snm>Filipp</snm><fnm>S.</fnm></au>
  </aug>
  <source>Phys. Rev. Applied</source>
  <publisher>American Physical Society</publisher>
  <pubdate>2019</pubdate>
  <volume>11</volume>
  <fpage>044092</fpage>
  <url>https://link.aps.org/doi/10.1103/PhysRevApplied.11.044092</url>
</bibl>

<bibl id="B53">
  <title><p>An initialization strategy for addressing barren plateaus in
  parametrized quantum circuits</p></title>
  <aug>
    <au><snm>Grant</snm><fnm>E</fnm></au>
    <au><snm>Wossnig</snm><fnm>L</fnm></au>
    <au><snm>Ostaszewski</snm><fnm>M</fnm></au>
    <au><snm>Benedetti</snm><fnm>M</fnm></au>
  </aug>
  <source>{Quantum}</source>
  <publisher>{Verein zur F{\"{o}}rderung des Open Access Publizierens in den
  Quantenwissenschaften}</publisher>
  <pubdate>2019</pubdate>
  <volume>3</volume>
  <fpage>214</fpage>
  <url>https://doi.org/10.22331/q-2019-12-09-214</url>
</bibl>

<bibl id="B54">
  <title><p>Large gradients via correlation in random parameterized quantum
  circuits</p></title>
  <aug>
    <au><snm>Volkoff</snm><fnm>T</fnm></au>
    <au><snm>Coles</snm><fnm>PJ</fnm></au>
  </aug>
  <source>Quantum Sci. Technol.</source>
  <pubdate>2021</pubdate>
  <volume>6</volume>
  <issue>2</issue>
  <fpage>025008</fpage>
</bibl>

<bibl id="B55">
  <title><p>Generalized Unitary Coupled Cluster Wave functions for Quantum
  Computation</p></title>
  <aug>
    <au><snm>Lee</snm><fnm>J</fnm></au>
    <au><snm>Huggins</snm><fnm>WJ</fnm></au>
    <au><snm>Head Gordon</snm><fnm>M</fnm></au>
    <au><snm>Whaley</snm><fnm>KB</fnm></au>
  </aug>
  <source>J. Chem. Theory Comput.</source>
  <pubdate>2018</pubdate>
  <volume>15</volume>
  <issue>1</issue>
  <fpage>311</fpage>
  <lpage>-324</lpage>
</bibl>

<bibl id="B56">
  <title><p>Orbital optimized unitary coupled cluster theory for quantum
  computer</p></title>
  <aug>
    <au><snm>Mizukami</snm><fnm>W</fnm></au>
    <au><snm>Mitarai</snm><fnm>K</fnm></au>
    <au><snm>Nakagawa</snm><fnm>YO</fnm></au>
    <au><snm>Yamamoto</snm><fnm>T</fnm></au>
    <au><snm>Yan</snm><fnm>T</fnm></au>
    <au><snm>Ohnishi</snm><fnm>Yy</fnm></au>
  </aug>
  <source>Phys. Rev. Research</source>
  <publisher>American Physical Society</publisher>
  <pubdate>2020</pubdate>
  <volume>2</volume>
  <fpage>033421</fpage>
  <url>https://link.aps.org/doi/10.1103/PhysRevResearch.2.033421</url>
</bibl>

<bibl id="B57">
  <title><p>Resource-Efficient Chemistry on Quantum Computers with the
  Variational Quantum Eigensolver and the Double Unitary Coupled-Cluster
  Approach</p></title>
  <aug>
    <au><snm>Metcalf</snm><fnm>M</fnm></au>
    <au><snm>Bauman</snm><fnm>NP</fnm></au>
    <au><snm>Kowalski</snm><fnm>K</fnm></au>
    <au><snm>Jong</snm><fnm>WA</fnm></au>
  </aug>
  <source>J. Chem. Theory Comput.</source>
  <pubdate>2020</pubdate>
  <volume>16</volume>
  <issue>10</issue>
  <fpage>6165</fpage>
  <lpage>-6175</lpage>
  <url>https://doi.org/10.1021/acs.jctc.0c00421</url>
  <note>PMID: 32915568</note>
</bibl>

<bibl id="B58">
  <title><p>Properties of coupled-cluster equations originating in excitation
  sub-algebras</p></title>
  <aug>
    <au><snm>Kowalski</snm><fnm>K</fnm></au>
  </aug>
  <source>J. Chem. Phys.</source>
  <pubdate>2018</pubdate>
  <volume>148</volume>
  <issue>9</issue>
  <fpage>094104</fpage>
  <url>https://doi.org/10.1063/1.5010693</url>
</bibl>

<bibl id="B59">
  <title><p>Hybrid quantum-classical hierarchy for mitigation of decoherence
  and determination of excited states</p></title>
  <aug>
    <au><snm>McClean</snm><fnm>JR</fnm></au>
    <au><snm>Kimchi Schwartz</snm><fnm>ME</fnm></au>
    <au><snm>Carter</snm><fnm>J</fnm></au>
    <au><snm>Jong</snm><fnm>WA</fnm></au>
  </aug>
  <source>Phys. Rev. A</source>
  <publisher>American Physical Society</publisher>
  <pubdate>2017</pubdate>
  <volume>95</volume>
  <fpage>042308</fpage>
  <url>https://link.aps.org/doi/10.1103/PhysRevA.95.042308</url>
</bibl>

<bibl id="B60">
  <title><p>Increasing the Representation Accuracy of Quantum Simulations of
  Chemistry without Extra Quantum Resources</p></title>
  <aug>
    <au><snm>Takeshita</snm><fnm>T</fnm></au>
    <au><snm>Rubin</snm><fnm>NC</fnm></au>
    <au><snm>Jiang</snm><fnm>Z</fnm></au>
    <au><snm>Lee</snm><fnm>E</fnm></au>
    <au><snm>Babbush</snm><fnm>R</fnm></au>
    <au><snm>McClean</snm><fnm>JR</fnm></au>
  </aug>
  <source>Phys. Rev. X</source>
  <publisher>American Physical Society</publisher>
  <pubdate>2020</pubdate>
  <volume>10</volume>
  <fpage>011004</fpage>
  <url>https://link.aps.org/doi/10.1103/PhysRevX.10.011004</url>
</bibl>

<bibl id="B61">
  <title><p>Determining eigenstates and thermal states on a quantum computer
  using quantum imaginary time evolution</p></title>
  <aug>
    <au><snm>Motta</snm><fnm>M</fnm></au>
    <au><snm>Sun</snm><fnm>C</fnm></au>
    <au><snm>Tan</snm><fnm>AT</fnm></au>
    <au><snm>O’Rourke</snm><fnm>MJ</fnm></au>
    <au><snm>Ye</snm><fnm>E</fnm></au>
    <au><snm>Minnich</snm><fnm>AJ</fnm></au>
    <au><snm>Brandão</snm><fnm>FG</fnm></au>
    <au><snm>Chan</snm><fnm>GKL</fnm></au>
  </aug>
  <source>Nat. Phys.</source>
  <pubdate>2019</pubdate>
  <volume>16</volume>
  <issue>2</issue>
  <url>https://par.nsf.gov/biblio/10194990</url>
</bibl>

<bibl id="B62">
  <title><p>A non-orthogonal variational quantum eigensolver</p></title>
  <aug>
    <au><snm>Huggins</snm><fnm>WJ</fnm></au>
    <au><snm>Lee</snm><fnm>J</fnm></au>
    <au><snm>Baek</snm><fnm>U</fnm></au>
    <au><snm>O'Gorman</snm><fnm>B</fnm></au>
    <au><snm>Whaley</snm><fnm>KB</fnm></au>
  </aug>
  <source>New J. Phys.</source>
  <publisher>{IOP} Publishing</publisher>
  <pubdate>2020</pubdate>
  <volume>22</volume>
  <issue>7</issue>
  <fpage>073009</fpage>
  <url>https://doi.org/10.1088/1367-2630/ab867b</url>
</bibl>

<bibl id="B63">
  <title><p>Quantum Filter Diagonalization: Quantum Eigendecomposition without
  Full Quantum Phase Estimation</p></title>
  <aug>
    <au><snm>Parrish</snm><fnm>RM</fnm></au>
    <au><snm>McMahon</snm><fnm>PL</fnm></au>
  </aug>
  <source>arXiv</source>
  <pubdate>2019</pubdate>
</bibl>

<bibl id="B64">
  <title><p>Quantum equation of motion for computing molecular excitation
  energies on a noisy quantum processor</p></title>
  <aug>
    <au><snm>Ollitrault</snm><fnm>PJ</fnm></au>
    <au><snm>Kandala</snm><fnm>A</fnm></au>
    <au><snm>Chen</snm><fnm>CF</fnm></au>
    <au><snm>Barkoutsos</snm><fnm>PK</fnm></au>
    <au><snm>Mezzacapo</snm><fnm>A</fnm></au>
    <au><snm>Pistoia</snm><fnm>M</fnm></au>
    <au><snm>Sheldon</snm><fnm>S</fnm></au>
    <au><snm>Woerner</snm><fnm>S</fnm></au>
    <au><snm>Gambetta</snm><fnm>JM</fnm></au>
    <au><snm>Tavernelli</snm><fnm>I</fnm></au>
  </aug>
  <source>Phys. Rev. Research</source>
  <publisher>American Physical Society</publisher>
  <pubdate>2020</pubdate>
  <volume>2</volume>
  <fpage>043140</fpage>
  <url>https://link.aps.org/doi/10.1103/PhysRevResearch.2.043140</url>
</bibl>

<bibl id="B65">
  <title><p>A Multireference Quantum {Krylov} Algorithm for Strongly Correlated
  Electrons</p></title>
  <aug>
    <au><snm>Stair</snm><fnm>NH</fnm></au>
    <au><snm>Huang</snm><fnm>R</fnm></au>
    <au><snm>Evangelista</snm><fnm>FA</fnm></au>
  </aug>
  <source>J. Chem. Theory Comput.</source>
  <pubdate>2020</pubdate>
  <volume>16</volume>
  <issue>4</issue>
  <fpage>2236</fpage>
  <lpage>2245</lpage>
  <url>https://doi.org/10.1021/acs.jctc.9b01125</url>
  <note>PMID: 32091895</note>
</bibl>

<bibl id="B66">
  <title><p>Quantum simulations employing connected moments
  expansions</p></title>
  <aug>
    <au><snm>Kowalski</snm><fnm>K</fnm></au>
    <au><snm>Peng</snm><fnm>B</fnm></au>
  </aug>
  <source>J. Chem. Phys.</source>
  <pubdate>2020</pubdate>
  <volume>153</volume>
  <issue>20</issue>
  <fpage>201102</fpage>
  <url>https://doi.org/10.1063/5.0030688</url>
</bibl>

<bibl id="B67">
  <title><p>Variational quantum solver employing the {PDS} energy
  functional</p></title>
  <aug>
    <au><snm>Peng</snm><fnm>B</fnm></au>
    <au><snm>Kowalski</snm><fnm>K</fnm></au>
  </aug>
  <source>arXiv</source>
  <pubdate>2021</pubdate>
</bibl>

<bibl id="B68">
  <title><p>The t expansion: a nonperturbative analytic tool for Hamiltonian
  systems</p></title>
  <aug>
    <au><snm>Horn</snm><fnm>D</fnm></au>
    <au><snm>Weinstein</snm><fnm>M</fnm></au>
  </aug>
  <source>Phys. Rev. D</source>
  <publisher>APS</publisher>
  <pubdate>1984</pubdate>
  <volume>30</volume>
  <issue>6</issue>
  <fpage>1256</fpage>
</bibl>

<bibl id="B69">
  <title><p>Connected moments expansion: a new tool for quantum many-body
  theory</p></title>
  <aug>
    <au><snm>Cioslowski</snm><fnm>J</fnm></au>
  </aug>
  <source>Phys. Rev. Lett.</source>
  <publisher>APS</publisher>
  <pubdate>1987</pubdate>
  <volume>58</volume>
  <issue>2</issue>
  <fpage>83</fpage>
</bibl>

<bibl id="B70">
  <title><p>Upper bounds for the free energy. A generalisation of the
  {Bogolubov} inequality and the Feynman inequality</p></title>
  <aug>
    <au><snm>Peeters</snm><fnm>FM</fnm></au>
    <au><snm>Devreese</snm><fnm>JT</fnm></au>
  </aug>
  <source>J. Phys. A</source>
  <publisher>IOP Publishing</publisher>
  <pubdate>1984</pubdate>
  <volume>17</volume>
  <issue>3</issue>
  <fpage>625</fpage>
</bibl>

<bibl id="B71">
  <title><p>Generalized variational principle in quantum mechanics</p></title>
  <aug>
    <au><snm>Soldatov</snm><fnm>AV</fnm></au>
  </aug>
  <source>Int. J. Mod. Phys. B</source>
  <publisher>World Scientific</publisher>
  <pubdate>1995</pubdate>
  <volume>9</volume>
  <issue>22</issue>
  <fpage>2899</fpage>
  <lpage>-2936</lpage>
</bibl>

<bibl id="B72">
  <title><p>Contracted Schr\"odinger equation: Determining quantum energies and
  two-particle density matrices without wave functions</p></title>
  <aug>
    <au><snm>Mazziotti</snm><fnm>DA</fnm></au>
  </aug>
  <source>Phys. Rev. A</source>
  <publisher>American Physical Society</publisher>
  <pubdate>1998</pubdate>
  <volume>57</volume>
  <fpage>4219</fpage>
  <lpage>-4234</lpage>
  <url>https://link.aps.org/doi/10.1103/PhysRevA.57.4219</url>
</bibl>

<bibl id="B73">
  <title><p>{Variational method for solving the contracted Schr{\"{o}}dinger
  equation through a projection of the N -particle power method onto the
  two-particle space}</p></title>
  <aug>
    <au><snm>Mazziotti</snm><fnm>DA</fnm></au>
  </aug>
  <source>J. Chem. Phys.</source>
  <pubdate>2002</pubdate>
  <volume>116</volume>
  <issue>4</issue>
  <fpage>1239</fpage>
  <lpage>-1249</lpage>
  <url>http://aip.scitation.org/doi/10.1063/1.1430257</url>
</bibl>

<bibl id="B74">
  <title><p>Exactness of wave functions from two-body exponential
  transformations in many-body quantum theory</p></title>
  <aug>
    <au><snm>Mazziotti</snm><fnm>DA</fnm></au>
  </aug>
  <source>Phys. Rev. A</source>
  <publisher>American Physical Society</publisher>
  <pubdate>2004</pubdate>
  <volume>69</volume>
  <fpage>012507</fpage>
  <url>https://link.aps.org/doi/10.1103/PhysRevA.69.012507</url>
</bibl>

<bibl id="B75">
  <title><p>Exact two-body expansion of the many-particle wave
  function</p></title>
  <aug>
    <au><snm>Mazziotti</snm><fnm>DA</fnm></au>
  </aug>
  <source>Phys. Rev. A</source>
  <publisher>American Physical Society</publisher>
  <pubdate>2020</pubdate>
  <volume>102</volume>
  <fpage>030802</fpage>
  <url>https://link.aps.org/doi/10.1103/PhysRevA.102.030802</url>
</bibl>

<bibl id="B76">
  <title><p>Anti-Hermitian Contracted Schr\"odinger Equation: Direct
  Determination of the Two-Electron Reduced Density Matrices of Many-Electron
  Molecules</p></title>
  <aug>
    <au><snm>Mazziotti</snm><fnm>DA</fnm></au>
  </aug>
  <source>Phys. Rev. Lett.</source>
  <publisher>American Physical Society</publisher>
  <pubdate>2006</pubdate>
  <volume>97</volume>
  <fpage>143002</fpage>
  <url>https://link.aps.org/doi/10.1103/PhysRevLett.97.143002</url>
</bibl>

<bibl id="B77">
  <title><p>Anti-Hermitian part of the contracted Schr\"odinger equation for
  the direct calculation of two-electron reduced density matrices</p></title>
  <aug>
    <au><snm>Mazziotti</snm><fnm>DA</fnm></au>
  </aug>
  <source>Phys. Rev. A</source>
  <publisher>American Physical Society</publisher>
  <pubdate>2007</pubdate>
  <volume>75</volume>
  <fpage>022505</fpage>
  <url>https://link.aps.org/doi/10.1103/PhysRevA.75.022505</url>
</bibl>

<bibl id="B78">
  <title><p>Direct calculation of excited-state electronic energies and
  two-electron reduced density matrices from the anti-Hermitian contracted
  Schr\"odinger equation</p></title>
  <aug>
    <au><snm>Gidofalvi</snm><fnm>G</fnm></au>
    <au><snm>Mazziotti</snm><fnm>DA</fnm></au>
  </aug>
  <source>Phys. Rev. A</source>
  <publisher>American Physical Society</publisher>
  <pubdate>2009</pubdate>
  <volume>80</volume>
  <fpage>022507</fpage>
  <url>https://link.aps.org/doi/10.1103/PhysRevA.80.022507</url>
</bibl>

<bibl id="B79">
  <title><p>Irreducible Brillouin conditions and contracted Schr{\"{o}}dinger
  equations for n -electron systems. I. The equations satisfied by the density
  cumulants</p></title>
  <aug>
    <au><snm>Mukherjee</snm><fnm>D</fnm></au>
    <au><snm>Kutzelnigg</snm><fnm>W</fnm></au>
  </aug>
  <source>J. Chem. Phys.</source>
  <pubdate>2001</pubdate>
  <volume>114</volume>
  <issue>5</issue>
  <fpage>2047</fpage>
  <lpage>-2061</lpage>
  <url>http://aip.scitation.org/doi/10.1063/1.1337058</url>
</bibl>

<bibl id="B80">
  <title><p>{Quantum Solver of Contracted Eigenvalue Equations for Scalable
  Molecular Simulations on Quantum Computing Devices}</p></title>
  <aug>
    <au><snm>Smart</snm><fnm>SE</fnm></au>
    <au><snm>Mazziotti</snm><fnm>DA</fnm></au>
  </aug>
  <source>Phys. Rev. Lett.</source>
  <pubdate>2021</pubdate>
  <volume>126</volume>
  <issue>7</issue>
  <fpage>070504</fpage>
</bibl>

<bibl id="B81">
  <title><p>{Resolving Correlated States of Benzyne on a Quantum Computer with
  an Error-Mitigated Quantum Contracted Eigenvalue Solver}</p></title>
  <aug>
    <au><snm>Smart</snm><fnm>SE</fnm></au>
    <au><snm>Boyn</snm><fnm>JN</fnm></au>
    <au><snm>Mazziotti</snm><fnm>DA</fnm></au>
  </aug>
  <pubdate>2021</pubdate>
  <url>http://arxiv.org/abs/2103.06876</url>
</bibl>

<bibl id="B82">
  <title><p>{An adaptive variational algorithm for exact molecular simulations
  on a quantum computer}</p></title>
  <aug>
    <au><snm>Grimsley</snm><fnm>HR</fnm></au>
    <au><snm>Economou</snm><fnm>SE</fnm></au>
    <au><snm>Barnes</snm><fnm>E</fnm></au>
    <au><snm>Mayhall</snm><fnm>NJ</fnm></au>
  </aug>
  <source>Nat. Commun.</source>
  <pubdate>2019</pubdate>
  <volume>10</volume>
  <issue>1</issue>
  <fpage>3007</fpage>
  <url>http://www.nature.com/articles/s41467-019-10988-2</url>
</bibl>

<bibl id="B83">
  <title><p>Generalized Unitary Coupled Cluster Wave functions for Quantum
  Computation</p></title>
  <aug>
    <au><snm>Lee</snm><fnm>J</fnm></au>
    <au><snm>Huggins</snm><fnm>WJ</fnm></au>
    <au><snm>Head Gordon</snm><fnm>M</fnm></au>
    <au><snm>Whaley</snm><fnm>KB</fnm></au>
  </aug>
  <source>J. Chem. Theory Comput.</source>
  <pubdate>2018</pubdate>
  <volume>15</volume>
  <issue>1</issue>
  <fpage>311</fpage>
  <lpage>-324</lpage>
</bibl>

<bibl id="B84">
  <title><p>{qubit-ADAPT-VQE: An adaptive algorithm for constructing
  hardware-efficient ansatze on a quantum processor}</p></title>
  <aug>
    <au><snm>Tang</snm><fnm>HL</fnm></au>
    <au><snm>Shkolnikov</snm><fnm>V O</fnm></au>
    <au><snm>Barron</snm><fnm>GS</fnm></au>
    <au><snm>Grimsley</snm><fnm>HR</fnm></au>
    <au><snm>Mayhall</snm><fnm>NJ</fnm></au>
    <au><snm>Barnes</snm><fnm>E</fnm></au>
    <au><snm>Economou</snm><fnm>SE</fnm></au>
  </aug>
  <source>arXiv</source>
  <pubdate>2019</pubdate>
</bibl>

<bibl id="B85">
  <title><p>Quantum computing methods for electronic states of the water
  molecule</p></title>
  <aug>
    <au><snm>Bian</snm><fnm>T</fnm></au>
    <au><snm>Murphy</snm><fnm>D</fnm></au>
    <au><snm>Xia</snm><fnm>R</fnm></au>
    <au><snm>Daskin</snm><fnm>A</fnm></au>
    <au><snm>Kais</snm><fnm>S</fnm></au>
  </aug>
  <source>Mol. Phys.</source>
  <publisher>Taylor & Francis</publisher>
  <pubdate>2019</pubdate>
  <volume>117</volume>
  <issue>15--16</issue>
  <fpage>2069</fpage>
  <lpage>2082</lpage>
  <url>https://doi.org/10.1080/00268976.2019.1580392</url>
</bibl>

<bibl id="B86">
  <title><p>Efficient symmetry-preserving state preparation circuits for the
  variational quantum eigensolver algorithm</p></title>
  <aug>
    <au><snm>Gard</snm><fnm>BT</fnm></au>
    <au><snm>Zhu</snm><fnm>L</fnm></au>
    <au><snm>Barron</snm><fnm>GS</fnm></au>
    <au><snm>Mayhall</snm><fnm>NJ</fnm></au>
    <au><snm>Economou</snm><fnm>SE</fnm></au>
    <au><snm>Barnes</snm><fnm>E</fnm></au>
  </aug>
  <source>npj Quantum Inf.</source>
  <pubdate>2020</pubdate>
  <volume>6</volume>
  <issue>1</issue>
  <fpage>10</fpage>
</bibl>

<bibl id="B87">
  <title><p>Learning in Modular Systems</p></title>
  <aug>
    <au><snm>Bradley</snm><fnm>DM</fnm></au>
    <au><snm>Bagnell</snm><fnm>JA</fnm></au>
    <au><snm>Bengio</snm><fnm>Y</fnm></au>
    <au><snm>Hebert</snm><fnm>M</fnm></au>
    <au><snm>De</snm><fnm>F</fnm></au>
    <au><snm>Torre</snm><fnm>L</fnm></au>
  </aug>
  <pubdate>2009</pubdate>
</bibl>

<bibl id="B88">
  <title><p>Failures of Gradient-Based Deep Learning</p></title>
  <aug>
    <au><snm>Shalev Shwartz</snm><fnm>S</fnm></au>
    <au><snm>Shamir</snm><fnm>O</fnm></au>
    <au><snm>Shammah</snm><fnm>S</fnm></au>
  </aug>
  <source>Proceedings of the 34th International Conference on Machine
  Learning</source>
  <publisher>International Convention Centre, Sydney, Australia:
  PMLR</publisher>
  <editor>Doina Precup and Yee Whye Teh</editor>
  <series><title><p>Proceedings of Machine Learning
  Research</p></title></series>
  <pubdate>2017</pubdate>
  <volume>70</volume>
  <fpage>3067</fpage>
  <lpage>-3075</lpage>
  <url>http://proceedings.mlr.press/v70/shalev-shwartz17a.html</url>
</bibl>

<bibl id="B89">
  <title><p>Field Guide to Dynamical Recurrent Networks</p></title>
  <aug>
    <au><snm>Kremer</snm><fnm>SC</fnm></au>
    <au><snm>Kolen</snm><fnm>JF</fnm></au>
  </aug>
  <publisher>Wiley-IEEE Press</publisher>
  <edition>1</edition>
  <pubdate>2001</pubdate>
</bibl>

<bibl id="B90">
  <title><p>Deep Learning</p></title>
  <aug>
    <au><snm>LeCun</snm><fnm>Y</fnm></au>
    <au><snm>Bengio</snm><fnm>Y</fnm></au>
    <au><snm>Hinton</snm><fnm>G</fnm></au>
  </aug>
  <source>Nature</source>
  <pubdate>2015</pubdate>
  <volume>521</volume>
  <issue>7553</issue>
  <fpage>436</fpage>
  <lpage>-444</lpage>
  <url>https://doi.org/10.1038/nature14539</url>
</bibl>

<bibl id="B91">
  <title><p>Batch Normalization: Accelerating Deep Network Training by Reducing
  Internal Covariate Shift</p></title>
  <aug>
    <au><snm>Ioffe</snm><fnm>S</fnm></au>
    <au><snm>Szegedy</snm><fnm>C</fnm></au>
  </aug>
  <source>Proceedings of the 32nd International Conference on Machine
  Learning</source>
  <publisher>Lille, France: PMLR</publisher>
  <editor>Francis Bach and David Blei</editor>
  <series><title><p>Proceedings of Machine Learning
  Research</p></title></series>
  <pubdate>2015</pubdate>
  <volume>37</volume>
  <fpage>448</fpage>
  <lpage>-456</lpage>
  <url>http://proceedings.mlr.press/v37/ioffe15.html</url>
</bibl>

<bibl id="B92">
  <title><p>A Fast Learning Algorithm for Deep Belief Nets</p></title>
  <aug>
    <au><snm>Hinton</snm><fnm>GE</fnm></au>
    <au><snm>Osindero</snm><fnm>S</fnm></au>
    <au><snm>Teh</snm><fnm>YW</fnm></au>
  </aug>
  <source>Neural Comput.</source>
  <pubdate>2006</pubdate>
  <volume>18</volume>
  <issue>7</issue>
  <fpage>1527</fpage>
  <lpage>1554</lpage>
  <url>https://doi.org/10.1162/neco.2006.18.7.1527</url>
  <note>PMID: 16764513</note>
</bibl>

<bibl id="B93">
  <title><p>Deep Residual Learning for Image Recognition</p></title>
  <aug>
    <au><snm>{He}</snm><fnm>K.</fnm></au>
    <au><snm>{Zhang}</snm><fnm>X.</fnm></au>
    <au><snm>{Ren}</snm><fnm>S.</fnm></au>
    <au><snm>{Sun}</snm><fnm>J.</fnm></au>
  </aug>
  <source>2016 IEEE Conference on Computer Vision and Pattern Recognition
  (CVPR)</source>
  <pubdate>2016</pubdate>
  <fpage>770</fpage>
  <lpage>-778</lpage>
</bibl>

<bibl id="B94">
  <title><p>Constrained Variational Quantum Eigensolver: Quantum Computer
  Search Engine in the {Fock space}</p></title>
  <aug>
    <au><snm>Ryabinkin</snm><fnm>IG</fnm></au>
    <au><snm>Genin</snm><fnm>SN</fnm></au>
    <au><snm>Izmaylov</snm><fnm>AF</fnm></au>
  </aug>
  <source>J. Chem. Theory Comput.</source>
  <publisher>American Chemical Society</publisher>
  <pubdate>2019</pubdate>
  <volume>15</volume>
  <issue>1</issue>
  <fpage>249</fpage>
  <lpage>-255</lpage>
  <url>https://doi.org/10.1021/acs.jctc.8b00943</url>
</bibl>

<bibl id="B95">
  <title><p>Efficient quantum algorithm for preparing molecular-system-like
  states on a quantum computer</p></title>
  <aug>
    <au><snm>Wang</snm><fnm>H</fnm></au>
    <au><snm>Ashhab</snm><fnm>S.</fnm></au>
    <au><snm>Nori</snm><fnm>F</fnm></au>
  </aug>
  <source>Phys. Rev. A</source>
  <publisher>American Physical Society</publisher>
  <pubdate>2009</pubdate>
  <volume>79</volume>
  <fpage>042335</fpage>
  <url>https://link.aps.org/doi/10.1103/PhysRevA.79.042335</url>
</bibl>

<bibl id="B96">
  <title><p>Analysis of a parametrically driven exchange-type gate and a
  two-photon excitation gate between superconducting qubits</p></title>
  <aug>
    <au><snm>Roth</snm><fnm>M</fnm></au>
    <au><snm>Ganzhorn</snm><fnm>M</fnm></au>
    <au><snm>Moll</snm><fnm>N</fnm></au>
    <au><snm>Filipp</snm><fnm>S</fnm></au>
    <au><snm>Salis</snm><fnm>G</fnm></au>
    <au><snm>Schmidt</snm><fnm>S</fnm></au>
  </aug>
  <source>Phys. Rev. A</source>
  <publisher>American Physical Society</publisher>
  <pubdate>2017</pubdate>
  <volume>96</volume>
  <fpage>062323</fpage>
  <url>https://link.aps.org/doi/10.1103/PhysRevA.96.062323</url>
</bibl>

<bibl id="B97">
  <title><p>Entanglement Generation in Superconducting Qubits Using Holonomic
  Operations</p></title>
  <aug>
    <au><snm>Egger</snm><fnm>D.J.</fnm></au>
    <au><snm>Ganzhorn</snm><fnm>M.</fnm></au>
    <au><snm>Salis</snm><fnm>G.</fnm></au>
    <au><snm>Fuhrer</snm><fnm>A.</fnm></au>
    <au><snm>M\"uller</snm><fnm>P.</fnm></au>
    <au><snm>Barkoutsos</snm><fnm>P</fnm></au>
    <au><snm>Moll</snm><fnm>N.</fnm></au>
    <au><snm>Tavernelli</snm><fnm>I.</fnm></au>
    <au><snm>Filipp</snm><fnm>S.</fnm></au>
  </aug>
  <source>Phys. Rev. Applied</source>
  <publisher>American Physical Society</publisher>
  <pubdate>2019</pubdate>
  <volume>11</volume>
  <fpage>014017</fpage>
  <url>https://link.aps.org/doi/10.1103/PhysRevApplied.11.014017</url>
</bibl>

<bibl id="B98">
  <title><p>Experimental error mitigation via symmetry verification in a
  variational quantum eigensolver</p></title>
  <aug>
    <au><snm>Sagastizabal</snm><fnm>R.</fnm></au>
    <au><snm>Bonet Monroig</snm><fnm>X.</fnm></au>
    <au><snm>Singh</snm><fnm>M.</fnm></au>
    <au><snm>Rol</snm><fnm>M. A.</fnm></au>
    <au><snm>Bultink</snm><fnm>C. C.</fnm></au>
    <au><snm>Fu</snm><fnm>X.</fnm></au>
    <au><snm>Price</snm><fnm>C. H.</fnm></au>
    <au><snm>Ostroukh</snm><fnm>V. P.</fnm></au>
    <au><snm>Muthusubramanian</snm><fnm>N.</fnm></au>
    <au><snm>Bruno</snm><fnm>A.</fnm></au>
    <au><snm>Beekman</snm><fnm>M.</fnm></au>
    <au><snm>Haider</snm><fnm>N.</fnm></au>
    <au><snm>O'Brien</snm><fnm>T. E.</fnm></au>
    <au><snm>DiCarlo</snm><fnm>L.</fnm></au>
  </aug>
  <source>Phys. Rev. A</source>
  <publisher>American Physical Society</publisher>
  <pubdate>2019</pubdate>
  <volume>100</volume>
  <fpage>010302</fpage>
  <url>https://link.aps.org/doi/10.1103/PhysRevA.100.010302</url>
</bibl>

<bibl id="B99">
  <title><p>Optimizing parametrized quantum circuits via noise-induced breaking
  of symmetries</p></title>
  <aug>
    <au><snm>Fontana</snm><fnm>E</fnm></au>
    <au><snm>Cerezo</snm><fnm>M.</fnm></au>
    <au><snm>Arrasmith</snm><fnm>A</fnm></au>
    <au><snm>Rungger</snm><fnm>I</fnm></au>
    <au><snm>Coles</snm><fnm>PJ</fnm></au>
  </aug>
  <source>arXiv</source>
  <pubdate>2020</pubdate>
</bibl>

<bibl id="B100">
  <title><p>Natural Evolutionary Strategies for Variational Quantum
  Computation</p></title>
  <aug>
    <au><snm>Anand</snm><fnm>A</fnm></au>
    <au><snm>Degroote</snm><fnm>M</fnm></au>
    <au><snm>Aspuru Guzik</snm><fnm>A</fnm></au>
  </aug>
  <source>arXiv</source>
  <pubdate>2020</pubdate>
</bibl>

<bibl id="B101">
  <title><p>Effect of barren plateaus on gradient-free optimization</p></title>
  <aug>
    <au><snm>Arrasmith</snm><fnm>A</fnm></au>
    <au><snm>Cerezo</snm><fnm>M.</fnm></au>
    <au><snm>Czarnik</snm><fnm>P</fnm></au>
    <au><snm>Cincio</snm><fnm>L</fnm></au>
    <au><snm>Coles</snm><fnm>PJ</fnm></au>
  </aug>
  <source>arXiv</source>
  <pubdate>2020</pubdate>
</bibl>

<bibl id="B102">
  <title><p>On barren plateaus and cost function locality in variational
  quantum algorithms</p></title>
  <aug>
    <au><snm>Uvarov</snm><fnm>A</fnm></au>
    <au><snm>Biamonte</snm><fnm>J</fnm></au>
  </aug>
  <source>arXiv</source>
  <pubdate>2020</pubdate>
</bibl>

<bibl id="B103">
  <title><p>Toward Trainability of Quantum Neural Networks</p></title>
  <aug>
    <au><snm>Zhang</snm><fnm>K</fnm></au>
    <au><snm>Hsieh</snm><fnm>MH</fnm></au>
    <au><snm>Liu</snm><fnm>L</fnm></au>
    <au><snm>Tao</snm><fnm>D</fnm></au>
  </aug>
  <source>arXiv</source>
  <pubdate>2020</pubdate>
</bibl>

<bibl id="B104">
  <title><p>Absence of Barren Plateaus in Quantum Convolutional Neural
  Networks</p></title>
  <aug>
    <au><snm>Pesah</snm><fnm>A</fnm></au>
    <au><snm>Cerezo</snm><fnm>M.</fnm></au>
    <au><snm>Wang</snm><fnm>S</fnm></au>
    <au><snm>Volkoff</snm><fnm>T</fnm></au>
    <au><snm>Sornborger</snm><fnm>AT</fnm></au>
    <au><snm>Coles</snm><fnm>PJ</fnm></au>
  </aug>
  <source>arXiv</source>
  <pubdate>2020</pubdate>
</bibl>

<bibl id="B105">
  <title><p>{Exact and approximate symmetry projectors for the electronic
  structure problem on a quantum computer}</p></title>
  <aug>
    <au><snm>Yen</snm><fnm>TC</fnm></au>
    <au><snm>Lang</snm><fnm>RA</fnm></au>
    <au><snm>Izmaylov</snm><fnm>AF</fnm></au>
  </aug>
  <source>J. Chem. Phys.</source>
  <pubdate>2019</pubdate>
  <volume>151</volume>
  <issue>16</issue>
  <fpage>164111</fpage>
</bibl>

<bibl id="B106">
  <title><p>Iterative Qubit Coupled Cluster Approach with Efficient Screening
  of Generators</p></title>
  <aug>
    <au><snm>Ryabinkin</snm><fnm>IG</fnm></au>
    <au><snm>Lang</snm><fnm>RA</fnm></au>
    <au><snm>Genin</snm><fnm>SN</fnm></au>
    <au><snm>Izmaylov</snm><fnm>AF</fnm></au>
  </aug>
  <source>J. Chem. Theory Comput.</source>
  <pubdate>2020</pubdate>
  <volume>16</volume>
  <issue>2</issue>
  <fpage>1055</fpage>
  <lpage>-1063</lpage>
</bibl>

<bibl id="B107">
  <title><p>Unitary Transformation of the Electronic Hamiltonian with an Exact
  Quadratic Truncation of the Baker-Campbell-Hausdorff Expansion</p></title>
  <aug>
    <au><snm>Lang</snm><fnm>RA</fnm></au>
    <au><snm>Ryabinkin</snm><fnm>IG</fnm></au>
    <au><snm>Izmaylov</snm><fnm>AF</fnm></au>
  </aug>
  <source>J. Chem. Theory Comput.</source>
  <pubdate>2021</pubdate>
  <volume>17</volume>
  <issue>1</issue>
  <fpage>66</fpage>
  <lpage>78</lpage>
</bibl>

</refgrp>
} 


\end{backmatter}
\end{document}